\documentclass[10pt,letterpaper]{article}
\usepackage{authblk}
\usepackage{amsmath}
\usepackage{graphicx}
\usepackage{subcaption}
\usepackage{gensymb}
\usepackage{nameref}
\usepackage{array}
\usepackage{color}
\usepackage{array}
\usepackage{wrapfig}
\usepackage{textcomp}
\usepackage{amsmath}
\usepackage{amssymb}
\usepackage{cancel}
\usepackage{graphicx}
\usepackage{MnSymbol,wasysym}
\usepackage{comment}
\usepackage{mathtools}
\usepackage{algorithm}
\usepackage{algpseudocode}
\usepackage{longtable}
\usepackage{multirow}
\usepackage{setspace}
\usepackage{etoolbox}
\usepackage{booktabs}
\usepackage{hyperref}
\usepackage[noabbrev]{cleveref}
\usepackage{xfrac}
\usepackage{nicefrac}
\usepackage{subcaption}
\usepackage{xcolor}
\usepackage{placeins}
\usepackage{gensymb}
\usepackage{authblk}
 \usepackage[margin=1in]{geometry}
\makeatletter
\renewcommand{\maketitle}{\bgroup\setlength{\parindent}{0pt}
\begin{flushleft}
  \textbf{\@title}

  \@author
\end{flushleft}\egroup
}

\makeatletter
\renewcommand{\maketitle}{\bgroup\setlength{\parindent}{0pt}
\begin{flushleft}
  \textbf{\@title}

  \@author
\end{flushleft}\egroup
}
\makeatother
\usepackage[margin=1in]{geometry}
\usepackage[none]{hyphenat} 
\usepackage{array}          

\newcolumntype{P}[1]{>{\raggedright\arraybackslash}p{#1}}

\title{Mathematical modeling of 1,2-propanediol utilization bacterial microcompartments {\it in vivo} activity}

\author[1]{Andre G. Archer}
\author[2]{Charlotte H. Abrahamson}
\author[3]{Brett J. Palmero}
\author[2]{Elizabeth R. Johnson}
\author[2]{Carolyn Mills}
\author[2, 3]{Nolan Kennedy}
\author[2,3,4]{Danielle Tullman-Ercek}
\author[1,4,*]{Niall Mangan}

\affil[1]{ Northwestern University, Department of Engineering Sciences and Applied Mathematics, Evanston, IL, USA}
\affil[2]{Northwestern University, Department of Chemical and Biological Engineering, Evanston, IL, USA}
\affil[3]{Northwestern University, Interdisciplinary Biological Sciences Program, Evanston, IL, USA}
\affil[4]{Northwestern University, Center for Synthetic Biology, Evanston, IL, USA}

\affil[*]{niall.mangan@northwestern.edu}

\begin{document}

\flushbottom
\maketitle
\begin{abstract}

On exposure to 1,2-propanediol (1,2-PD), {\it Salmonella enterica} serovar Typhimurium LT2 produces 1,2-PD utilization (Pdu) microcompartments (MCPs), nanoscale protein-bound shells that encapsulate metabolic enzymes. MCPs serve as a bioengineering platform to study reaction organization and enhance flux through specific pathways. However, a recently published assay of purified wild-type (WT) MCPs reported metabolic activity that differed markedly from that observed {\it in vivo}. Using kinetic modeling, we attribute these discrepancies to {\it in vivo} cell growth and to the cytosolic presence of MCP-associated enzymes and promiscuous alcohol dehydrogenases, which are not present in the purified MCPs. Assays of purified MCPs in {\it E. coli} lysate, together with a LT2 growth assay in which the native Pdu MCP-associated alcohol dehydrogenase, PduQ, was knocked out, support the conclusion that exogenous Pdu cytosolic enzyme activity can narrow the gap between {\it in vitro} and {\it in vivo} experiments. Our modeling further suggests that MCP-localized enzymes contribute little to {\it in vivo} metabolic flux downstream of PduCDE. We therefore propose a revised {\it in vivo} model of WT growth on 1,2-PD in which PduCDE is fully encapsulated, while much of the downstream Pdu activity occurs in the cytosol.


\end{abstract}

%
%
\thispagestyle{empty}

\section*{Introduction}


{\it Salmonella enterica} serovar Typhimurium LT2 produces bacterial microcompartments (MCPs) upon exposure to 1,2-propanediol (1,2-PD). 1,2-{\bf P}ropane{\bf d}iol {\bf u}tilization (Pdu) MCPs are 100–150 nanometers in diameter and are composed of eight shell proteins (PduABB'NJKTU) \cite{Chowdhury2014, Kennedy2022, Mills2022}. The protein shells encapsulate a catalytic core consisting of nine unique proteins (PduCDEGHLOPQSVW) \cite{Yang2020}. Among these, the catalytic enzymes PduCDELPQW directly participate in 1,2-PD fermentation \cite{Chowdhury2014}. The Pdu core also encodes cobalamin adenylation \cite{Johnson2004, Cheng2010, Fan2008} and diol dehydratase reactivation \cite{Bobik1999}, reactions that are essential for optimal 1,2-PD fermentation. Biologically, MCPs are hypothesized to provide {\it Salmonella} with a growth advantage by enabling 1,2-PD metabolism while limiting cellular exposure to propionaldehyde, a toxic intermediate of 1,2-PD fermentation \cite{Sampson2008}.

\begin{figure}[!ht]\centering
\includegraphics[width=1.\linewidth]{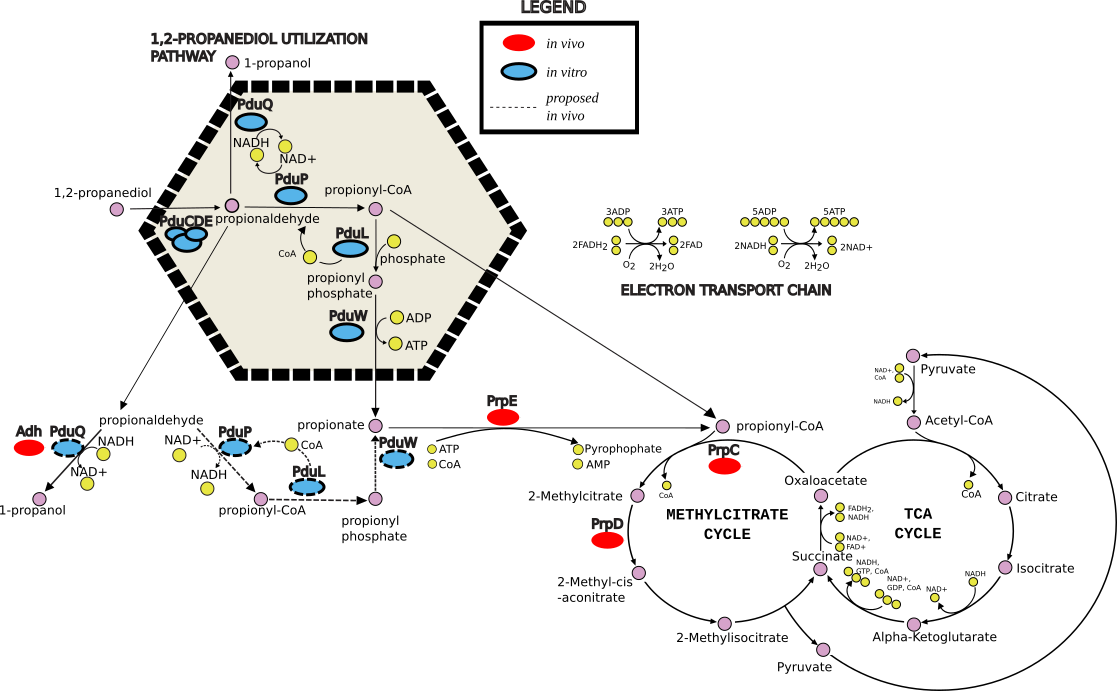}
\caption{Proposed reaction model of the Pdu MCPs with complete encapsulation of PduCDE and partial encapsulation of PduP, PduQ, PduL, and PduW. In this model, 1,2-propanediol is converted to propionyl-CoA, propionate, and 1-propanol. Propionate and 1-propanol act as shunts for excess carbon flux and are reconverted to propionyl-CoA. Propionyl-CoA is further metabolized to generate ATP via the 2-methylcitrate cycle, the tricarboxylic acid (TCA) cycle, and the electron transport chain \cite{Horswill1999a, Horswill2001b, Horswill1999a,Palacios2003, Dolan2018, Noster2019, Deshpande2022}. The findings presented in this study support partial encapsulation of PduP, PduQ, PduL, and PduW.}
\label{fig:Fig1}
\end{figure}

Pdu-mediated 1,2-PD fermentation consists of a five-reaction sequence (Figure \ref{fig:Fig1}). PduCDE first dehydrates 1,2-PD to propionaldehyde using adenosylcobalamin (AdoB\textsubscript{12}) as a cofactor \cite{Bobik1997, Toraya2000}. Propionaldehyde is then either converted to propionyl-coenyme A (propionyl-CoA) by PduP while converting nicotinamide adenine dinucleotide (NAD+) to nicotinamide adenine dinucleotide hydride (NADH) or to 1-propanol by PduQ while converting NADH to NAD+ \cite{Leal2003, Cheng2012}.  Propionyl-CoA subsequently serves as a carbon source for central carbon metabolism via the 2-methylcitrate cycle (Figure \ref{fig:Fig1}) \cite{Palacios2003}. Reported metabolite dynamics indicate that 1-propanol is reassimilated under 1,2-PD-limited conditions and thus likely acts as a shunt for excess carbon flux through the Pdu pathway \cite{Kennedy2022, Mills2022}.  PduL and PduW convert propionyl-CoA to propionate while generating one adenosine triphosphate (ATP) \cite{Liu2007, Palacios2003}. Similar to 1-propanol, propionate is observed to act as a temporary sink for excess flux and is reconverted to propionyl-CoA under 1,2-PD-limited conditions \cite{Kennedy2022, Mills2022}.

A recently designed assay measured the dynamics of a functional 1,2-PD enzymatic pathway in purified Pdu MCPs \cite{Palmero2025}. However, 1,2-PD metabolite pathway dynamics in purified MCPs were substantially different from those observed upon growth of wild-type (WT) LT2 {\it Salmonella} on 1,2-PD as a sole carbon source \cite{Sampson2008, Kennedy2022, Mills2022}. First, in growth assays, WT LT2 consumed 1,2-PD more slowly than corresponding MCP isolates: WT depleted 1,2-PD within 18 hours, whereas MCP isolates were previously observed to consume it within the first hour (Figure \ref{fig:Fig2}). Second, in WT LT2 cells, the Pdu pathway was more efficient than MCP isolates, in that a larger proportion of the 1,2-PD was converted to  1-propanol and propionate. At the point of peak propionaldehyde concentration in WT LT2 growth, 1\% of the originally-supplied 1,2-PD was converted to propionaldehyde while 42\% was converted to 1-propanol and propionate. In contrast, MCP isolates accumulated $\sim$75-fold higher levels of propionaldehyde and produced 6.4-fold and 1.5-fold lower levels of 1-propanol and propionate, respectively. Finally, WT LT2 ultimately consumed 1-propanol and propionate once 1,2-PD had been consumed, whereas purified MCPs showed no consumption of these metabolites. The goal of this work is to understand these differences. Our analysis suggests that enzymatic activity in the cytosol contributes more significantly {\it in vivo} than previously suspected.

It is critical to understand how MCPs contribute to {\it in vivo} growth dynamics for both the development of MCPs as metabolic engineering platforms and the identification of new antibiotic targets against {\it Salmonella} infections. Existing literature points to cytosolic enzyme activity as a potential source of the observed discrepancies. The encapsulation mechanisms of PduCDE, PduP, and PduL have been shown to mediate green fluorescent protein (GFP) encapsulation with varying efficiency, implying that PduCDE, PduP, and PduL localize to both MCPs and the cytosol at different rates \cite{Nichols2020}. However, the presence of Pdu enzymes in the cytosol has yet to be conclusively confirmed. Additionally, {\it Salmonella} expresses several housekeeping enzymes, including alcohol dehydrogenases (AdhE and AdhP), phosphotransacetylase (Pta), and acetate kinase (AckA) \cite{Huseby2013}. The precise contribution of these housekeeping enzymes to Pdu metabolite dynamics remains undetermined; however, they have been shown to sustain suboptimal growth on 1,2-PD and propionate \cite{Palacios2003, Liu2007}. 

To explore what levels of MCP and cytosolic enzyme activity are consistent with experimental observations of metabolite dynamics {\it in vivo} and {\it in vitro}, we constructed and calibrated a kinetic model of WT {\it Salmonella} growth on 1,2-PD (Figure \ref{fig:Fig2}). We have previously used mechanistic models to investigate the interactions between compartmentalization and metabolite dynamics {\it in vivo} \cite{Archer2025, Jakobson2017, Jakobson2018, Mills2022, Kennedy2022}. In this work, our {\it in vivo} model captures the spatial organization of Pdu MCPs and previously un-modeled cytosolic 2-methylcitrate cycle reactions and cell growth over time. Mass-action kinetics were used to describe all but one reaction. Kinetic parameters were constrained using quasi–steady-state relationships between mass-action and Michaelis–Menten measurements similar to our work in \cite{Archer2025, Palmero2025}. MCP enzyme number and the number of MCPs per cell were constrained to values reported in the literature \cite{Yang2020, Kennedy2022}.

Forward simulations of our {\it in vivo} model using calibrated {\it in vitro} parameters identified the specific sources of the discrepancies in metabolite profiles across {\it in vivo} and {\it in vitro} experiments. Importantly, we were able to decouple the factors that caused varying 1,2-PD consumption levels across assays from those that contributed to variation in 1-propanol and propionate metabolite profiles. First, we found that differences in the 1,2-PD consumption rate across {\it in vivo} and {\it in vitro} assays are attributable to biomass differences, due to cell growth, rather than differences in the PduCDE diol dehydratase across the two experimental systems. Because PduCDE behavior is consistent across {\it in vivo} and {\it in vitro} systems, we can infer that PduCDE is the primary diol dehydratase {\it in vivo} (which we know is the case {\it in vitro}), and that it is predominantly, if not completely, contained within the MCP. However, our model simulations with this initial model could not fully capture {\it in vivo} 1-propanol and propionate dynamics. 

This prompted us to pursue simultaneous calibration of our {\it in vivo} model with the published {\it in vivo} model \cite{Archer2025}. This calibration identified cytosolic coenzyme-A–acylating aldehyde dehydrogenase (ALDH), alcohol dehydrogenase (ADH), and phosphotransacetylase activity {\it in vivo} as the likely sources of differences in 1-propanol and propionate metabolite profiles. Analysis of model flux dynamics further indicated that much of the ALDH, ADH, and phosphotransacetylase activity during Pdu MCP metabolism must occur in the cytosol. Previously published work reports that PduP and PduL account for most of the ALDH and phosphotransacetylase activity \cite{Liu2007, Leal2003}, respectively, implying that cytosolic PduP and PduL are required to reproduce the {\it in vivo} metabolite dynamics observed experimentally. Overall, these results suggest that the lower Pdu pathway (transformations downstream of PduCDE) is primarily localized to the cytosol rather than the MCP. To validate the hypothesis generated by our model that exogenous enzymes are required to reconcile {\it in vitro} and {\it in vivo} metabolite activity, we carried out {\it in vitro} experiments with purified MCPs in {\it E. coli} lysate, where this lysate is expected to have a similar enzymatic composition to the {\it Salmonella} cytosol.

Unlike the case with PduP and PduL, growth assay data of $\Delta$pduQ knockout presented in this paper indicates that there is likely a promiscuous alcohol dehydrogenase contributing to 1-propanol assimilation and dissimilation. Given that the promiscuous cytosolic alcohol dehydrogenase has not been identified and its kinetics unknown, including the cytosolic alcohol dehydrogenase reaction in the model resulted in fits with the majority of the WT 1-propanol dynamics attributed to the unconstrained enzyme. The temporal profile of metabolites during the $\Delta$pduQ growth assay strongly suggest that $\Delta$pduQ suffers from a redox imbalance due to the lack of PduQ. We show that fully encapsulated PduQ contributes little to the total alcohol dehydrogenase activity and, thus, conclude that cytosolic expression of PduQ is necessary to alleviate the redox imbalance.

\section*{Results}


%

\subsection*{WT metabolite activity differed markedly from MCP isolates}

Bacterial microcompartments are hypothesized to provide a growth advantage to {\it Salmonella} by sequestering toxic propionaldehyde and enhancing flux toward downstream metabolites. Recently, Palmero et al. 2025 and Archer et al. 2025 reported metabolite time courses of purified WT MCPs exposed to 1,2-PD and showed that MCP isolates rapidly consume 1,2-PD while producing only limited amounts of 1-propanol and propionyl-phosphate. This behavior contrasts with {\it in vivo} assays, in which 1,2-PD is metabolized or converted into extracellular metabolites which the cell subsequently uses for growth \cite{Mills2022,Kennedy2022}. To confirm this previous observation, we conducted an {\it in vivo} assay of WT grown on 55 mM 1,2-PD and 150 nM Ado-B\textsubscript{12} (Figure \ref{fig:Fig2}B and SI Figure \ref{fig:Mode1PduPPduQAdhPduLPduWZoomed}B). Consistent with prior observations, WT exposed to excess Ado-B\textsubscript{12} exhibited uninterrupted growth on 1,2-PD (Figure \ref{fig:Fig2}Bv).

Cultures consumed all available 1,2-PD within 18 h, during which WT grew at a mean rate of 0.203 optical density units per second and produced net propionaldehyde, propionate, and 1-propanol. Once the 1,2-PD pool was depleted, WT reassimilated the remaining extracellular metabolites. 1-propanol reached the highest peak concentration and was consumed more slowly than the other intermediates: WT depleted 18 mM of 1-propanol within 33 h of its peak, compared with 0.5 mM propionaldehyde and 3 mM propionate within 3 h and 6 h, respectively. After exhausting all available metabolites, WT LT2 growth slowed and plateaued at 4.7 optical density units.

 We identified at least three notable differences between previously reported data on MCP isolates (hereafter referred to as the ``{\it in vitro} assay") and the {\it in vivo} assay reported in this paper (Figure \ref{fig:Fig2}A–B; Table \ref{table:ComparisonInVivoInVitro}) \cite{Palmero2025,Archer2025}. First, WT LT2 consumed 1,2-PD more slowly than MCP isolates, depleting 1,2-PD within 18 h compared within 1 h {\it in vitro}--a discrepancy that is likely due to biomass accumulation during the lag phase (the first 12 h) growth {\it in vivo} (Figure \ref{fig:Fig2}Bi, Bv). Second, the {\it in vivo} assay produced 1-propanol and propionate more efficiently than the {\it in vitro} assay. Peak {\it in vivo} 1-propanol and propionate were 6- and 1.5-fold higher, respectively, while peak propionaldehyde was 80-fold lower relative to the {\it in vitro} assay (Figure \ref{fig:Fig2}Aii–iv, Bii–iv). This yielded a 480-fold and 120-fold increase in the propionate-to-propionaldehyde and 1-propanol-to-propionaldehyde ratios, respectively. Finally, WT consumed both propionate and 1-propanol {\it in vivo}, whereas purified MCPs did not (Figure \ref{fig:Fig2}Aiii–iv, Biii–iv). While it is true that several metabolites, including propionate, are assimilated into the downstream methylcitrate cycle {\it in vivo}, we will show that these differences between {\it in vivo} and {\it in vitro} metabolite kinetic profiles extend beyond what can be explained by ``pull" from the methylcitrate cycle. Specifically, our results indicate that the observed discrepancies are likely explained by assay conditions, including the presence of promiscuous enzymes, aerobic versus microaerobic environments, and biomass accumulation.
\begin{table}\centering
\begin{tabular}{||P{3.5cm}P{1.75cm}P{1.5cm}P{4cm}P{4cm}||} 
 \hline
 Discrepancy & {\it in vitro} & {\it in vivo} & Proposed Explanation & Methods \\ [0.5ex] 
 \hline\hline
 1,2-PD consumption time & $\leq 1$ hour & $\leq 18$ hours & Biomass accumulation lag {\it in vivo} & {\it in vitro}/{\it in vivo} modeling\\
 \hline
 Maximum propionaldehyde &  40.8 mM & 0.551 mM  & Increased PduP and PduQ/ADH biomass {\it in vivo} & {\it in vitro}/{\it in vivo} modeling \\
  Maximum propionate & 2.02 mM &  3.04 mM  & Increased PduP and PduL biomass & {\it in vitro}/{\it in vivo} modeling \\
  Maximum 1-propanol & 2.87 mM &  17.9 mM & Increased PduQ/ADH biomass {\it in vivo} & {\it in vitro}/{\it in vivo} modeling and $\Delta$pduQ growth curve \\
 \hline
  1-propanol metabolism  & Little to no activity & Activity observed &   &  \\
   Propionate metabolism  & Little to no activity & Activity observed & PduW and PrpE activity {\it in vivo} & Prior research  \\ 
 [1ex] 
 \hline
\end{tabular}
\caption{Differences between {\it in vitro} and {\it in vivo} assays, proposed explanations for their origins, and methodology or evidence supporting these hypotheses.}
\label{table:ComparisonInVivoInVitro}
\end{table}
\subsection*{{\it In vivo} model with {\it in vitro} posterior parameters accurately predicts 1,2-PD dynamics}


Building on the Pdu MCP kinetic model of Archer et al. (2024), we developed a mass-action kinetic model of WT LT2 {\it in vivo} growth to reconcile differences between {\it in vivo} and {\it in vitro} assay results. Our model included important features of the {\it in vivo} assay: an increase in biomass over time, 2 to 10 MCPs per cell \cite{Kennedy2022}, Pdu MCP spatial organization within the cytosol, MCP and cytosol interactions, and downstream Prp enzymes in the cytosol. Unlike Archer et al. 2024, our {\it in vivo} model did not include inactivation of PduQ due to oxidation because the {\it in vivo} assay was conducted under microaerobic conditions. However, we did include inhibition of PduP by coenzyme A (CoA), which was important for achieving consistent {\it in vitro} fits in previous studies \cite{Archer2025}, since cytosolic CoA can bind to and inhibit PduP \cite{CHOHNAN1998, Bennett2009, Park2016}. A full description of the model is provided in Methods sections on the \nameref{sec:InVitroMathematicalModel} and the \nameref{sec:InVivoMathematicalModel}.

We tested whether our parameterized model accurately captured all potential {\it in vivo} effects by running {\it in vivo} model simulations with {\it in vitro} calibrated Pdu kinetic rates and MCP permeabilities as parameter estimates (Figure \ref{fig:Fig7}A). We then compared the simulation results with our {\it in vivo} experimental data. To ensure an accurate description of biomass growth and cytosolic activity, the simulations also used calibrated WT LT2 growth parameters and published cytosolic parameters, including Prp enzyme kinetic rates and the cytosolic NAD\textsuperscript{+}:NADH ratio. Details of the WT growth model and calibration are provided in section \nameref{ssec:ODGrowth}, and the {\it in vivo} parameter priors are described in section \nameref{ssec:parameterconstraints}.

\begin{figure}[!p]
\includegraphics[width=\linewidth]{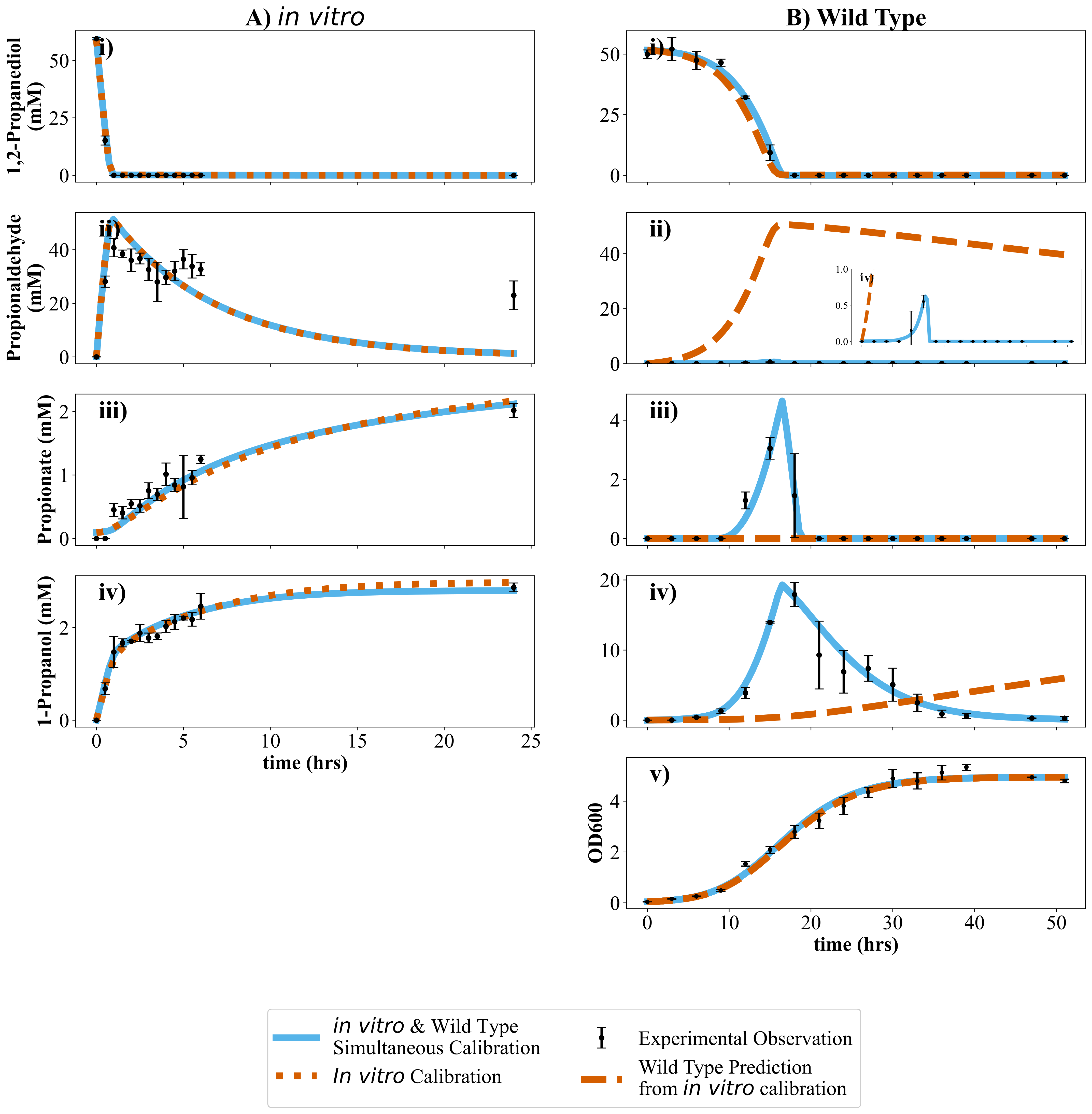}
    \caption{Model simulations of A) {\it in vitro} and B) {\it in vivo} metabolite dynamics for i) 1,2-propanediol, ii) propionaldehyde, iii) propionate, and iv) 1-propanol. Experimental time-series data were compared with three model configurations: {\it in vitro} calibrated parameters (short red dashed lines, Ai–iv), {\it in vitro} calibrated parameters with {\it in vivo} priors (long red dashed lines, Ai–iv), and jointly calibrated {\it in vitro} + WT parameters (solid blue lines, Ai–iv and Bi–vi).}
    \label{fig:Fig2}
\end{figure}

Forward simulations of the {\it in vivo} model with {\it in vitro} calibrated parameters reproduced 1,2-PD dynamics (Figure \ref{fig:Fig2}Bi) but failed to accurately capture downstream metabolite dynamics (Figure \ref{fig:Fig2}Bii–iv). Accurate prediction of 1,2-PD indicates that PduCDE kinetics were well captured {\it in vitro}, and that biomass increase during growth was the only adjustment needed to match {\it in vivo} consumption--indeed, without biomass scaling, 1,2-PD consumption stalled and no longer fit the dataset (SI Figure \ref{fig:Mode1WTPredictionNoGrowth}). Because MCP-localized PduCDE activity alone was sufficient to reproduce the 1,2-PD time series, we conclude that PduCDE within the MCP is the dominant source of diol dehydratase activity and the cytosol contributes negligible additional PduCDE activity {\it in vivo}. The consistency in 1,2-PD dynamics also suggests that the MCP purification process did not significantly impact 1,2-PD activity. However, forward simulations failed to reproduce propionaldehyde, propionate, and 1-propanol dynamics (Figure \ref{fig:Fig2}Bii–iv). The main limitation was the model’s inability to accurately capture observed propionaldehyde consumption. In the simulations, more than 95\% of the 1,2-PD mass was converted to propionaldehyde, while less than 5\% was converted to extracellular metabolites and directed into central metabolism.

One might assume that weak prior constraints on mass-action kinetics contributed to poor {\it in vivo} metabolite predictions, since only 34 of 70 mass-action parameters were constrained by available Michaelis–Menten measurements. However, this argument fails when applied to 1-propanol. All PduQ kinetic parameters are constrained to available measurements from the literature, yet our {\it in vivo} model fails to predict 1-propanol production accurately given the concentrations of enzyme estimated from {\it in vitro} fits. Similarly, weak constraints on MCP permeability estimates are unlikely to have affected {\it in vivo} metabolite predictions. Our simulations included samples with MCP permeabilities both above and below 10\textsuperscript{-6.5} m/s, the threshold for MCP entrainment to the external volume \cite{Archer2025} (SI Figure \ref{fig:Mode2PduPPduQAdhPduLPduWZoomed}). These findings suggest instead that structural deficiencies in the model, such as additional Pdu activity beyond the MCP contribution, contribute to the poor model predictions. 

\subsection*{Simultaneous fit to {\it in vitro} and wild type data sets suggest that enzymes downstream of PduCDE are active in the cytosol}
 
We conducted simultaneous calibration to both {\it in vivo} and {\it in vitro} datasets to test whether a consistent parameter set could describe both models or whether additional differences must be accounted for to achieve a good joint fit. The models shared Pdu MCP permeability, Pdu enzyme numbers, and kinetic rate parameters (see section \ref{ssec:ModelInferencePipeline} for details). Thus, we assume that the purified MCPs are representative of MCP enzymatic activity, and MCP-localized enzymes are not substantially lost or deactivated in the purification process, with exception of PduQ, which, as mentioned above, is assumed to be oxidized and thus partially inactivated for {\it in vitro} but not {\it in vivo} experiments \cite{Cheng2012}. The additional structure of the cell and downstream methylcitrate metabolism {\it in vivo} were the main differences between the initial {\it in vivo} model and the {\it in vitro} model. Calibration failed to identify a parameter set consistent with both datasets (SI Figure \ref{fig:Mode1PduW}, \ref{fig:Mode2PduW}). Moreover, the {\it in vivo} model could not be independently calibrated to reproduce the observed {\it in vivo} time series, indicating that MCPs constrained to literature measurements could not generate sufficient activity to match experimental data (SI Figure \ref{fig:Mode1WTPduW}, \ref{fig:Mode2WTPduW}). We therefore conclude that the initial {\it in vivo} model contained deficiencies that prevented reproduction of the experimental data.

 Given existing evidence that some Pdu pathway enzymes may exist in both the compartment lumen and the cytosol \cite{Nichols2020}, we tested whether adding cytosolic PduP, PduQ, PduL, and a promiscuous alcohol dehydrogenase (ADH) to the {\it in vivo} model could produce better joint fits to data. We iteratively updated the {\it in vivo} model until both {\it in vivo} and {\it in vitro} models could fit their respective datasets. Both models achieved acceptable fits only when the {\it in vivo} model included cytosolic PduP, PduL, and ADH (Figure \ref{fig:Fig2}; Table \ref{table:ModelIteration}). Cytosolic PduQ was not required to reproduce the {\it in vivo} time series due to limited constraints on promiscuous ADH model kinetics, which allowed ADH and PduQ to serve the same function (SI Fig \ref{fig:WTdPduQdADH}). However, we retained cytosolic PduQ in the final model due to experimental evidence, presented later in this study, supporting cytosolic localization. PduW was also modeled with a cytosolic component, consistent with prior work showing localization of the enzyme to both the MCP and cytosol \cite{Yang2020}.

\begin{table}\centering
\begin{tabular}{||P{2.5cm} P{2.25cm} P{2.25cm} P{2.5cm} P{2.25cm} P{2.5cm}||} 
 \hline
 Enzyme & Iteration \#1 & Iteration \#2 & Iteration \#3 & Iteration \#4 & Iteration \#5 (Final Model) \\ [0.5ex] 
 \hline\hline 
  PduCDE & M & M  & M & M & M\\
 PduP & M & M/C & M/C & M/C & M/C \\
 PduQ & M & M & M/C & M/C&  M/C\\
 ADH &  &  &  &  C & C  \\
 PduL & M & M  &  M & M & M/C \\
 PduW &  M/C & M/C  & M/C & M/C  & M/C \\
  \hline\hline
Log-posterior & 1713986 & 9718 & 2084 & 172 & 138 \\
 \hline
 Summary of WT model fit & did not consume 1,2-propanediol &  did not produce 1-propanol and propionate & did not produce sufficient 1-propanol; did not produce propionate & did not produce propionate & qualitatively described the data\\
 \hline
\end{tabular}
\caption{Summary of updates to the {\it in vivo} model for simultaneous description of {\it in vitro} and {\it in vivo} datasets. PduP, PduQ, ADH, and PduL were iteratively added to the cytosol of the {\it in vivo} model until both {\it in vivo} and {\it in vitro} models could simultaneously reproduce their respective datasets. The final {\it in vivo} model comprised fully encapsulated PduCDE, a cytosolic promiscuous alcohol dehydrogenase, and PduP, PduQ, PduL, and PduW localized to both the cytosol and MCP. Although not required for model fit, cytosolic PduQ was included based on experimental evidence supporting its cytosolic expression. We assume PduW to have a cytosolic component due to prior experimental evidence \cite{Yang2020}. Abbreviations: M = localized in MCP; C = localized in cytosol; M/C = localized in both MCP and cytosol.}
\label{table:ModelIteration}
\end{table}

\subsubsection*{Prior evidence indicates that PduW has a cytosolic component}

PduW was included as a cytosolic enzyme in all iterations of our model due to strong prior experimental evidence \cite{Yang2020, Kennedy2022} (Table \ref{table:ModelIteration}). PduW was previously shown to have MCP and cytosolic expression \cite{Yang2020}, and to be necessary for necessary for WT fermentative growth on 1,2-PD \cite{Palacios2003}. Therefore, WT PduW activity is likely partially due to cytosolic PduW expression. However, with no available kinetic measurements of PduW, we are not able to determine the independent MCP and cytosolic PduW contributions to propionate dynamics via modeling.

\subsubsection*{Cytosolic PduP is necessary to enable high PduCDE flux without propionaldehyde accumulation}

Cytosolic PduP was necessary to include in the model to compensate for low PduP activity within the MCP. When simultaneously calibrated to {\it in vitro} and {\it in vivo} data, the {\it in vivo} model with MCP-localized PduP slowly consumed 1,2-PD and accumulated propionaldehyde  (SI Figure \ref{fig:Mode1PduW}, \ref{fig:Mode2PduW}).
This accumulation can be explained by the activity differential between PduCDE and PduP: in the purified MCP model the $V_{\max}$ of PduP (6-9 mol/s) is lower than that of PduCDE (111-288 mol/s). As a result, purified MCPs lack sufficient PduP activity to consume PduCDE flux, leading to propionaldehyde buildup in models without cytosolic PduP.

It could be argued that the additional cytosolic ALDH activity arises from a promiscuous enzyme rather than cytosolic PduP. However, prior work shows that PduP accounts for the majority of aldehyde dehydrogenase activity during 1,2-PD metabolism \cite{Leal2003}. Thus, PduP is the most likely source of the additional cytosolic activity required to fit the {\it in vivo} dataset, and we included it in model iteration \#2 and beyond (Table \ref{table:ModelIteration}).



\subsubsection*{Promiscuous alcohol dehydrogenase, independent of PduQ, is necessary to produce and consume 1-propanol}

A promiscuous ADH was required to reproduce the 1-propanol time series, as PduQ activity alone was unable to explain alcohol dehydrogenase activity {\it in vivo}. Calibration assuming complete PduQ encapsulation, consistent in both models, resulted in little to no 1-propanol production (SI Figure \ref{fig:Mode1PduPPduW}). Updating the {\it in vivo} model to include cytosolic PduQ enabled production and consumption of 1-propanol, leading to better agreement with experimental observations (Model iteration 3 in Table \ref{table:ModelIteration}, SI Figure \ref{fig:Mode1PduPPduQPduLExpandedPduW}). However, the model underestimated propionaldehyde dynamics and predicted cytosolic PduQ concentrations exceeding the typical range of cytosolic enzyme concentrations which is on the order of $\sim10^{-4}-10^{-1}$ mM \cite{Albe1990, Dourado2021}. Restricting PduQ concentration to 0.5 mM left the {\it in vivo} model unable to produce sufficient 1-propanol (SI Figure \ref{fig:Mode1PduPPduQPduW}, \ref{fig:Mode2PduPPduQPduW}).

Because PduQ kinetics are highly constrained (7 of 8 mass-action parameters), we conclude that its kinetics are inconsistent with our {\it in vivo} measurements and that PduQ cannot be the sole contributor to 1-propanol activity {\it in vivo}. To account for this discrepancy, we updated the model in iteration 4 and included an alcohol dehydrogenase with unmeasured kinetics (Table \ref{table:ModelIteration}, SI Figure \ref{fig:Mode1PduPPduQAdhPduW}, \ref{fig:Mode2PduPPduQAdhPduW}). We retained cytosolic PduQ in the final model, supported by experimental and modeling evidence presented here, showing that cytosolic PduQ is necessary for optimal 1-propanol production and consumption {\it in vivo} (Section \nameref{ssec:PduQpropanolmetabolism}). However, because we cannot completely differentiate between ADH and PduQ activity given the data available, ADH may be overcompensating in our final {\it in vivo} model fits.

\subsubsection*{Cytosolic PduL is likely necessary to consume propionate}

The {\it in vivo} model with only encapsulated PduL struggled to identify parameter regions that could generate propionate (Iteration \#4 in Table \ref{table:ModelIteration}, SI Figure \ref{fig:Mode1PduPPduQAdhPduW}, \ref{fig:Mode2PduPPduQAdhPduW}). Before investigating changes to the model itself, we investigated whether convergence to regions with nonzero propionate concentrations was hindered by the low contribution of propionate to the cost function. By decreasing the propionate data standard deviation, thereby increasing its contribution to the cost function, the model reproduced the propionate time series but at the expense of reducing the fit to other compounds (SI Figure \ref{fig:Mode1PduPPduQAdhPduW_PropionateTightened}). This led us to conclude that models with only encapsulated PduL and parameter constraints could not reproduce the propionate time series because improving propionate fit reduced the overall quality of fit across other metabolites. Thus, encapsulated PduL with parameter constraints could not produce activity consistent with {\it in vivo} metabolite dynamics. We therefore concluded that cytosolic phosphotransacetylase activity must compensate for the inability of MCP PduL to reproduce {\it in vivo} propionate dynamics. Accordingly, we added cytosolic PduL to the final model (Table \ref{table:ModelIteration}, Figure \ref{fig:Fig2}).

As with PduP, one could argue that this activity does not necessarily arise from cytosolic PduL, but from some other cytosolic phosphotransacetylase. Indeed, {\it Salmonella} expresses the housekeeping phosphotransacetylase, Pta, which can also produce propionate \cite{Liu2007}. However, Liu et al. (2007) showed that Pta has a low preference for propionyl-phosphate over acetyl-phosphate and that PduL contributes 37\% more to total WT phosphotransacetylase activity than Pta {\it in vivo}. Thus, PduL must also be present in the cytosol, as it accounts for the majority of {\it in vivo} phosphotransacetylase activity, and MCP-encapsulated PduL alone cannot fully explain the observed propionate-related activity.

We note that only the reverse kinetics and thermodynamics of PduL have been reported in the literature and were used to constrain PduL kinetics in our model \cite{Liu2007}. With no forward kinetic constraints, it is unsurprising that our {\it in vivo} model with cytosolic PduL could reproduce the propionate time series. While PduL is the primary enzyme contributing to propionate-related activity, Pta also makes a non-negligible contribution to propionate metabolism \cite{Liu2007}. Further experimental and modeling work is needed to confirm the relative contributions of PduL and Pta to propionate dynamics. As in the case of PduQ, forward kinetic constraints would provide a clearer estimate of PduL’s contribution to propionate production relative to Pta.




\subsection*{MCP enzyme activity contributes little to downstream metabolite flux}

In our final model (Iteration \#5 in Table \ref{table:ModelIteration}), MCP-encapsulated PduCDE accounted for 100\% of the 1,2-PD flux dynamics. Consequently, cytosolic PduCDE expression was not required. Among all enzymes, PduCDE exhibited the highest $K^{f}_{\text{cat}}$, $K_{eq}$, and MCP enyzme number. As a result, its reaction velocity ($V_{\text{max}} = 111$–$287$ mol/s) was sufficient to overcome the MCP barrier and generate the cytosolic flux needed to consume extracellular 1,2-PD and produce propionaldehyde.

\begin{figure}[!ht]
\includegraphics[width=\linewidth]{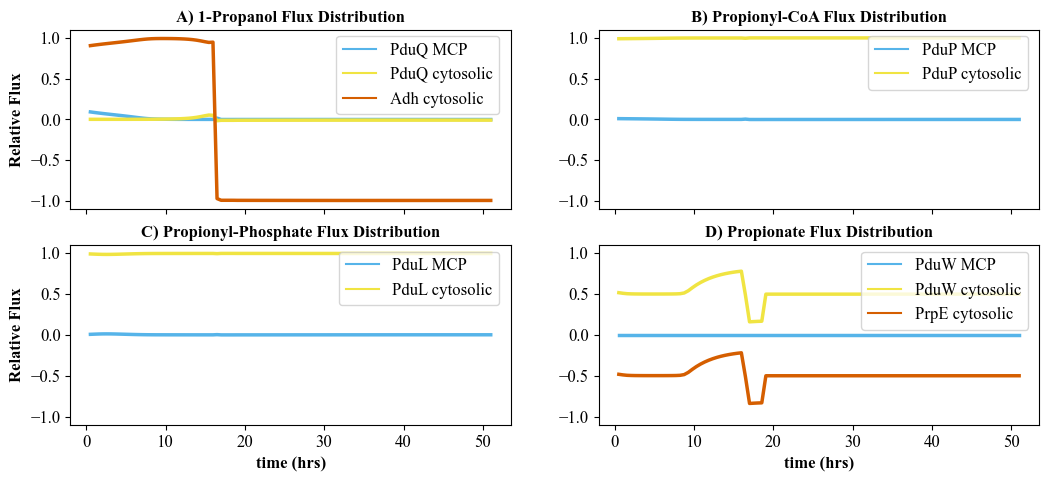}
    \caption{Panel of plots showing MCP and cytosolic enzyme flux contribution to metabolite dynamics. A) Contribution of MCP PduQ, cytosolic PduQ, and cytosolic alcohol dehydrogenase to the total cytosolic 1-propanol. B) Comparison of MCP PduP and cytosolic PduP Propionyl-CoA flux. C) Comparison of MCP PduL and cytosolic PduL Propionyl-Phosphate flux. D) Contribution of MCP PduW, cytosolic PduW, and cytosolic PrpE to the total cytosolic propionate flux.}
\label{fig:Fig3}
\end{figure}

To better understand the enzymatic contributions to downstream metabolite profiles {\it in vivo}, we computed flux contributions of Pdu enzymes. Cytosolic PduP, ADH, PduL, and PduW fluxes dominated over MCP contributions (Figure \ref{fig:Fig3}A–D). This is consistent with the observation that models containing only MCP expression could not reproduce experimentally-observed {\it in vivo} time series. The limited MCP contribution can be explained by the interaction of enzyme properties: slow kinetics and low enzyme numbers. These factors constrained MCP activity of non-PduCDE enzymes and reduced the MCP contribution to extracellular metabolite production. Cytosolic alcohol dehydrogenase and PduP/L activity were therefore required to elevate the 1-propanol- and propionate-to-propionaldehyde ratios from {\it in vitro} to {\it in vivo} levels (Table \ref{table:ComparisonInVivoInVitro}).

We note that our model could not accurately resolve the cytosolic PduQ contribution to 1-propanol activity. With limited constraints on alcohol dehydrogenase kinetics, the model assigned much of the 1-propanol flux to a promiscuous alcohol dehydrogenase (Figure \ref{fig:Fig2}C). Similarly, the model attributed propionate production to the PduL–PduW pathway and propionate consumption to the PrpE pathway, (Figure \ref{fig:Fig2}E). However, prior research has shown that both pathways are valid routes for propionate consumption \cite{Palacios2003}. Limited constraints on PduW kinetics may have contributed to the model predicting little to no PduW propionate consumption flux.

\subsection*{Lysate conditions improved extracellular metabolite yield from purified MCPs}

\begin{figure}[!ht]
\centering
\includegraphics[width=\linewidth]{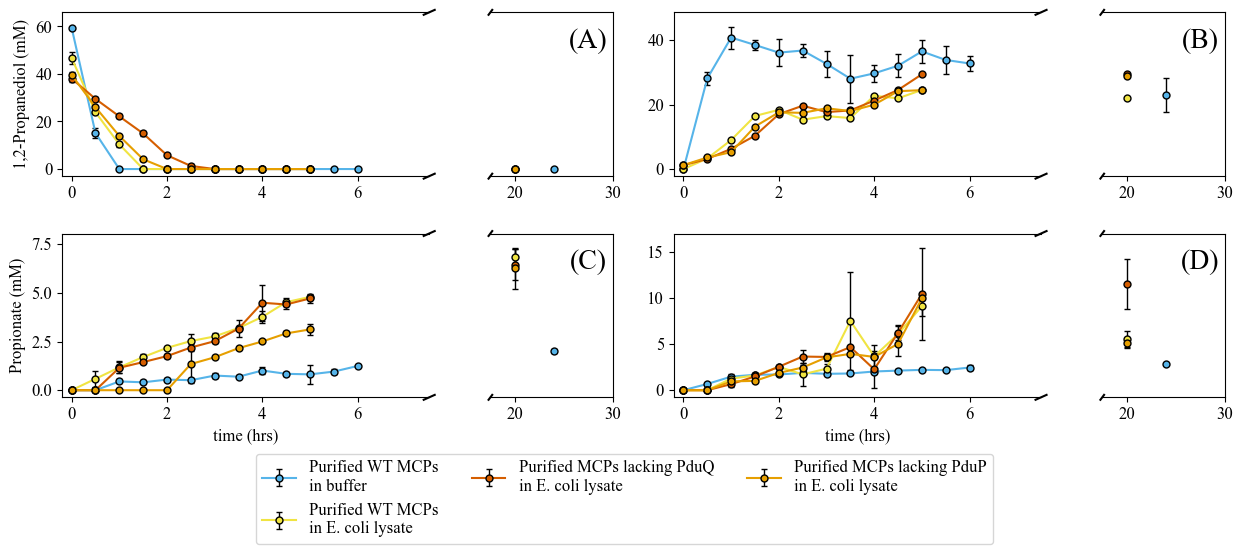}         
\caption{Metabolic activity of WT MCPs assayed with 1,2-propanediol in buffer or {\it E. coli} lysate, compared to MCPs deficient in the PduP/PduQ catalytic proteins assayed with 1,2-propanediol in {\it E. coli} lysate. A) 1,2-propanediol, B) propionaldehyde, C) propionate, and D) 1-propanol concentrations were measured over a 20–24 h period. Propionate and 1-propanol profiles are consistent with the hypothesis that exogenous enzymes are required to elevate metabolite dynamics to {\it in vivo} levels. } 
\label{fig:Fig4}
\end{figure}

Taken together, our results show that cytosolic enzyme activity improved the maximum 1-propanol- and propionate-to-propionaldehyde concentration ratios {\it in vivo}. To confirm that exogenous enzymes can enhance 1-propanol and propionate yield, we conducted an {\it in vitro} assay of purified MCPs in {\it E. coli} lysate, which contains the soluble native enzymes from the {\it E. coli} cell. While not a perfect analog for the {\it Samonella} cytosol, {\it E. coli} shares a broad metabolic similarity with {\it Samonella} \cite{strain2009genome}, and the methylcitrate cycle utilizing propionate is active in both systems \cite{horswill1999salmonella, london1999carbon}. Relative to buffer, wild-type purified MCPs in {\it E. coli} lysate produced approximately threefold higher peak propionate and 1-propanol concentrations (Figure \ref{fig:Fig4}). This supports our hypothesis that exogenous enzymes are required to reconcile the discrepancy between {\it in vitro} and {\it in vivo} activity. We note, however, that propionaldehyde consumption stalled as in the buffer. Given the potential discrepancies between {\it E. coli} lysate and {\it Samonella} cytosol, additional studies are needed to determine the factors that led to propionaldehyde accumulation, but it is likely due to bottlenecks in downstream pathway activity. 

\subsection*{Purified MCPs lacking downstream enzymes in lysate indicate that PduQ, PduL and PduP contribute minimally to total metabolite flux}

We conducted an {\it in vitro} assay of purified MCPs lacking PduQ and PduL enzymes in {\it E. coli} lysate. The metabolic activity observed in these assays confirmed that encapsulated enzymes downstream of PduCDE contribute little to overall metabolite activity. MCPs lacking PduQ and PduL exhibited 1-propanol and propionate production profiles similar to those of WT MCPs in lysate (Figure \ref{fig:Fig3}, SI Figure \ref{fig:all_reactant_partition_reactants_in_vitro_WT_dPduL}). However, the MCP lysate assay lacking PduQ showed that PduQ is required for 1-propanol consumption; in its absence, 1-propanol consumption was not observed.

The propionate profile of MCPs lacking PduP differed slightly from WT MCP lysate dynamics (SI Figure \ref{fig:Fig3}). MCPs lacking PduP exhibited a substantially longer lag in propionate production (2 h vs. 0.5 h in WT) with a notable increase in curve concavity. However, after 20 h, the final propionate yield was comparable to WT. These results suggest that encapsulated PduP may have an impact, albeit marginal, on metabolite dynamics.


\subsection*{$\Delta$pduQ knockout {\it in vivo} assay confirms promicious alcohol dehydrogenase activity}  

\begin{figure}[!ht]\centering
\includegraphics[width=0.85\linewidth]{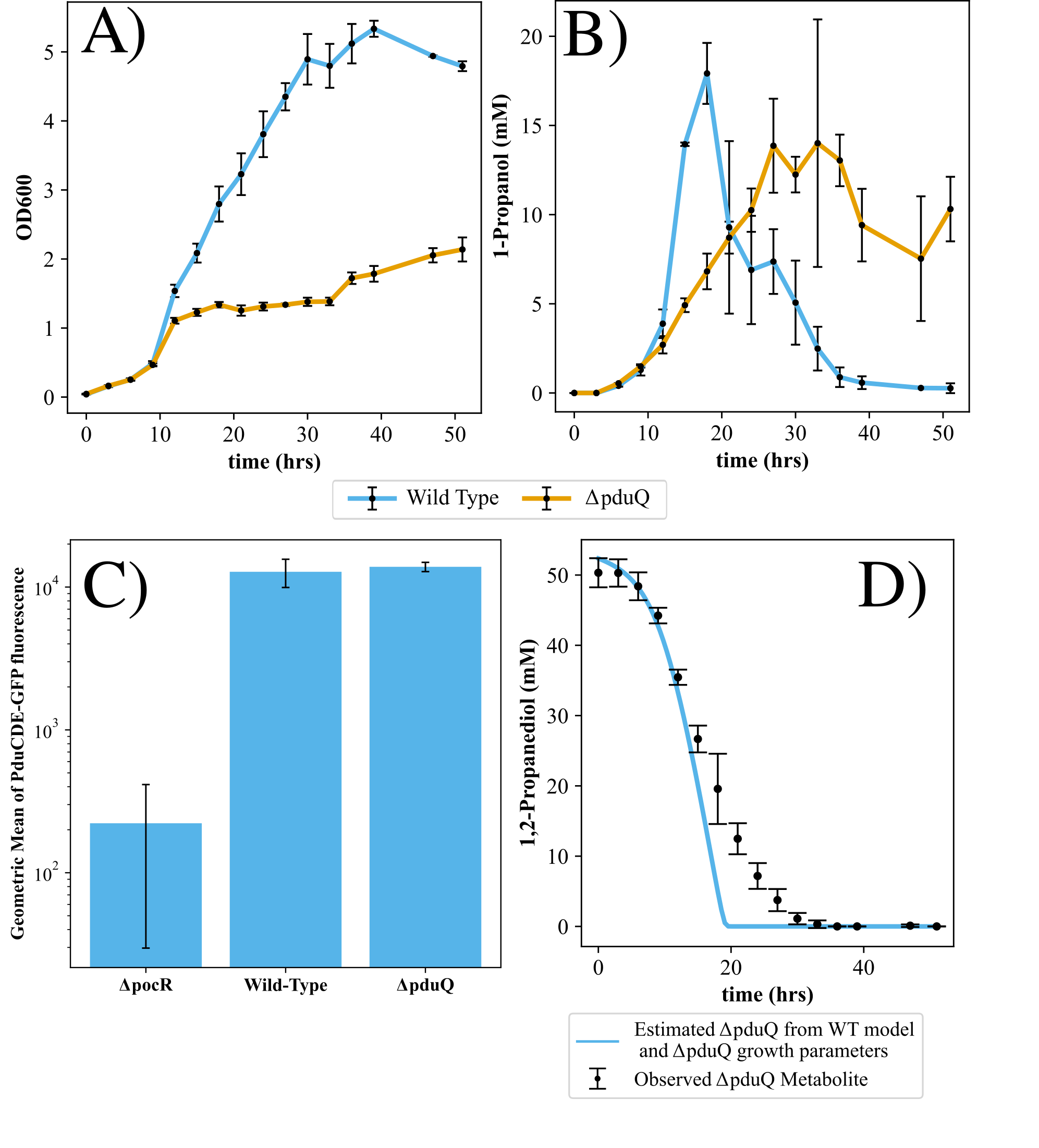}
\caption{Metabolic activity of $\Delta$pduQ. A–B) Growth curves (A) and 1-propanol activity (B) of WT and $\Delta$pduQ {\it in vivo}. C) PduCDE–GFP fluorescence in $\Delta$PocR, WT, and $\Delta$pduQ strains. D) Comparison of {\it in silico} predictions of $\Delta$pduQ 1,2-propanediol dynamics, generated using {\it in vitro} and {\it in vivo} calibrated model parameters, with experimental measurements.   }
\label{fig:Fig5}
\end{figure}

A growth assay of $\Delta$pduQ confirmed the presence of a promiscuous alcohol dehydrogenase. Like WT, $\Delta$pduQ was grown on 55 mM 1,2-PD and 150 nM Ado-B\textsubscript{12}. $\Delta$pduQ produced on average 80\% of the WT maximum 1-propanol concentration as WT LT2 (Figure \ref{fig:Fig5}B). The WT maximum (18 $\pm$ 2 mM) was within one standard deviation of the $\Delta$pduQ maximum (14 $\pm$ 7 mM). Despite reaching a similar maximum in 1-propanol concentration, $\Delta$pduQ produced 1-propanol more slowly than WT, requiring 15 h longer to reach peak concentration. Like WT, $\Delta$pduQ was also able to consume 1-propanol; however, it consumed only 26.4\% of the maximum, and consumption stalled within 18 h of peak production. In contrast, WT LT2 consumed 95\% of excess 1-propanol within 18 h of its peak. These observations suggest that PduQ contributes more to 1-propanol consumption than to production.

\subsection*{PduQ is necessary for optimal 1-propanol metabolism}
\label{ssec:PduQpropanolmetabolism}
While PduQ is not essential for activity on 1-propanol, our results indicate that it is necessary for optimal 1-propanol metabolism and growth. As discussed above, $\Delta$pduQ produced and consumed 1-propanol less efficiently than WT (Figure \ref{fig:Fig5}B). $\Delta$pduQ also grew markedly worse on 1,2-PD, reaching a maximum OD 2.5-fold lower than WT and exhibiting a growth arrest between 15 and 33 h. Thus, the absence of PduQ severely impaired both 1-propanol metabolism and growth. Consistent with our findings, Cheng et al. 2012 reported growth impairment in $\Delta$pduQ that was rescued by ectopic expression of PduQ.

The slow growth of the $\Delta$pduQ strain coincided with reduced 1,2-PD consumption and extracellular metabolite accumulation. However, slow growth alone is not causal for reduced metabolite activity. We predicted $\Delta$pduQ 1,2-PD consumption using {\it in vitro}/{\it in vivo} calibrated parameters and the $\Delta$pduQ growth curve (Figure \ref{fig:Fig5}D). This estimate serves as a surrogate for PduCDE activity in $\Delta$pduQ, as $\Delta$pduQ and WT were experimentally shown to have similar PduCDE expression levels (Figure \ref{fig:Fig5}C). Compared with experimental observations, the model overestimated the rate of 1,2-PD consumption (Figure \ref{fig:Fig5}D), indicating that decreased biomass alone cannot explain reduced $\Delta$pduQ 1,2-PD consumption. A notable divergence between model predictions and experimental data occurred at 15 h, coinciding with the period of growth arrest. This suggests that hidden variables affecting both growth and 1,2-PD consumption were not captured by the model.


We hypothesize that PduQ is present in both the cytosol and within the MCP, and that its removal impairs NADH-to-NAD\textsuperscript{+} recycling in the cytosol, resulting in poor growth and decreased 1,2-PD activity. PduQ must be cytosolic to influence NADH:NAD\textsuperscript{+} balance because, like PduP and PduL, MCP-localized PduQ would contribute little flux to the cytosolic environment. Removing cytosolic PduQ would therefore lead to NADH accumulation as more PduCDE flux is directed through the PduP branch. Limited NAD\textsuperscript{+} availability due to NADH accumulation would explain stalled 1-propanol consumption in $\Delta$pduQ. A disrupted cytosolic NADH:NAD\textsuperscript{+} ratio would also create metabolic bottlenecks in the TCA cycle, thereby slowing ATP production through the downstream electron transport chain and ultimately impairing growth. Reduced ATP availability would in turn decrease Ado-B\textsubscript{12} turnover, ATP-dependent diol hydratase reactivation, and ultimately PduCDE activity, explaining the slow 1,2-PD uptake observed in $\Delta$pduQ.

Our redox hypothesis is consistent with the observation by Cheng et al. 2012 that $\Delta$pduQ growth improved after either decreasing AdoB\textsubscript{12} from 150 nM to 40 nM or removing the MCP shell \cite{Cheng2012}. Under our hypothesis, lowering B\textsubscript{12} reduces PduCDE flux and thereby lessens the burden on downstream redox processes, leading to improved growth compared with excess AdoB\textsubscript{12} conditions. Disrupting the MCP shell would also improve $\Delta$pduQ growth by allowing NADH-dependent PduS to freely utilize excess NADH, restoring the NADH:NAD\textsuperscript{+} ratio and enhancing AdoB\textsubscript{12} turnover without the restriction of shell encapsulation. 

\subsection*{{\it in vitro} fit with rapid PduP-CoA formation is likely inconsistent with the WT {\it in vivo} time series}

Previous modeling of the purified MCP system revealed two data-consistent parameter regimes: one characterized by low NADH/NAD\textsuperscript{+} permeability and slow PduP inhibition (mode 1), and the other by high NADH/NAD\textsuperscript{+} permeability and rapid PduP inhibition (mode 2) \cite{Archer2025}. Simultaneous calibration to both {\it in vitro} and {\it in vivo} data yielded consistent model fits only under mode 1 constraints (Figure \ref{fig:Fig2}Ai, \ref{fig:Fig2}Bi). Mode 2–restricted chains failed to adequately fit both datasets (SI Figure \ref{fig:Mode2PduPPduQAdhPduLPduW}). Differences in their ability to reproduce experimental dynamics can be attributed to underlying parameter regimes of each mode: MCP cofactor permeability and the PduP inhibition rate \cite{Archer2025}. 

While the MCP cofactor permeability is important to explain the \textit{in vitro} dynamics that are dominated by the MCP-localized reactions, it likely has a marginal impact on the fit to {\it in vivo} activity. The cofactor-dependent activity of PduP/Q is primarily in the cytosol (Figure \ref{fig:Fig2}A,D) and is, therefore, controlled by the cytosolic NADH/NAD\textsuperscript{+}  balance. Since the cytosolic PduP/Q flux dominates, the transit of co-factors in and out of the compartment has almost no effect on the measured concentrations. 

The high PduP inhibition rates of mode 2 likely impacted {\it in vivo} and {\it in vitro} consistency. In mode 2, PduP is more rapidly inhibited by CoA than in mode 1. In the presence of cytosolic CoA, this rapid inhibition would lead to strong PduP inhibition and slow propionyl-CoA and CoA turnover, thus explaining the impaired propionate production observed in mode 2 fits to the {\it in vivo} and {\it in vitro} data (SI Figure \ref{fig:Mode2PduPPduQAdhPduLPduW}). This indicates that high PduP inhibition rates in mode 2 may not be consistent with MCP activity. In Archer et al. 2025, PduP inhibition was proposed as a mechanism to describe low downstream metabolite production in purified MCPs. However, the present work potentially eliminates this mechanism as a route of MCP enzyme inactivation and instead leaves PduQ inhibition via oxygen inactivation as the primary factor contributing to reduced downstream metabolite production.

\section*{Discussion}

We conducted an assay of WT, MCP-dependent {\it Salmonella} growth on 1,2-PD and identified notable differences between our {\it in vivo} and previously published {\it in vitro} assay results. These differences included slower 1,2-PD consumption {\it in vivo}, higher extracellular metabolite production {\it in vivo}, and increased extracellular metabolite consumption {\it in vivo}. To investigate the sources of these discrepancies, we developed and calibrated a mechanistic model of WT {\it in vivo} growth on 1,2-PD. As in Archer et al. 2025, we employed mass-action kinetics with Michaelis–Menten constraints to describe reaction rates.

We found that slower 1,2-PD consumption {\it in vivo} was explained by reduced biomass accumulation. Using calibrated parameters for Pdu MCP permeability and enzyme kinetics, our {\it in vivo} model accurately predicted 1,2-PD dynamics when biomass growth was accounted for (Figure \ref{fig:Fig2}Bi). From this {\it in silico} experiment, we showed that MCP-encapsulated PduCDE alone can account for all diol dehydratase activity {\it in vivo}. However, our uncalibrated {\it in silico} model also demonstrated that other cytosolic enzymes are required to increase extracellular metabolite yields (Figure \ref{fig:Fig2}Bii-vi). Predictions with no cytosolic Pdu enzymes overestimated propionaldehyde production and produced little to no 1-propanol or propionate, indicating that purified MCPs lack the downstream activity needed to process PduCDE flux (Figure \ref{fig:Mode1PduW}).

By simultaneously fitting our {\it in vivo} model and the Archer et al. 2025 {\it in vitro} model to their respective datasets, we find that increased {\it in vivo} production of 1-propanol and propionate is likely due to the presence of Pdu-relevant enzymes in the cytosol. A consistent fit was only possible when a promiscuous alcohol dehydrogenase and all Pdu enzymes except PduCDE were added to the cytosol in our model (Figure \ref{fig:Fig2}B). With the exception of PduQ, prior research supports the interpretation that this additional activity originates from Pdu enzymes themselves and not promiscuous cytosolic enzymes \cite{Leal2003, Liu2007, Palacios2003, Yang2020}. Our {\it in vivo} model required an additional promiscuous alcohol dehydrogenase to reproduce the 1-propanol time series (Figure \ref{fig:Mode1PduPPduQPduW} and \ref{fig:Mode1PduPPduQAdhPduW}), and its presence was confirmed by a pduQ knockout assay (Figure \ref{fig:Fig5}B). 


Analysis of metabolite fluxes in our calibrated {\it in vivo} Pdu model fit suggest that much of the observed metabolite activity must be attributed to cytosolic enzymes (Figure \ref{fig:Fig3}A-D). An {\it in vitro} lysate assay further demonstrated that encapsulated PduL, PduQ, and, to a lesser extent, PduP contributed negligibly to total flux (Figure \ref{fig:Fig4}A-D). Future experiments could assess the effects of encapsulated and cytosolic PduP, PduL and PduQ in MCPs {\it in vivo} to determine if they play roles in MCP structure or pathway flux beyond what could be resolved by our model. 

These findings led us to propose a new {\it in vivo} mathematical model of the Pdu system in which PduCDE is fully encapsulated and the downstream enzymes of PduCDE are partially encapsulated (Figure~\ref{fig:Fig1}). This model is consistent with existing Pdu kinetic and enzyme number measurements and provides a framework for reconciling {\it in vitro} and {\it in vivo} observations of MCP activity. Importantly, the model is constrained by the currently available kinetic and enzyme number measurements, and future refinement of these measurements may enable alternative encapsulation scenarios to be evaluated.

We note that our study could not fully explain why the {\it in vivo} assay consumed 1-propanol and propionate while the {\it in vitro} assay did not. Our results point to cytosolic PduQ and a promiscuous alcohol dehydrogenase as potential contributors to 1-propanol consumption, but without additional kinetic constraints on the promiscuous enzyme, we cannot distinguish the specific contributions of each. Similarly, prior research implicates the PduL–PduW pathway and PrpE as sources of propionate consumption \cite{Palacios2003}, but given the limited kinetic constraints on PduW, the source of {\it in vivo} propionate consumption remains unresolved. Future work could include knockout studies to resolve the contribution of specific alcohol dehydrogenases, phosphotranscylases and acetate kinases to Pdu metabolites dynamics.

Our work indicates that MCPs primarly serve to encapsulate PduCDE and process 1,2-propanediol. We hypothesize that encapsulation of downstream enzymes provides a structural role (for example, helping to maintain MCP integrity) \cite{Nichols2020} and marginal catalytic functionality. To harness the catalytic and functional capability of MCPs, we recommend 1) focusing on engineering and optimizing the functionality of PduCDE, 2) understanding B\textsubscript{12} MCP permeability, and 3) researching PduS, PduO and PduGH encapsulation and mechanism of action. This would provide a better understanding of how MCPs support PduCDE turnover. For example, it remains an open question: can PduS and PduO independently react to produce AdoB\textsubscript{12}, despite low PduS k\textsubscript{cat} for cob(II)alamin reduction, or is there an unknown mechanism of cobalamin adenylation, such as PduS-PduO interaction, that reduces the cob(II)alamin/cob(I)alamin redox potential  \cite{Mera2010}? Together, our results indicate that taking full advantage of MCP function in engineering applications requires a paradigm shift from downstream enzyme encapsulation to the kinetic optimization, cofactor handling and redox coupling of the PduCDE-centered core pathway.

\section*{Methods}
\subsection*{{\it in vitro} Mathematical Model}\label{sec:InVitroMathematicalModel}
As in Archer et al. 2024, we model the assay of purified compartments as a multi-compartment ordinary differential equation system with MCP and external volume component, see equation \ref{Eq:Chapter6_InVitroModel}. Each component contains a diffusion term, $P_{\text{MCP}, X}\times \text{number of MCPs} \times \frac{\text{Surface Area of MCP}}{\text{ volume}}$ and a reaction term, $R_{X}(X)$.

\begin{equation}
\label{Eq:Chapter6_InVitroModel}
\begin{aligned}
\frac{dX_{\text{MCP}}}{dt} &= R^X_{\text{MCP}}(X_{\text{MCP}}) - P_{\text{MCP}, X} \times \frac{\text{Surface Area of MCP}}{\text{external volume}}( X_{\text{ext}} - X_{\text{MCP}})\\
\frac{dX_{\text{ext}}}{dt} &= R^X_{\text{ext}}(X_{\text{MCP}}) \\
&\qquad+ P_{\text{MCP}, X}\times \text{number of MCPs} \times \frac{\text{Surface Area of MCP}}{\text{external volume}}\times(X_{\text{ext}}-X_{\text{MCP}})
\end{aligned}
\end{equation}
where $P_{\text{cell}}^{X}$ and $P_{\text{MCP}}^{X}$ are the permeability of $X$ to the cell membrane and MCP, respectively, $S_{\text{cell/MCP}}$ is the surface area of the cell/MCP and $V_{\text{cell/MCP}}$ is the volume of the cell/MCP.

\begin{figure}[!ht]\centering
\includegraphics[width=0.95\linewidth]{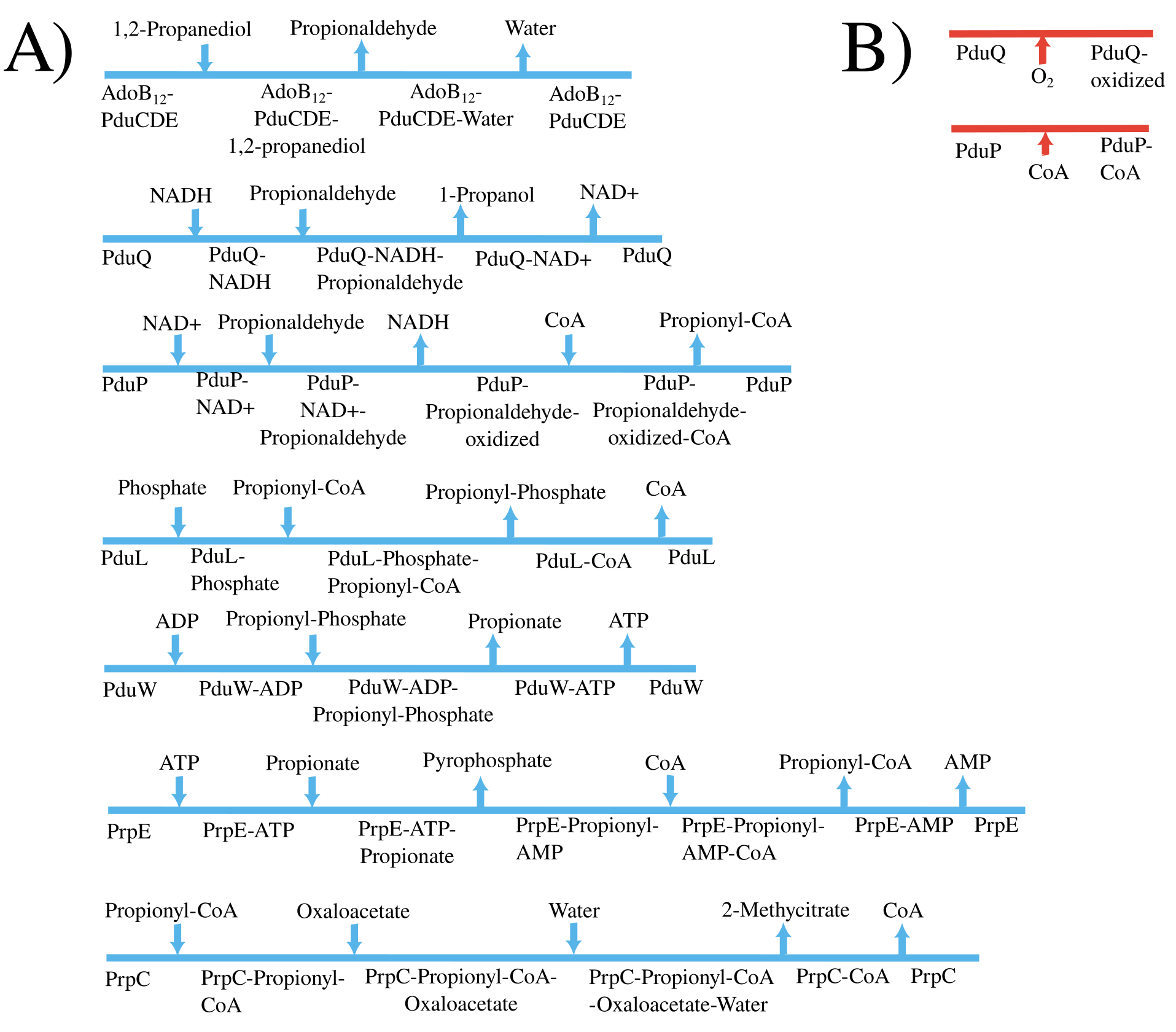}
\caption{Mass-action decomposition of Pdu and Prp enzyme reactions and B) inhibition of PduQ and PduP. Reactions proceed from left to right, with the sequence of substrates, intermediate complexes, and products indicated below the line. A downward-facing arrow indicates a binding event, and an upward-facing arrow indicates an unbinding event. A) From top to bottom: PduCDE conversion of 1,2-propanediol to propionaldehyde \cite{Toraya2000}; PduQ reduction of propionaldehyde to 1-propanol; PduP oxidation of propionaldehyde to propionyl-CoA; PduL conversion of propionyl-CoA to propionyl-phosphate \cite{Smith1980}; PduW conversion of propionyl-phosphate to propionate; PrpE conversion of propionate to propionyl-CoA \cite{Horswill2002}; and PrpC conversion of propionyl-CoA to 2-methylcitrate. We assume that PduQ, PduL, PduW, and PrpC sequentially bind cofactors first and organic substrates second. We further assume that AckA in the {\it in vitro} assay and ADH in the {\it in vivo} assay follow the reaction order of PduW and PduQ, respectively. B) From top to bottom: mass-action decomposition of PduQ oxidation and PduP–CoA inhibition \cite{Smith1980}. We assume PduQ oxidation follows first-order elimination. Note that PduQ oxidation was not modeled {\it in vivo}. }
\label{Fig6}
\end{figure}

$R^{X}_{\text{MCP/ext}}(X_{\text{MCP/ext}})$ describes the kinetics of the enzyme acting on the reactant, $X$, in the MCP/external volume. The PduCDE, PduP, PduQ, PduL, and PduW catalyzed reactions take place in the MCP component. AckA catalyzed reaction takes place in the external volume. We use mass-action kinetics to model the enzyme kinetics. The binding order of the mass-action reactions is outlined in Figure \ref{Fig6}.

\subsection*{{\it in vivo} Mathematical Model}
\label{sec:InVivoMathematicalModel}

We use a multi-compartment ordinary differential equation system to model {\it in vivo} dynamics, 

\begin{equation}
\label{Eq:InVivoModel}
\begin{aligned}
\frac{dX_{\text{MCP}}}{dt} &= R^X_{\text{MCP}}(X_{\text{MCP}}) +  P_{\text{MCP}, X}\times\frac{ S_{\text{MCP}}}{V_{\text{MCP}}}\times ( X_{\text{cytosol}} - X_{\text{MCP}})\\
\frac{dX_{\text{cytosol}}}{dt} &= R^X_{\text{cytosol}}(X_{\text{cytosol}}) -P_{\text{cell}, X}\times\frac{S_{\text{cell}}}{V_{\text{cell}}} \times( X_{\text{cytosol}}-X_{\text{ext}})\\
&\qquad- P_{\text{MCP}, X}\times \text{number of MCPs} \times \frac{S_{\text{MCP}}}{V_{\text{cell}}}( X_{\text{cytosol}} - X_{\text{MCP}})\\
\frac{dX_{\text{ext}}}{dt} &= \mathcal{F}(N(t)) \times P_{\text{cell}, X} \times \frac{S_{\text{cell}}}{V_{\text{ext}}}\times(X_{\text{cytosol}}-X_{\text{ext}})\\
\frac{dN}{dt} &= r \left( 1 - \frac{N}{K}\right) N
\end{aligned}
\end{equation}
where P\textsubscript{\text{cell/MCP,X}} is the permeability of X to the cell membrane/MCP, respectively, S\textsubscript{\text{MCP}} is the surface area of the cell/MCP, V\textsubscript{\text{MCP/ext}} is the volume of the MCP/media, $\mathcal{F}(N(t))$ is the number of cells, and X=1,2-propanediol, propionaldehyde, 1-propanol, propionyl-CoA, propionate and 2-methylcitrate. The model contains an MCP, cytosol, and external volume component. MCPs and cells are assumed to be well mixed in the cytosol and external volume, respectively. Thus, we model only a representative MCP and cell while accounting for all constituent interactions with their encasing volume via diffusion terms. All reactants contain a reaction term, R\textsubscript{X}(X), to account for mass flux through the substrate-enzyme network. We model the reaction terms for PduCDELPQW and PrpEC reactions using mass-action kinetics, and PrpD reaction using Michaelis-Menten kinetics. 

Cell growth, $\mathcal{F}(N(t))$, is modelled as a function of the sigmoid, $N(t)$. This provides the model with sufficient flexibility to fit growth arrest and recovery events. We take $\mathcal{F}(N_1(t))=\sigma\times (b + N(t))^2$ where $b$ is defined in the OD Growth Parameters subsection and $\sigma$ is the factor conversion for OD to the number of cells. Cells are assumed to contain a fixed, unknown number of MCPs. We assume that cytosolic cofactor concentrations (OAA, NADH/NAD+) are held at a static homeostatic ratio. We model CoA to have an initial concentration of 0.5 mM \cite{Wolfe2005}.


\subsection*{Forward Translation}

We use calibrated {\it in vitro} model parameters to predict {\it in vivo} metabolite dynamics. First, we calibrated an {\it in vitro} model to the {\it in vitro} dataset of purified MCPs using pyMC \cite{Archer2025}. We treated all Michaelis–Menten parameters that spanned a tight interval as constants in our Bayesian calibration. Second, we fit the square-root transformed WT OD time series to the sigmoid function, equation \ref{Eq:sigmoid}, using pyMC. As the {\it in vitro} posterior samples did not contain information on cytosolic parameters, we sample these parameters from the respective prior distributions outlined in Parameter Constraints and Prior Distributions. With the combined sigmoid, {\it in vitro} posterior, and {\it in vivo} prior samples, we run forward simulations of the WT {\it in vivo} model (equation \ref{Eq:InVivoModel}). See Figure \ref{fig:Fig7}b.

\subsection*{Model Fitting}
\subsubsection*{Model Inference Pipeline}\label{ssec:ModelInferencePipeline}

We use pyMC, a Python-based Bayesian inference package, to calibrate our mathematical models to external metabolite concentrations. Bayesian inference defines a posterior, $p(\theta, \text{data}) \propto \ell(\theta, \text{data})p(\theta)$, that weighs between prior parameter belief integration, $p(\theta)$, and data fitting, $\ell(\theta, \text{data})$. We describe prior distribution formulation for parameters with existing literature values in the Parameter and Prior Distributions Section. We use uninformative truncated normal distributions for understudied parameters. Given the large order of magnitude range and many unknown parameters, the prior distributions act as a regularizer with bounded support. We define a Chi-square likelihood function, $\ell(\theta, \text{data})$,

$$\log \ell(\theta, \hat{\mu}, \hat{\sigma}) = -\sum_{\substack{X \in \{\text{1,2-PD, propionaldehyde,}\\\text{ 1-propanol, propionate}\}}}\,\,\sum_{i=1}^{N} \frac{(X_{\text{ext}}(t_i, \theta) - \hat{\mu}_{X,i})^2}{2\hat{\sigma}_{X,i}^2},$$

where $X_{\text{ext}}(t_i, \theta)$ is the external model solution of equation \ref{Eq:Chapter6_InVitroModel} and \ref{Eq:InVivoModel}, $X\in \{$1,2-PD, Propionaldehyde, 1-propanol, propionate$\}$, at time $t_i$,  and   $\hat{\mu}_{X,i}$ and $\hat{\sigma}_{X,i}$ are the sample mean and sample standard deviation computed using biological replicates at time $t_i$. 

\begin{figure}[!ht]\centering
\includegraphics[width=0.8\linewidth]{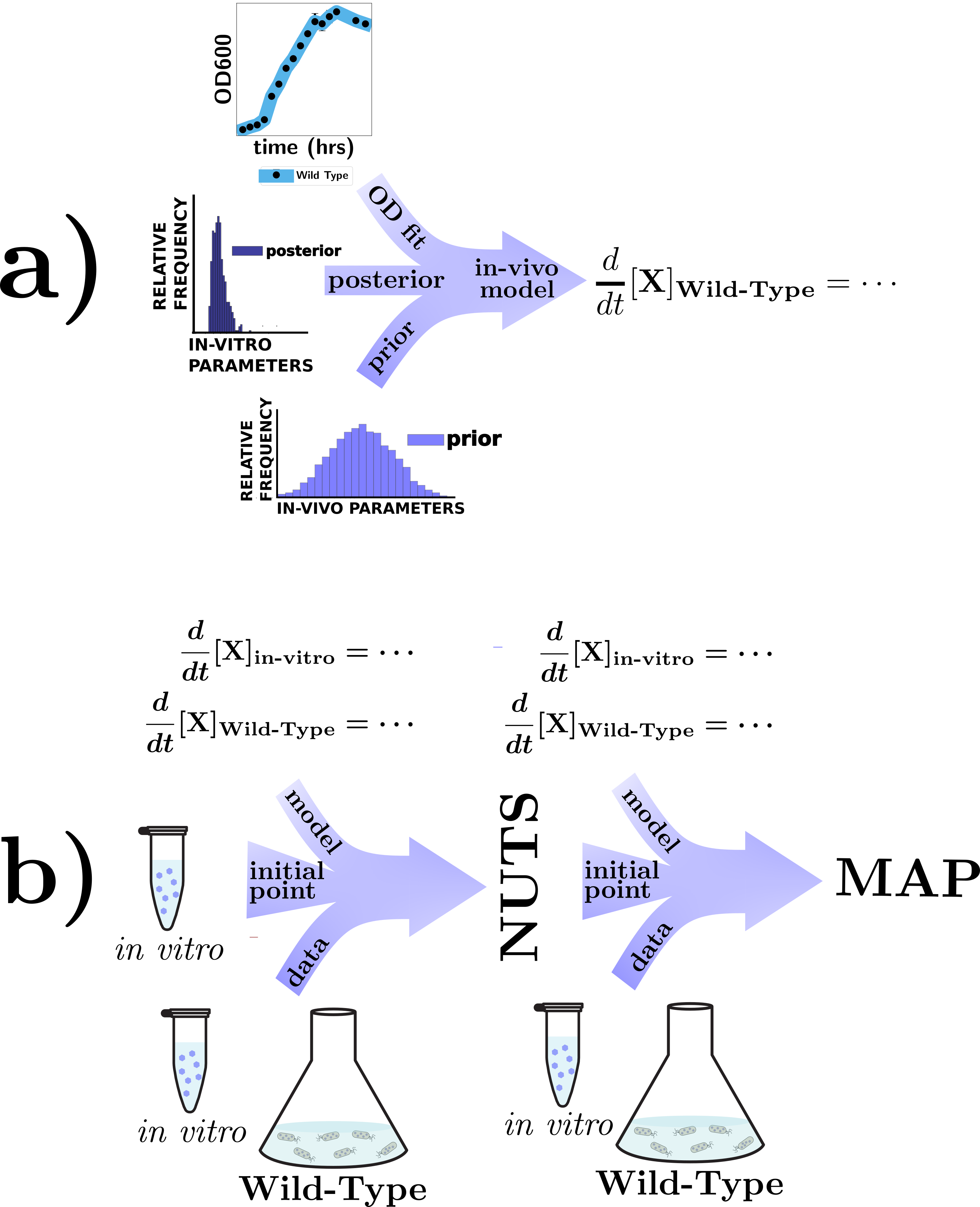}
\caption{A) Translation pipeline from the {\it in vitro} posterior to WT metabolite prediction. Translation requires cytosolic priors and ODE-based growth parameters for estimation. B) Inference pipeline implemented using PyMC \cite{pymc2023}. First, the {\it in vitro} model is calibrated to observed metabolite data. The posterior parameter distribution is then used to initialize a new Bayesian parameter estimation for fitting the {\it in vitro} and WT models to a parameter set in the neighborhood of the in-vitro/in-vivo MAP. Finally, the output from this convergence is used as the initial estimate for MAP inference in PyMC.}
\label{fig:Fig7}
\end{figure}

As in Archer et al. 2024, we use the pyMC implementation of No U-Turns Hamiltonian Monte Carlo (NUTS) to sample from $p_{\text{{\it in vitro}}}$. Sampling from $p_{\text{{\it in vitro}, WT}}$ would prohibitively expensive. Thus, we instead simulateous calibrate the {\it in vitro} and WT models by estimating the maximum a posteri (MAP). However, the pyMC MAP search algorithm method failed to converge due to stiffness of our ODE and poor prior estimates. To improve convergence, we use NUTS to burn-in chains in the neighbourhood of the MAP and initialize pyMC MAP search algorithm method at the final NUTS sample (Fig \ref{fig:Fig7}b). 

 
 
\subsubsection*{OD Growth Parameters}\label{ssec:ODGrowth}

We fit the OD time series to a sigmoid model independent of metabolite calibration. Using pyMC, we calibrate the model
\begin{equation}
\label{Eq:sigmoid}
N(t) = b + \frac{L}{1 + e^{r\times(t^{\ast} -t)}} 
\end{equation}
to the square root transformed OD time series. The $t_1$ and $t_2$ parameters in equation \ref{Eq:sigmoid} are related to the initial condition of the logistic differential equations in equation \ref{Eq:InVivoModel}:
\begin{align*}
N(0) &=\frac{ L}{1 + e^{r t^{\ast}}}
\end{align*}
We also have that $K=L$. We take the posterior sigmoid parameter mean as a deterministic input in the Bayesian model.  

\subsection*{Parameter Constraints and Prior Distributions}\label{ssec:parameterconstraints}
\subsubsection*{Kinetic Parameters}

We incorporate Michaelis-Menten parameters in our mass-action kinetic parameter sampling. Following the previous Archer et al. 2024, we first solve the mass-action quasi-steady-state approximation of each enzyme. This allows us to derive expressions for Michaelis-Menten parameters in terms of mass-action kinetic parameters (SI Table \ref{Table:Chapter6MMFormulae}). We also derive expressions for the K\textsubscript{eq}, a thermodynamic measure of enzyme reversibility, in terms of mass-action kinetics. We then simultaneously solve expressions of measured Michaelis-Menten parameters and K\textsubscript{eq} for mass-action kinetic parameters.

As the number of known macroscopic kinetic parameters does not equal the number of mass-action parameters, we have two mass-action parameter groups: free parameters, underdetermined from prior measurements, and leading parameters, completely determined by the simultaneous solve of prior measurement expressions and free parameters (SI Table \ref{Table:Chapter6ConstrainedKineticswithFactors}). Thus in our pyMC model, we sample free mass-action parameters and K\textsubscript{eq} from our posterior while computing leading mass-action parameters from their respective expressions and ensuring strict positivity (SI Table  \ref{Table:Chapter6FreeVariableKinetics}). We take all Michaelis-Menten parameters that span a tight interval to be constant. 

\subsubsection*{MCP Parameters}
MCP properties were informed with experimental findings. The number of Pdu enzymes per MCP was taken to be uniform over the range identified by Yang et al. 2020 \cite{Yang2020}. We modeled the number of MCPs per cell as a transformed alpha distribution over the interval, $[2, 10]$ \cite{Kennedy2022}. The MCP metabolite permeabilities were taken as a truncated normal distribution over 10\textsuperscript{-14}-10\textsuperscript{-3} m/s. The MCP cofactor permeabilities were taken as a truncated normal distribution over 10\textsuperscript{-14}-10\textsuperscript{-3} m/s or 10\textsuperscript{-10}-10\textsuperscript{-3} m/s.

\subsubsection*{Cell/Cytosolic Parameters}

Cytosolic enzyme concentrations are taken to be uniform over 10\textsuperscript{-3} and 0.5 mM. We took the NADH:NAD+ ratio to be truncated normal over the range, 0.3-1.3 as the {\it in vivo} assays held under microaerobic conditions \cite{Grose2006}.  The oxaloacetate concentration was taken to be uniform over 10\textsuperscript{-4} - 10\textsuperscript{-3} mM \cite{Park2016}. NAD(H) pool, ATP:ADP ratio, total adenosine pool, AMP, phosphate, pyrophosphate and water were fixed at 0.75 mM \cite{Grose2006}, 30 \cite{Bennett2009,Park2016,Tran1998}, 2.5 mM \cite{Mempin2013,Bennett2009,Park2016}, 10\textsuperscript{-1} mM \cite{Bennett2009,Park2016}, 20 mM \cite{Nesmeyanova2000,Park2016,Nausch2004}, $\approx 0$ \cite{Wolfe2005,Baykov2013} and 4 $\times$ 10\textsuperscript{4} mM, respectively. 

1,2-Propanediol and 1-propanol membrane permeability were determined to be 2 $\times$ 10\textsuperscript{-6} \cite{Orbach1980} and 6.5 $\times$ 10\textsuperscript{-5} m/s \cite{Brahm1983}. As the Pdu operon encodes an aquaporin, PduF, for 1,2-propanediol diffusion, 1,2-PD membrane permeability is likely greater than 2 $\times$ 10\textsuperscript{-6} m/s. The cell membrane permeability of propionaldehyde, propionyl-phosphate, and propionate is not known to be estimated or measured. We find that cytosolic membrane permeability $>$ 10\textsuperscript{-6} m/s entrains cytosolic dynamics to external dynamics. All reactant cell membrane permeability except membrane impenetrable, CoA and propionyl-CoA \cite{Pietrocola2015, Visser2006}, were then fixed to 10\textsuperscript{-4} m/s. 

\subsection*{Compartment expression and purification}

Pdu MCPs were purified using a differential centrifugation method as previously described  \cite{Sinha2012, Nichols2019}. We inoculated single colonies of strains into 5 mL of lysogeny broth (LB) liquid media which were incubated at 30 \degree C, 225 rpm for 24 hours. The overnight culture was subcultured 1:1,000 into 200 mL of No Carbon Essential (NCE) media comprising 29 mM potassium phosphate monobasic, 34 mM potassium phosphate dibasic, 17 mM sodium ammonium hydrogen phosphate, supplemented with 50 $\mu$M ferric citrate, 1 mM magnesium sulfate, 42 mM succinate as a carbon source, and 55 mM 1,2-propanediol as the {\it pdu} operon inducer in a 1 L Erlenmeyer flask. The NCE subculture was grown at 37 \degree C at 225 rpm until the cultures reached an OD\textsubscript{600} of 1.0-1.5. 200 mL of cells were spun down at 5,000 x g for 5 minutes at 4 \degree C. The supernatant was decanted into bleach and the pellet was resuspended in lysis buffer (32 mM Tris-HCl, 200 mM potassium chloride, 5 mM magnesium chloride, 0.6\% (v/v) 1,2-propanediol, 0.6\% (w/w) octylthioglucoside, 5 mM $\beta$-mercaptoethanol, 0.8 mg/mL lysozyme (Thermo Fisher Scientific), 0.04 units/mL DNase I (New England Biolabs, Inc.) pH 7.5–8.0. The resuspended cells underwent a 30-minute incubation in this lysis buffer at room temperature, with gentle rocking at 60 rpm. Following the lysis period, the lysate was kept on ice for 5 minutes and then clarified by centrifugation (12,000 x g, 5 minutes, 4 \degree C) twice. To isolate the MCPs from the clarified lysate, centrifugation was performed at 21,000 x g for 20 minutes at 4 \degree C in a swinging bucket rotor. The supernatant was discarded, and the pellet was washed with buffer (32 mM Tris-HCl, 200 mM KCl, 5 mM MgCl\textsubscript{2}, 0.6\% (v/v) 1,2-propanediol, 0.6\% (w/w) OTG, pH 7.5–8.0). Subsequent MCP pelleting was achieved by spinning at 21,000 x g for 20 minutes at 4 \degree C in a swinging bucket rotor. The supernatant was removed, and the MCP pellet was resuspended in buffer (50 mM Tris-HCl, 50 mM KCl, 5 mM MgCl\textsubscript{2}, pH 8.0) and stored at 4 \degree C until needed. The concentration of the purified Pdu MCPs was determined using a bicinchoninic acid assay (Thermo Scientific).  

\subsection*{{\it in vitro} assay with lysate}

Cell-free reactions were performed in 30 $\mu$L in 2 mL Eppendorf tubes and incubated at 30 °C as previously described \cite{Dudley2016}. The standard reaction contained the following components: 200 mM glucose, acetate salts (8mM magnesium acetate, 10mM ammonium acetate, 134 mM potassium acetate), 50 ug/mL kanamycin, 100 mM Bis-Tris, 1,2-propanediol (0.4\%), and 20uM Ado B12. All reagents and chemicals were purchased from Sigma Aldrich.

Extract concentration for all CFME reactions was 10 mg/mL total protein. The relative levels of each MCP were adjusted to maintain a total MCP concentration of .067 mg/mL. Reactions were quenched by precipitating proteins using 30 $\mu$L of 10\% trichloroacetic acid and centrifuging at 21,000 x g for 10 minutes at 4 °C. The resulting supernatant was then stored at -80 °C until analysis by HPLC.

\subsection*{Strain generation}
Strain modifications were incorporated by applying the lambda red recombineering technique, as previously documented \cite{Datta2006}. Within the target genomic locus, a cassette comprising a chloramphenicol resistance gene ({\it cat}) and a sucrose sensitivity gene ({\it sacB}), amplified from the TUC01 genome with primers that have homology to the target locus, was introduced. Subsequent replacement of the {\it catsacB} cassette involved the introduction of either single-stranded DNA for knockout purposes or a PCR product of the entire gene with homology to the target locus. Gene incorporation was validated by assessing sucrose sensitivity on sucrose plates and with subsequent verification through Sanger sequencing (Genewiz). 

\subsection*{{\it in vivo} growth assay}

Growth curves were performed as previously described \cite{Kennedy2022}. Single colonies from an LB plate were inoculated into 5 mL of LB media and incubated at overnight at 37 \degree C and 225 rpm for 12-16 hours. The overnight culture was subcultured to an OD\textsubscript{600} of 0.05 in 250 mL of NCE media (29 mM potassium phosphate monobasic, 34 mM potassium phosphate dibasic, 17 mM sodium ammonium hydrogen phosphate) supplemented with 50 $\mu$M ferric citrate, 1 mM magnesium sulfate, 150 nM adenosylcobalamin (Santa Cruz Biotechnology), and 55 mM 1,2-propanediol. These cultures were grown at 37 \degree C and 225 rpm in 250 mL Erlenmeyer flasks. 

500 $\mu$L of cultures was taken for each time point. The OD\textsubscript{600} of the 500 $\mu$L sample was taken using a BioTek Synergy HTX multi-mode plate reader. After the 9 hour time point, the sample for the OD600 measurements was being diluted 1:5 in fresh NCE media to remain in the linear range of the machine. The remaining sample was pelleted at 13,000 x g for 5 minutes. The supernatant was collected and filtered through Corning\textsuperscript{TM} Costar\textsuperscript{TM} Spin-X LC filters. The filtered supernatant was collected and frozen at -20 \degree C for later HPLC analysis.

\subsection*{Metabolite quantification}
Pdu metabolites were quantified using an Agilent 1260 HPLC system. The injection volume was 5 $\mu$L, and the analytical column used was a Rezex\textsubscript{TM} ROA-Organic Acid H+ (8\%) LC Column (Phenomenex) at 35 \degree C with 5 mM sulfuric acid as the mobile phase at a flow rate of 0.4 mL/min for 45 minutes. A refractive index detector (RID) was used to detect metabolites, and we compared the resulting peaks from the RID to peak areas of known dilutions of 1,2-propanediol, propionaldehyde, 1-propanol and propionate as described previously\cite{Lee2017}.    

\subsection*{Phase contrast and fluorescence microscopy}

Fluorescence patterns of fluorescent reporters were assessed through phase-contrast and fluorescent microscopy techniques. Imaging of the cells was performed on a Nikon Eclipse Ni-U upright microscope, utilizing a 100X oil immersion objective and an Andor Clara digital camera. Image acquisition was executed using the NIS Elements Software (Nikon). To capture GFP fluorescence, a C-FL Endow GFP HYQ bandpass filter was employed with a 500 ms exposure time. Cells from a single colony were inoculated into LB-media and incubated at 37 \degree C, 225 rpm for 16-20 hours. This overnight culture diluted in NCE media supplemented with 50 $\mu$M ferric citrate, 1 mM magnesium sulfate, 42 mM succinate as a carbon source, and 55 mM 1,2-propanediol as the pdu operon inducer. This diluted culture was allowed to grow at 37 °C, 225 rpm. After 16 hours, 1 mL of the cell culture was centrifuged at 4,000 xg for 90 seconds. 800 $\mu$L of the supernatant was removed, and the remaining cell pellet was then resuspended in the leftover 200 $\mu$L of supernatant. Subsequently, 1.48 $\mu$L of cells were pipetted onto Fisherbrand\textsuperscript{TM} frosted microscope slides (Thermo Fisher Scientific Cat\# 12-550-343) and sandwiched between the slide and a 22 $\times$ 22 mm, \#1.5 thickness coverslip (VWR Cat\# 16004-302). Prior to use, both the microscopy slides and slide covers were cleaned with 70\% ethanol.  

\section*{Acknowledgements}

This work was funded in part by the US Army Contracting Command - Rock Island (grant  W52P1J-21-9-3023 to D.T.E. and N.M.M.), the Army Research Office (grant W911NF-19-1-0298 to D.T.E.), the Department of Energy (grant DE-SC0019337 to D.T.E. and N.M.M. ). BJP was partially funded by a National Science Foundation Graduate training grant (grant DGE-2021900) via the Northwestern University Synthetic Biology Across the Scales Training Program.


\bibliographystyle{unsrt}

\bibliography{diss.bib}

@article{Horswill1999a,
  title={Salmonella typhimurium LT2 Catabolizes Propionate via the 2-Methylcitric Acid Cycle},
  author={Alexander R. Horswill and Jorge C. Escalante-Semerena},
  journal={Journal of Bacteriology},
  year={1999},
  volume={181},
  pages={5615 - 5623}
}

@article{Huseby2013,
author = {Douglas L. Huseby and John R. Roth},
title = {Evidence that a Metabolic Microcompartment Contains and Recycles Private Cofactor Pools},
journal = {Journal of Bacteriology},
volume = {195},
number = {12},
pages = {2864-2879},
year = {2013},
doi = {10.1128/jb.02179-12},

URL = {https://journals.asm.org/doi/abs/10.1128/jb.02179-12},
eprint = {https://journals.asm.org/doi/pdf/10.1128/jb.02179-12}
}

@article{Palacios2003,
author = {Palacios, Sergio and Starai, Vincent J. and Escalante-Semerena, Jorge C.},
doi = {10.1128/JB.185.9.2802-2810.2003},
issn = {00219193},
journal = {Journal of Bacteriology},
number = {9},
pages = {2802--2810},
pmid = {12700259},
title = {{Propionyl coenzyme A is a common intermediate in the 1,2-propanediol and propionate catabolic pathways needed for expression of the prpBCDE operon during growth of Salmonella enterica on 1,2-propanediol}},
volume = {185},
year = {2003}
}

@article{Liu2007,
author = {Liu, Yu and Leal, Nicole A. and Sampson, Edith M. and Johnson, Celeste L.V. and Havemann, Gregory D. and Bobik, Thomas A.},
doi = {10.1128/JB.01151-06},
issn = {00219193},
journal = {Journal of Bacteriology},
number = {5},
pages = {1589--1596},
pmid = {17158662},
title = {{PduL is an evolutionarily distinct phosphotransacylase involved in B 12-dependent 1,2-propanediol degradation by Salmonella enterica serovar typhimurium LT2}},
volume = {189},
year = {2007}
}

@article{Nichols2020,
title = {A genomic integration platform for heterologous cargo encapsulation in 1,2-propanediol utilization bacterial microcompartments},
journal = {Biochemical Engineering Journal},
volume = {156},
pages = {107496},
year = {2020},
issn = {1369-703X},
doi = {https://doi.org/10.1016/j.bej.2020.107496},
url = {https://www.sciencedirect.com/science/article/pii/S1369703X20300115},
author = {Taylor M. Nichols and Nolan W. Kennedy and Danielle Tullman-Ercek}
}

@article{Orbach1980,
    author = {Orbach, E and Finkelstein, A},
    title = "{The nonelectrolyte permeability of planar lipid bilayer membranes.}",
    journal = {Journal of General Physiology},
    volume = {75},
    number = {4},
    pages = {427-436},
    year = {1980},
    month = {04},
    issn = {0022-1295},
    doi = {10.1085/jgp.75.4.427},
    url = {https://doi.org/10.1085/jgp.75.4.427},
    eprint = {https://rupress.org/jgp/article-pdf/75/4/427/595765/427.pdf},
}

@article{Yang2020,
author = {Yang, Mengru and Simpson, Deborah and Wenner, Nicolas and Brownridge, Philip and Harman, Victoria and Hinton, Jay and Beynon, Rob and Liu, Luning},
year = {2020},
month = {04},
pages = {1976},
title = {Decoding the stoichiometric composition and organisation of bacterial metabolosomes},
volume = {11},
journal = {Nature Communications},
doi = {10.1038/s41467-020-15888-4}
}

@article{Dudley2016,
author = {Dudley, Quentin M. and Anderson, Kim C. and Jewett, Michael C.},
title = {Cell-Free Mixing of Escherichia coli Crude Extracts to Prototype and Rationally Engineer High-Titer Mevalonate Synthesis},
journal = {ACS Synthetic Biology},
volume = {5},
number = {12},
pages = {1578-1588},
year = {2016},
doi = {10.1021/acssynbio.6b00154},
    note ={PMID: 27476989},

URL = { 
    
        https://doi.org/10.1021/acssynbio.6b00154
    
    

},
eprint = { 
    
        https://doi.org/10.1021/acssynbio.6b00154
    
    

}

}

@article{Kennedy2022,
author = {Nolan W. Kennedy and Carolyn E. Mills and Charlotte H. Abrahamson and Andre G. Archer and Sasha Shirman and Michael C. Jewett and Niall M. Mangan and Danielle Tullman-Ercek},
title = {Linking the Salmonella Enterica 1,2-Propanediol Utilization Bacterial Microcompartment Shell to the Enzymatic Core via the Shell Protein PduB},
journal = {Journal of Bacteriology},
volume = {204},
number = {9},
pages = {e00576-21},
year = {2022},
doi = {10.1128/jb.00576-21},

URL = {https://journals.asm.org/doi/abs/10.1128/jb.00576-21},
eprint = {https://journals.asm.org/doi/pdf/10.1128/jb.00576-21}
}

@article{Smith1980,
title = {Purification, properties, and kinetic mechanism of coenzyme A-linked aldehyde dehydrogenase from Clostridium kluyveri},
journal = {Archives of Biochemistry and Biophysics},
volume = {203},
number = {2},
pages = {663-675},
year = {1980},
issn = {0003-9861},
doi = {https://doi.org/10.1016/0003-9861(80)90224-6},
url = {https://www.sciencedirect.com/science/article/pii/0003986180902246},
author = {Linda Tombras Smith and Nathan O. Kaplan}
}

@article{Mempin2013,
author = {Mempin, Roberto and Tran, Helen and Chen, Connie and Gong, Hao and Ho, Katharina and Lu, Sangwei},
year = {2013},
month = {12},
pages = {301},
title = {Release of extracellular ATP by bacteria during growth},
volume = {13},
journal = {BMC microbiology},
doi = {10.1186/1471-2180-13-301}
}

@article{Tran1998,
author = {Tran, Quang Hon and Unden, Gottfried},
title = {Changes in the proton potential and the cellular energetics of Escherichia coli during growth by aerobic and anaerobic respiration or by fermentation},
journal = {European Journal of Biochemistry},
volume = {251},
number = {1-2},
pages = {538-543},
doi = {https://doi.org/10.1046/j.1432-1327.1998.2510538.x},
url = {https://febs.onlinelibrary.wiley.com/doi/abs/10.1046/j.1432-1327.1998.2510538.x},
eprint = {https://febs.onlinelibrary.wiley.com/doi/pdf/10.1046/j.1432-1327.1998.2510538.x},
year = {1998}
}

@article{CHOHNAN1998,
    author = {CHOHNAN, Shigeru and IZAWA, Hiroaki and NISHIHARA, Hirofumi and TAKAMURA, Yoshichika},
    title = "{Changes in Size of Intracellular Pools of Coenzyme A and Its Thioesters in Escherichia coli K-12 Cells to Various Carbon Sources and Stresses}",
    journal = {Bioscience, Biotechnology, and Biochemistry},
    volume = {62},
    number = {6},
    pages = {1122-1128},
    year = {1998},
    month = {01},
    issn = {0916-8451},
    doi = {10.1271/bbb.62.1122},
    url = {https://doi.org/10.1271/bbb.62.1122},
    eprint = {https://academic.oup.com/bbb/article-pdf/62/6/1122/35014558/bbb1122.pdf},
}

@article{Albe1990,
title = {Cellular concentrations of enzymes and their substrates},
journal = {Journal of Theoretical Biology},
volume = {143},
number = {2},
pages = {163-195},
year = {1990},
issn = {0022-5193},
doi = {https://doi.org/10.1016/S0022-5193(05)80266-8},
url = {https://www.sciencedirect.com/science/article/pii/S0022519305802668},
author = {Kathy R. Albe and Margaret H. Butler and Barbara E. Wright}
}

@article{pymc2023,
  title={PyMC: A Modern and Comprehensive Probabilistic Programming Framework in Python},
  author={Abril-Pla Oriol and Andreani Virgile and Carroll Colin and Dong Larry and Fonnesbeck Christopher J. and Kochurov Maxim and Kumar Ravin and Lao Jupeng and Luhmann Christian C. and Martin Osvaldo A. and Osthege Michael and Vieira Ricardo and Wiecki Thomas and Zinkov Robert},
  journal = {{PeerJ} Computer Science},
  publisher = {{PeerJ}},
  volume={9},
  pages={e1516},
  year={2023},
  doi={10.7717/peerj-cs.1516}
}

@article{Baykov2013,
author = {Alexander A. Baykov and Anssi M. Malinen and Heidi H. Luoto and Reijo Lahti},
title = {Pyrophosphate-Fueled Na<sup>+</sup> and H<sup>+</sup> Transport in Prokaryotes},
journal = {Microbiology and Molecular Biology Reviews},
volume = {77},
number = {2},
pages = {267-276},
year = {2013},
doi = {10.1128/MMBR.00003-13},

URL = {https://journals.asm.org/doi/abs/10.1128/MMBR.00003-13},
eprint = {https://journals.asm.org/doi/pdf/10.1128/MMBR.00003-13}
}

@article{Nesmeyanova2000,
author = {Nesmeyanova, M},
year = {2000},
month = {04},
pages = {309-14},
title = {Polyphosphates and Enzymes of Polyphosphate Metabolism in Escherichia coli},
volume = {65},
journal = {Biochemistry. Biokhimiia}
}

@article{Husain2008,
author = {Husain, Maroof and Bourret, Travis and Mccollister, Bruce and Jones-Carson, Jessica and Laughlin, James and Vazquez-Torres, Andres},
year = {2008},
month = {04},
pages = {7682-9},
title = {Nitric Oxide Evokes an Adaptive Response to Oxidative Stress by Arresting Respiration},
volume = {283},
journal = {The Journal of biological chemistry},
doi = {10.1074/jbc.M708845200}
}

@article{Grose2006,
author = {Julianne H. Grose  and Lisa Joss  and Sidney F. Velick  and John R. Roth },
title = {Evidence that feedback inhibition of NAD kinase controls responses to oxidative stress},
journal = {Proceedings of the National Academy of Sciences},
volume = {103},
number = {20},
pages = {7601-7606},
year = {2006},
doi = {10.1073/pnas.0602494103},
URL = {https://www.pnas.org/doi/abs/10.1073/pnas.0602494103},
eprint = {https://www.pnas.org/doi/pdf/10.1073/pnas.0602494103}}

@article{Park2016,
author = {Park, Junyoung and Rubin, Sara and Xu, Yi-Fan and Amador, Daniel and Fan, Jing and Shlomi, Tomer and Rabinowitz, Joshua},
year = {2016},
month = {05},
pages = {},
title = {Metabolite concentrations, fluxes and free energies imply efficient enzyme usage},
volume = {12},
journal = {Nature chemical biology},
doi = {10.1038/nchembio.2077}
}

@article{Bennett2009,
author = {Bennett, Bryson and Kimball, Elizabeth and Gao, Melissa and Osterhout, Robin and Van Dien, Steve and Rabinowitz, Joshua},
year = {2009},
month = {07},
pages = {593-9},
title = {Absolute Metabolite Concentrations and Implied Enzyme Active Site Occupancy in Escherichia coli},
volume = {5},
journal = {Nature chemical biology},
doi = {10.1038/nchembio.186}
}

@article{Liu2019,
author = {Liu, Yang and Landick, Robert and Raman, Srivatsan},
title = {A Regulatory NADH/NAD+ Redox Biosensor for Bacteria},
journal = {ACS Synthetic Biology},
volume = {8},
number = {2},
pages = {264-273},
year = {2019},
doi = {10.1021/acssynbio.8b00485},

URL = { 
        https://doi.org/10.1021/acssynbio.8b00485
    
},
eprint = { 
        https://doi.org/10.1021/acssynbio.8b00485
    
}

}

@article{Wolfe2005,
author = {Alan J. Wolfe},
title = {The Acetate Switch},
journal = {Microbiology and Molecular Biology Reviews},
volume = {69},
number = {1},
pages = {12-50},
year = {2005},
doi = {10.1128/MMBR.69.1.12-50.2005},

URL = {https://journals.asm.org/doi/abs/10.1128/MMBR.69.1.12-50.2005},
eprint = {https://journals.asm.org/doi/pdf/10.1128/MMBR.69.1.12-50.2005}

}

@article{Nausch2004,
author = {Nausch, Monika and Nausch, G},
year = {2004},
month = {12},
pages = {237-245},
title = {Bacterial utilization of phosphorus pools after nitrogen and carbon amendment and its relation to alkaline phosphatase activity},
volume = {37},
journal = {Aquatic Microbial Ecology - AQUAT MICROB ECOL},
doi = {10.3354/ame037237}
}

@article{Wimpenny1972,
author = {Julian W. T. Wimpenny and Anne Firth},
title = {Levels of Nicotinamide Adenine Dinucleotide and Reduced Nicotinamide Adenine Dinucleotide in Facultative Bacteria and the Effect of Oxygen},
journal = {Journal of Bacteriology},
volume = {111},
number = {1},
pages = {24-32},
year = {1972},
doi = {10.1128/jb.111.1.24-32.1972},

URL = {https://journals.asm.org/doi/abs/10.1128/jb.111.1.24-32.1972},
eprint = {https://journals.asm.org/doi/pdf/10.1128/jb.111.1.24-32.1972}
}

@article{Zhou2011,
author = {Yongjin Zhou and Lei Wang and Fan Yang and Xinping Lin and Sufang Zhang and Zongbao K. Zhao},
title = {Determining the Extremes of the Cellular NAD(H) Level by Using an Escherichia coli NAD<sup>+</sup>-Auxotrophic Mutant },
journal = {Applied and Environmental Microbiology},
volume = {77},
number = {17},
pages = {6133-6140},
year = {2011},
doi = {10.1128/AEM.00630-11},

URL = {https://journals.asm.org/doi/abs/10.1128/AEM.00630-11},
eprint = {https://journals.asm.org/doi/pdf/10.1128/AEM.00630-11}
}

@article{Leonardo1996,
author = {M R Leonardo and Y Dailly and D P Clark},
title = {Role of NAD in regulating the adhE gene of Escherichia coli},
journal = {Journal of Bacteriology},
volume = {178},
number = {20},
pages = {6013-6018},
year = {1996},
doi = {10.1128/jb.178.20.6013-6018.1996},

URL = {https://journals.asm.org/doi/abs/10.1128/jb.178.20.6013-6018.1996},
eprint = {https://journals.asm.org/doi/pdf/10.1128/jb.178.20.6013-6018.1996}
}

@article{Andersen1977,
author = {Andersen, K.B. and Meyenburg, Kasper},
year = {1977},
month = {07},
pages = {4151-6},
title = {Charges of nicotinamide adenine-nucleotides and adenylate energy-charge as regulatory parameters of metabolism in Escherichia Coli},
volume = {252},
journal = {The Journal of biological chemistry},
doi = {10.1016/S0021-9258(17)40245-6}
}

@article{Sampson2008,
author = {Edith M. Sampson  and Thomas A. Bobik },
title = {Microcompartments for {B}$_{12}$-Dependent 1,2-Propanediol Degradation Provide Protection from DNA and Cellular Damage by a Reactive Metabolic Intermediate},
journal = {Journal of Bacteriology},
volume = {190},
number = {8},
pages = {2966-2971},
year = {2008},
doi = {10.1128/JB.01925-07},
URL = {https://journals.asm.org/doi/abs/10.1128/JB.01925-07},
eprint = {https://journals.asm.org/doi/pdf/10.1128/JB.01925-07}
}

@article{Horswill2001b,
author = {Horswill, A. R. and Escalante-Semerena, J. C.},
doi = {10.1021/bi015503b},
issn = {00062960},
journal = {Biochemistry},
month = {apr},
number = {15},
pages = {4703--4713},
pmid = {11294638},
title = {{In vitro conversion of propionate to pyruvate by Salmonella enterica enzymes: 2-methylcitrate dehydratase (PrpD) and aconitase enzymes catalyze the conversion of 2-methylcitrate to 2-methylisocitrate}},
volume = {40},
year = {2001}
}

@article{Dourado2021,
    doi = {10.1371/journal.pbio.3001416},
    author = {Dourado, Hugo AND Mori, Matteo AND Hwa, Terence AND Lercher, Martin J.},
    journal = {PLOS Biology},
    publisher = {Public Library of Science},
    title = {On the optimality of the enzyme–substrate relationship in bacteria},
    year = {2021},
    month = {10},
    volume = {19},
    url = {https://doi.org/10.1371/journal.pbio.3001416},
    pages = {1-18},
    number = {10},

}

@misc{Archer2025,
      title={Uncertainty Quantification of Bacterial Microcompartment Permeability}, 
      author={Andre Archer and Brett J. Palmero and Charlotte Abrahamson and Carolyn E. Mills and Nolan W. Kennedy and Danielle Tullman-Ercek and Niall M. Mangan},
      year={2025},
      eprint={2509.23445},
      archivePrefix={arXiv},
      primaryClass={q-bio.QM},
      url={https://arxiv.org/abs/2509.23445}, 
}

@article {Palmero2025,
	author = {Palmero, Brett J and Catlett, Christina and Abrahamson, Charlotte and Shirman, Sasha and Mills, Carolyn and Jewett, Michael Christopher and Mangan, Niall and Tullman-Ercek, Danielle},
	title = {Evaluation of Bacterial Microcompartment Cofactor Recycling and Permeability with a Model Guided In Vitro Assay},
	elocation-id = {2025.08.20.671389},
	year = {2025},
	doi = {10.1101/2025.08.20.671389},
	publisher = {Cold Spring Harbor Laboratory},
	URL = {https://www.biorxiv.org/content/early/2025/08/23/2025.08.20.671389},
	eprint = {https://www.biorxiv.org/content/early/2025/08/23/2025.08.20.671389.full.pdf},
	journal = {bioRxiv}
}

@article{Smit1975,
author = {Smit, John and Kamio, Y and Nikaido, Hiroshi},
year = {1975},
month = {12},
pages = {942-58},
title = {Outer membrane of {\it salmonella typhimurium}: chemical analysis and freeze fracture studies with lipopolysaccharide mutants},
volume = {124},
journal = {Journal of bacteriology},
doi = {10.1128/JB.124.2.942-958.1975}
}

@article{Cheng2012,
    doi = {10.1371/journal.pone.0047144},
    author = {Cheng, Shouqiang AND Fan, Chenguang AND Sinha, Sharmistha AND Bobik, Thomas A.},
    journal = {PLOS ONE},
    publisher = {Public Library of Science},
    title = {The PduQ Enzyme Is an Alcohol Dehydrogenase Used to Recycle NAD+ Internally within the Pdu Microcompartment of Salmonella enterica},
    year = {2012},
    month = {10},
    volume = {7},
    url = {https://doi.org/10.1371/journal.pone.0047144},
    pages = {1-11},
    number = {10},

}

@article{Mills2022,
author = {Mills, Carolyn E. and Waltmann, Curt and Archer, Andre G. and Kennedy, Nolan W. and Abrahamson, Charlotte H. and Jackson, Alexander D. and Roth, Eric W. and Shirman, Sasha and Jewett, Michael C. and Mangan, Niall M. and {Olvera de la Cruz}, Monica and Tullman-Ercek, Danielle},
doi = {10.1038/s41467-022-31279-3},
issn = {20411723},
journal = {Nature Communications},
number = {1},
pmid = {35768404},
publisher = {Springer US},
title = {{Vertex protein PduN tunes encapsulated pathway performance by dictating bacterial metabolosome morphology}},
volume = {13},
year = {2022}
}

@article{Bobik1999,
author = {Bobik, Thomas A. and Havemann, Gregory D. and Busch, Robert J. and Williams, Donna S. and Aldrich, Henry C.},
doi = {10.1128/jb.181.19.5967-5975.1999},
issn = {00219193},
journal = {Journal of Bacteriology},
number = {19},
pages = {5967--5975},
pmid = {10498708},
title = {{The propanediol utilization (pdu) operon of Salmonella enterica serovar Typhimurium LT2 includes genes necessary for formation of polyhedral organelles involved in coenzyme B12-dependent 1,2-propanediol degradation}},
volume = {181},
year = {1999}
}

@article{Sinha2012,
author = {Sinha, Sharmistha and Cheng, Shouqiang and Fan, Chenguang and Bobik, Thomas A.},
doi = {10.1128/JB.06529-11},
issn = {00219193},
journal = {Journal of Bacteriology},
month = {apr},
number = {8},
pages = {1912--1918},
pmid = {22343294},
title = {{The PduM protein is a structural component of the microcompartments involved in coenzyme B12-dependent 1,2-propanediol degradation by salmonella enterica}},
volume = {194},
year = {2012}
}

@article{Dolan2018,
author = {Dolan, Stephen K and Wijaya, Andre and Geddis, Stephen M and Spring, David R and Silva-rocha, Rafael and Welch, Martin},
doi = {10.1099/mic.0.000604},
pages = {251--259},
title = {{Loving the poison : the methylcitrate cycle and bacterial pathogenesis}},
year = {2018}
}

@article{Noster2019,
author = {Janina Noster and Nicole Hansmeier and Marcus Persicke and Tzu-Chiao Chao and Rainer Kurre and Jasmin Popp and Viktoria Liss and Tatjana Reuter and Michael Hensel},
title = {Blocks in Tricarboxylic Acid Cycle of Salmonella enterica Cause Global Perturbation of Carbon Storage, Motility, and Host-Pathogen Interaction},
journal = {mSphere},
volume = {4},
number = {6},
pages = {10.1128/msphere.00796-19},
year = {2019},
doi = {10.1128/msphere.00796-19},

URL = {https://journals.asm.org/doi/abs/10.1128/msphere.00796-19},
eprint = {https://journals.asm.org/doi/pdf/10.1128/msphere.00796-19}
}

@book{Deshpande2022,
	Title = {Biochemistry, Oxidative Phosphorylation},
	Author = {Deshpande, Ojas A. and Mohiuddin, Shamim S.},
	Publisher = {StatPearls Publishing, Treasure Island (FL)},
	Year = {2022},
	URL = {http://europepmc.org/books/NBK553192},
}

@article{Bobik1997,
author = {Bobik, T.A. and Xu, YP and Jeter, R and Otto, K and Roth, John},
year = {1997},
month = {12},
pages = {6633-9},
title = {Propanediol utilization genes (pdu) of Salmonella Typhimurium: Three genes for the propanediol dehydratase},
volume = {179},
journal = {Journal of bacteriology},
doi = {10.1128/jb.179.21.6633-6639.1997}
}

@article{Chowdhury2014,
author = {Chowdhury, Chiranjit and Sinha, Sharmistha and Chun, Sunny and Yeates, Todd O. and Bobik, Thomas A.},
doi = {10.1128/mmbr.00009-14},
issn = {1092-2172},
journal = {Microbiology and Molecular Biology Reviews},
number = {3},
pages = {438--468},
pmid = {25184561},
title = {{Diverse Bacterial Microcompartment Organelles}},
volume = {78},
year = {2014}
}

@article{Horswill2002,
author = {Horswill, Alexander R. and Escalante-Semerena, Jorge C.},
title = {Characterization of the Propionyl-CoA Synthetase (PrpE) Enzyme of Salmonella enterica: Residue Lys592 Is Required for Propionyl-AMP Synthesis},
journal = {Biochemistry},
volume = {41},
number = {7},
pages = {2379-2387},
year = {2002},
doi = {10.1021/bi015647q},
    note ={PMID: 11841231},

URL = { 
        https://doi.org/10.1021/bi015647q
    
},
eprint = { 
        https://doi.org/10.1021/bi015647q
    
}

}

@article{Beber2021,
author = {Beber, Moritz and Gollub, Mattia and Mozaffari, Dana and Shebek, Kevin and Flamholz, Avi I and Milo, Ron and Noor, Elad},
year = {2021},
month = {11},
pages = {},
title = {eQuilibrator 3.0: a database solution for thermodynamic constant estimation},
volume = {50},
journal = {Nucleic Acids Research},
doi = {10.1093/nar/gkab1106}
}

@article{Toraya2000,
author = {Toraya, T.},
doi = {10.1007/s000180050502},
issn = {1420682X},
journal = {Cellular and Molecular Life Sciences},
number = {1},
pages = {106--127},
pmid = {10949584},
title = {{Radical catalysis of B12 enzymes: Structure, mechanism, inactivation, and reactivation of diol and glycerol dehydratases}},
volume = {57},
year = {2000}
}

@article{Leal2003,
author = {Leal, Nicole A. and Havemann, Gregory D. and Bobik, Thomas A.},
doi = {10.1007/s00203-003-0601-0},
issn = {03028933},
journal = {Archives of Microbiology},
number = {5},
pages = {353--361},
pmid = {14504694},
title = {{PduP is a coenzyme-a-acylating propionaldehyde dehydrogenase associated with the polyhedral bodies involved in B12-dependent 1,2-propanediol degradation by Salmonella enterica serovar Typhimurium LT2}},
volume = {180},
year = {2003}
}

@article{Jakobson2018,
author = {Jakobson, Christopher and Tullman-Ercek, Danielle and Mangan, Niall},
year = {2018},
month = {05},
pages = {},
title = {Spatially organizing biochemistry: Choosing a strategy to translate synthetic biology to the factory},
volume = {8},
journal = {Scientific Reports},
doi = {10.1038/s41598-018-26399-0}
}

@article{Jakobson2017,
    doi = {10.1371/journal.pcbi.1005525},
    author = {Jakobson, Christopher M. AND Tullman-Ercek, Danielle AND Slininger, Marilyn F. AND Mangan, Niall M.},
    journal = {PLOS Computational Biology},
    publisher = {Public Library of Science},
    title = {A systems-level model reveals that 1,2-Propanediol utilization microcompartments enhance pathway flux through intermediate sequestration},
    year = {2017},
    month = {05},
    volume = {13},
    url = {https://doi.org/10.1371/journal.pcbi.1005525},
    pages = {1-24},
    number = {5},

}

@article{Fan2008,
author = {Fan, Chenguang and Bobik, Thomas},
year = {2008},
month = {05},
pages = {11322-9},
title = {The PduX Enzyme of Salmonella enterica Is an L-Threonine Kinase Used for Coenzyme B12 Synthesis},
volume = {283},
journal = {The Journal of biological chemistry},
doi = {10.1074/jbc.M800287200}
}

@article{Mera2010,
author = {Mera, Paola E. and Escalante-Semerena, Jorge C.},
doi = {10.1074/jbc.M109.059485},
issn = {00219258},
journal = {Journal of Biological Chemistry},
number = {5},
pages = {2911--2917},
pmid = {19933577},
publisher = {{\^{A}}{\textcopyright} 2010 ASBMB. Currently published by Elsevier Inc; originally published by American Society for Biochemistry and Molecular Biology.},
title = {{Dihydroflavin-driven adenosylation of 4-coordinate Co(II) corrinoids: Are cobalamin reductases enzymes or electron transfer proteins?}},
url = {http://dx.doi.org/10.1074/jbc.M109.059485},
volume = {285},
year = {2010}
}

@article{Johnson2004,
author = {Johnson, Celeste L.V. and Buszko, Marian L. and Bobik, Thomas A.},
doi = {10.1128/JB.186.23.7881-7887.2004},
issn = {00219193},
journal = {Journal of Bacteriology},
number = {23},
pages = {7881--7887},
pmid = {15547259},
title = {{Purification and initial characterization of the Salmonella enterica PduO ATP:Cob(I)alamin adenosyltransferase}},
volume = {186},
year = {2004}
}

@article{Volkmer2011,
    doi = {10.1371/journal.pone.0023126},
    author = {Volkmer, Benjamin AND Heinemann, Matthias},
    journal = {PLOS ONE},
    publisher = {Public Library of Science},
    title = {Condition-Dependent Cell Volume and Concentration of Escherichia coli to Facilitate Data Conversion for Systems Biology Modeling},
    year = {2011},
    month = {07},
    volume = {6},
    url = {https://doi.org/10.1371/journal.pone.0023126},
    pages = {1-6},
    number = {7},

}

@article{Cheng2010,
author = {Cheng, Shouqiang and Bobik, Thomas A.},
doi = {10.1128/JB.00575-10},
issn = {00219193},
journal = {Journal of Bacteriology},
number = {19},
pages = {5071--5080},
pmid = {20656910},
title = {{Characterization of the PduS cobalamin reductase of Salmonella enterica and its role in the Pdu microcompartment}},
volume = {192},
year = {2010}
}

@article{Visser2006,
    author = {Visser, Wouter F. and van Roermund, Carlo W. T. and Ijlst, Lodewijk and Waterham, Hans R. and Wanders, Ronald J. A.},
    title = "{Metabolite transport across the peroxisomal membrane}",
    journal = {Biochemical Journal},
    volume = {401},
    number = {2},
    pages = {365-375},
    year = {2006},
    month = {12},
    issn = {0264-6021},
    doi = {10.1042/BJ20061352},
    url = {https://doi.org/10.1042/BJ20061352},
    eprint = {https://portlandpress.com/biochemj/article-pdf/401/2/365/646970/bj4010365.pdf},
}

@article{Pietrocola2015,
title = {Acetyl Coenzyme A: A Central Metabolite and Second Messenger},
journal = {Cell Metabolism},
volume = {21},
number = {6},
pages = {805-821},
year = {2015},
issn = {1550-4131},
doi = {https://doi.org/10.1016/j.cmet.2015.05.014},
url = {https://www.sciencedirect.com/science/article/pii/S1550413115002260},
author = {Federico Pietrocola and Lorenzo Galluzzi and José Manuel Bravo-San Pedro and Frank Madeo and Guido Kroemer}
}

@article{Brahm1983,
author = {Brahm, J.},
doi = {10.1085/jgp.81.2.283},
issn = {15407748},
journal = {Journal of General Physiology},
number = {2},
pages = {283--304},
pmid = {6842175},
title = {{Permeability of human red cells to a homologous series of aliphatic alcohols: Limitations of the continuous flow-tube method}},
volume = {81},
year = {1983}
}

@article{Datta2006,
author = {Datta, Simanti and Costantino, Nina and Court, Donald},
year = {2006},
month = {10},
pages = {109-15},
title = {A set of recombineering plasmids for Gram-negative bacteria},
volume = {379},
journal = {Gene},
doi = {10.1016/j.gene.2006.04.018}
}

@incollection{Nichols2019,
title = {Chapter Seven - Cargo encapsulation in bacterial microcompartments: Methods and analysis},
editor = {Claudia Schmidt-Dannert and Maureen B. Quin},
series = {Methods in Enzymology},
publisher = {Academic Press},
volume = {617},
pages = {155-186},
year = {2019},
booktitle = {Metabolons and Supramolecular Enzyme Assemblies},
issn = {0076-6879},
doi = {https://doi.org/10.1016/bs.mie.2018.12.009},
url = {https://www.sciencedirect.com/science/article/pii/S0076687918305056},
author = {Taylor M. Nichols and Nolan W. Kennedy and Danielle Tullman-Ercek}
}

@article{Lee2017,
author = {Lee, Marilyn and Jakobson, Christopher and Tullman-Ercek, Danielle},
year = {2017},
month = {06},
pages = {},
title = {Evidence for Improved Encapsulated Pathway Behavior in a Bacterial Microcompartment through Shell Protein Engineering},
volume = {6},
journal = {ACS Synthetic Biology},
doi = {10.1021/acssynbio.7b00042}
}

@article{strain2009genome,
  title={Genome Scale Reconstruction of aSalmonellaMetabolic Model},
  author={STRAIN, COLI},
  journal={THE JOURNAL OF BIOLOGICAL CHEMISTRY},
  volume={284},
  number={43},
  pages={29480--29488},
  year={2009}
}

@article{horswill1999salmonella,
  title={Salmonella typhimurium LT2 catabolizes propionate via the 2-methylcitric acid cycle},
  author={Horswill, Alexander R and Escalante-Semerena, Jorge C},
  journal={Journal of bacteriology},
  volume={181},
  number={18},
  pages={5615--5623},
  year={1999},
  publisher={American Society for Microbiology}
}

@article{london1999carbon,
  title={Carbon-13 nuclear magnetic resonance study of metabolism of propionate by Escherichia coli},
  author={London, Robert E and Allen, Devon L and Gabel, Scott A and DeRose, Eugene F},
  journal={Journal of bacteriology},
  volume={181},
  number={11},
  pages={3562--3570},
  year={1999},
  publisher={American Society for Microbiology}
}

\pagebreak
\section*{Supplemental Information}
\subsection*{Figures}
\renewcommand{\thefigure}{S\arabic{figure}}
\setcounter{figure}{0}
\begin{figure}[!htp]
\centering
\includegraphics[width=\linewidth]{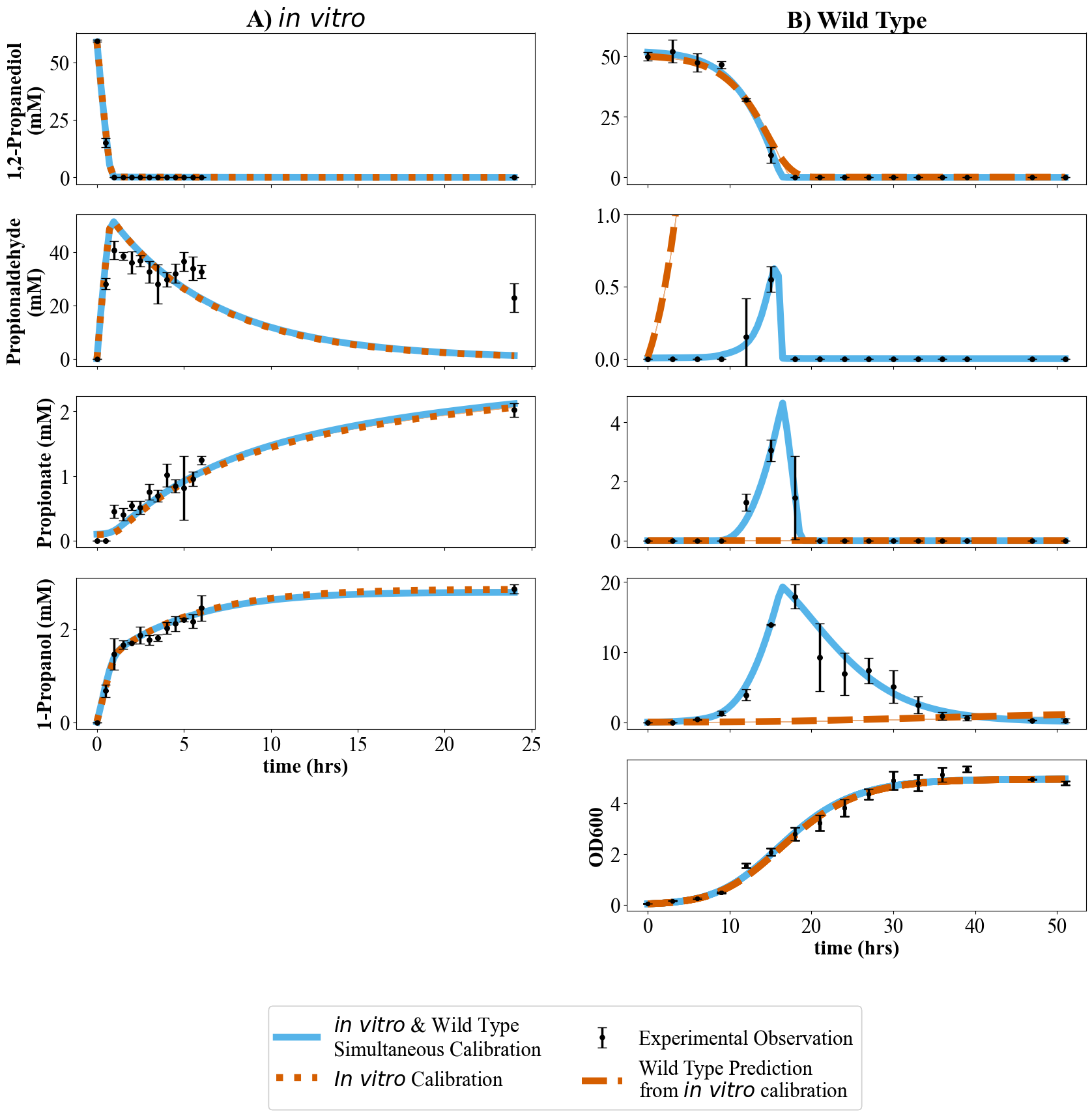}
\caption{Mode 1 fit to in-vitro and WT data with propionaldehyde plot zoomed in 0 to 1 mM. Model assumes PduQ, PduP, PduL, and PduW are located in both the cytosol and MCP, all other Pdu enzymes are localized to the MCP and the presense of promiscous alcohol dehydrogenase.} 
\label{fig:Mode1PduPPduQAdhPduLPduWZoomed}
\end{figure}

\begin{figure}[!htp]
\centering
\includegraphics[width=\linewidth]{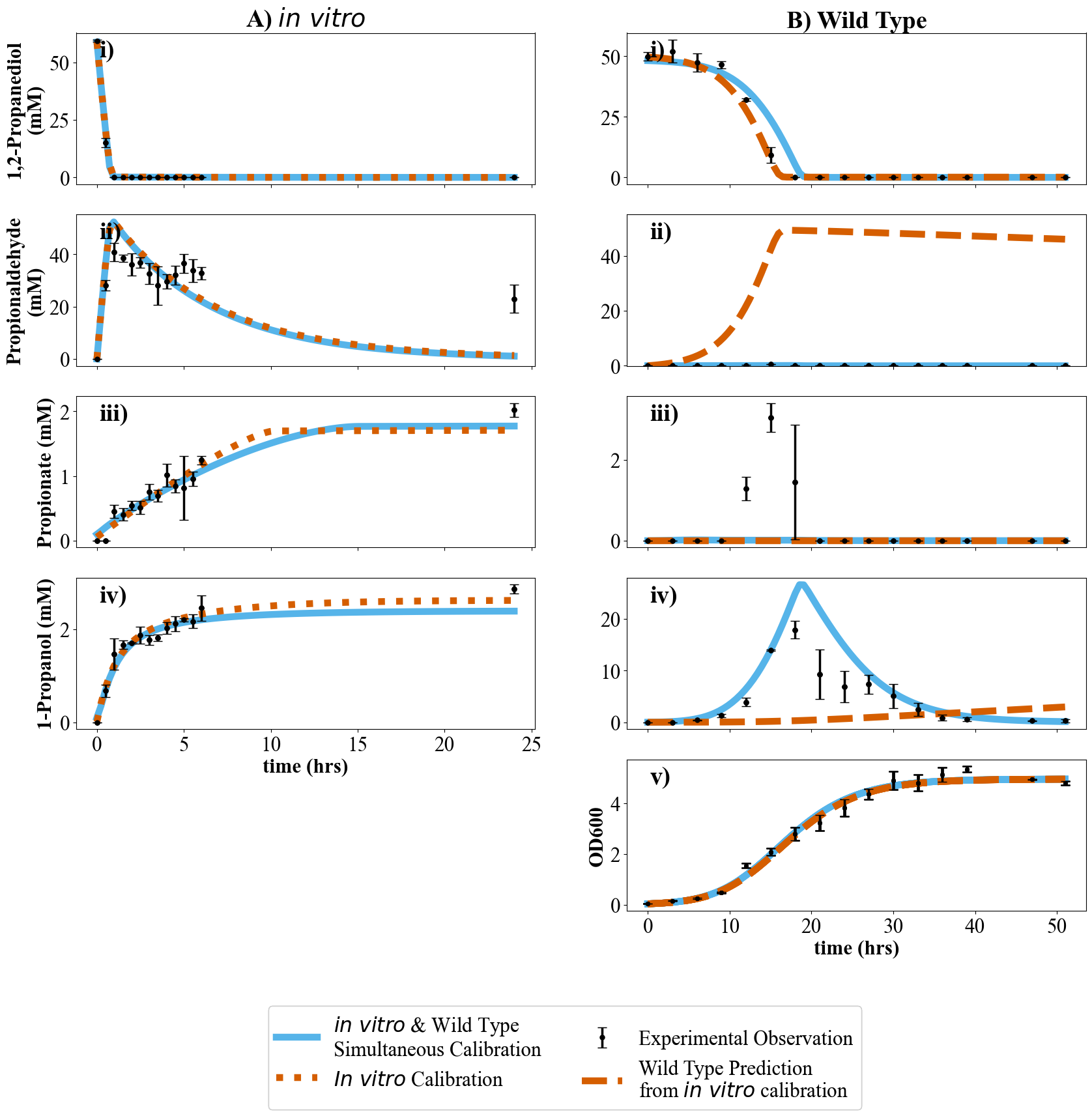}
\caption{Mode 2 fit to in-vitro and WT data with propionaldehyde. Model assumes PduQ, PduP, PduL, and PduW are located in both the cytosol and MCP, all other Pdu enzymes are localized to the MCP and the presense of promiscous alcohol dehydrogenase.} 
\label{fig:Mode2PduPPduQAdhPduLPduW}
\end{figure}

\begin{figure}[!htp]
\centering
\includegraphics[width=\linewidth]{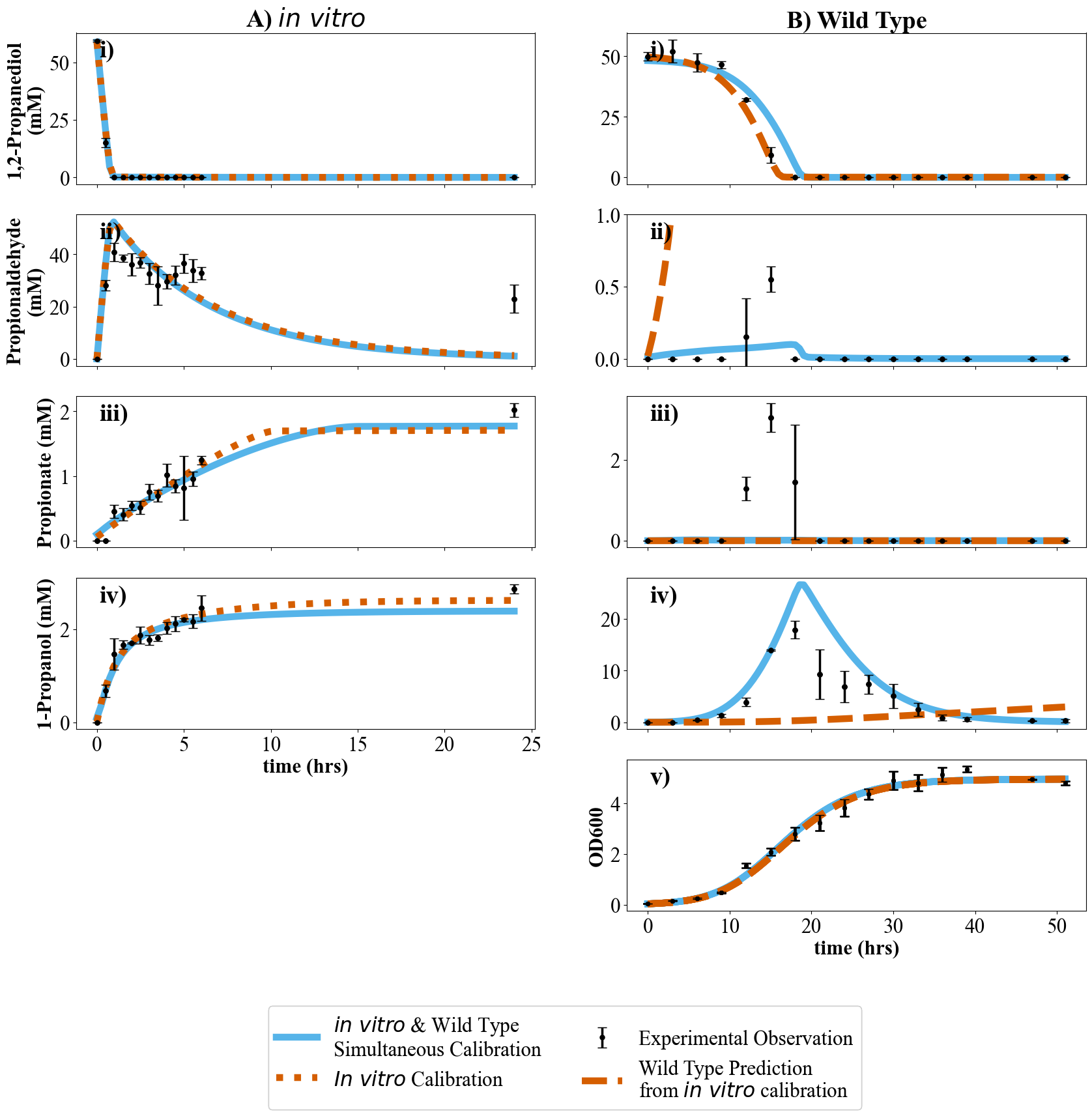}
\caption{Mode 2 fit to in-vitro and WT data with propionaldehyde plot zoomed in 0 to 1 mM. Model assumes PduQ, PduP, PduL, and PduW are located in both the cytosol and MCP, all other Pdu enzymes are localized to the MCP and the presense of promiscous alcohol dehydrogenase.} 
\label{fig:Mode2PduPPduQAdhPduLPduWZoomed}
\end{figure}

\begin{figure}[!htp]
\centering
\includegraphics[width=0.75\linewidth]{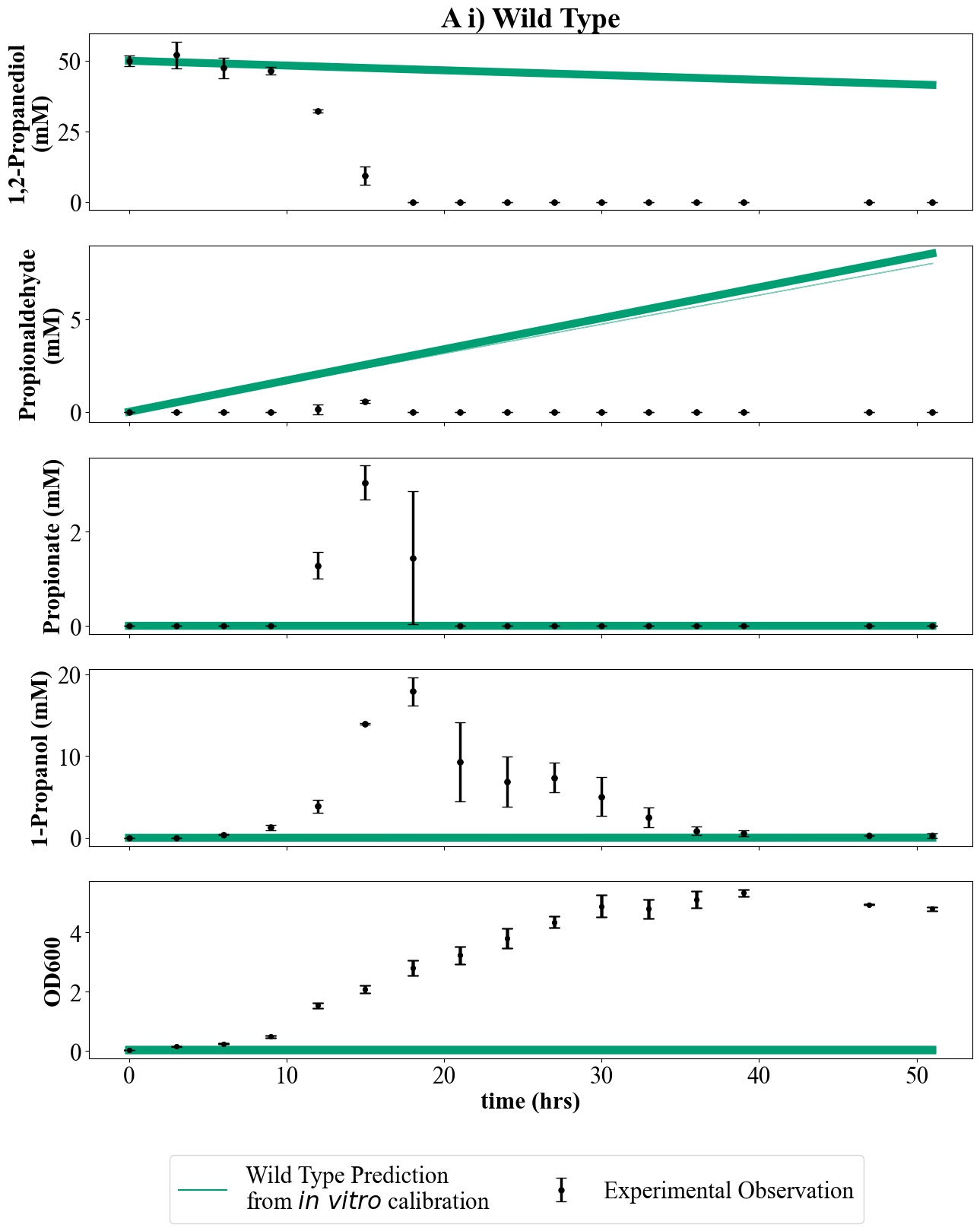}
\caption{WT predicted dynamics using {\it in vitro} posterior parameters, {\it in vivo} prior parameters and OD growth rate set to 0.} 
\label{fig:Mode1WTPredictionNoGrowth}
\end{figure}

\begin{figure}[!htp]
\centering
\includegraphics[width=\linewidth]{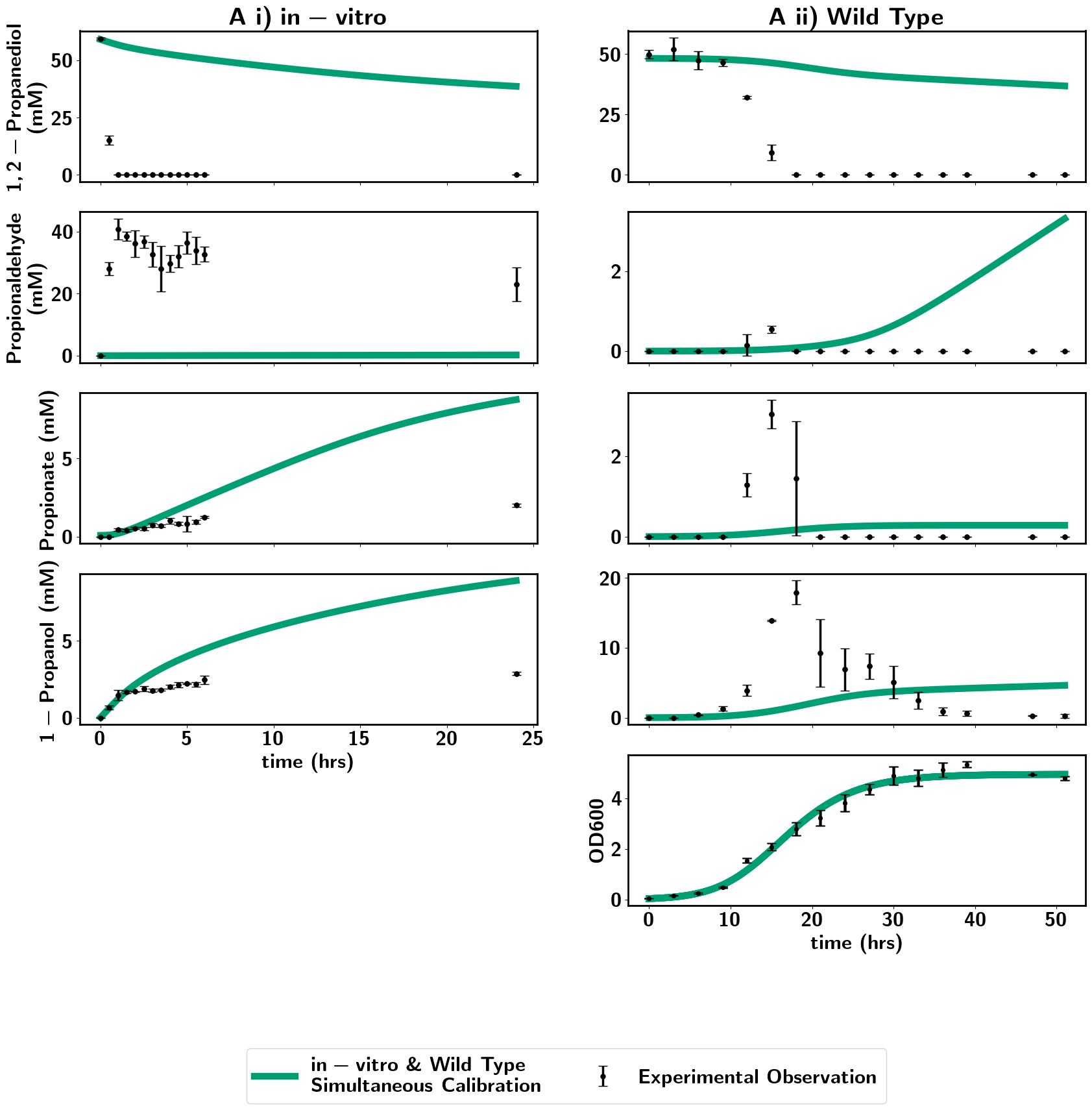}
\caption{Mode 1 fit to in-vitro and WT data. Model assumes PduW are located in both the cytosol and MCP, and all other Pdu enzymes are localized to the MCP.} 
\label{fig:Mode1PduW}
\end{figure}

\begin{figure}[!htp]
\centering
\includegraphics[width=\linewidth]{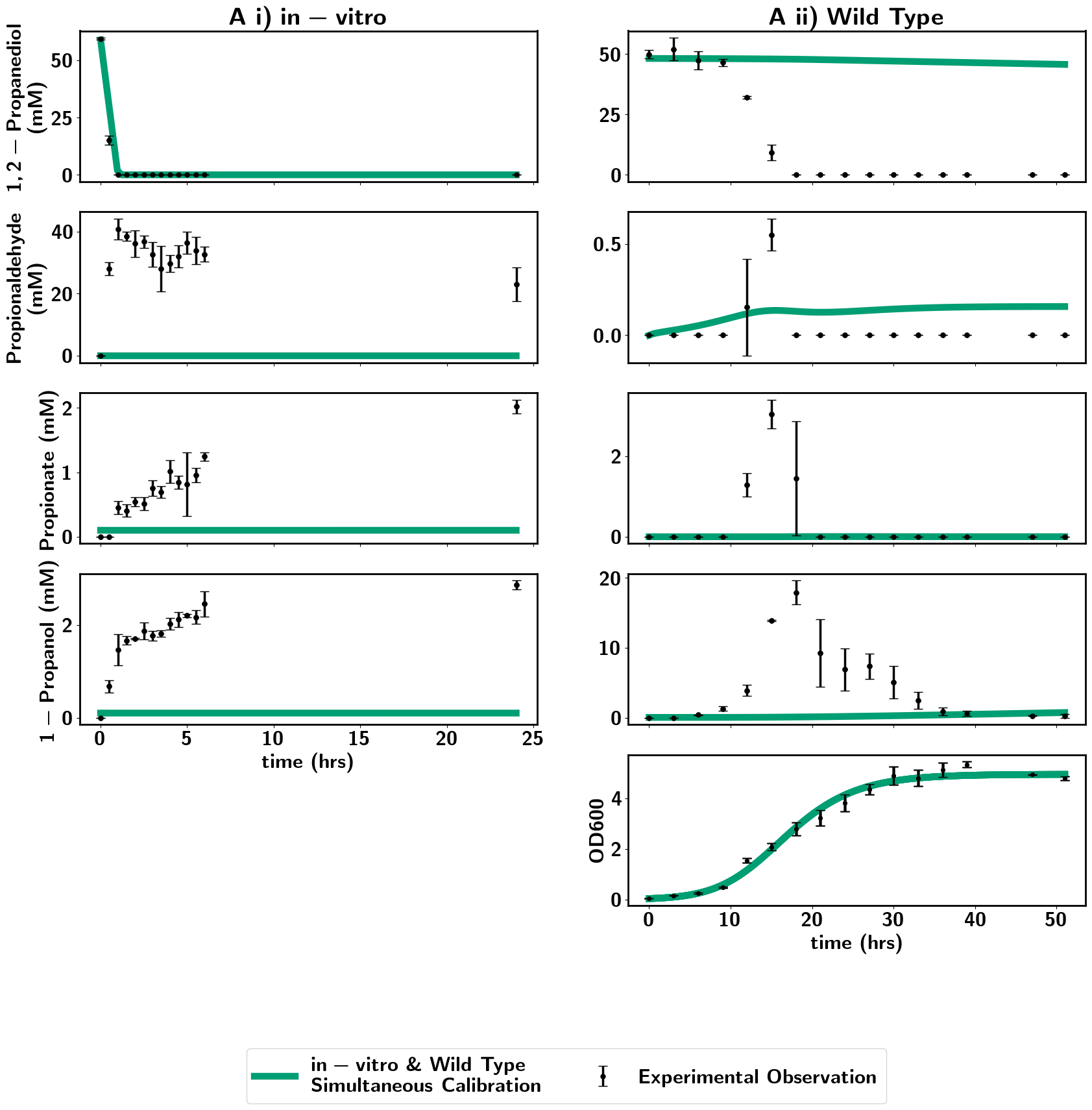}
\caption{Mode 2 fit to in-vitro and WT data. Model assumes PduW are located in both the cytosol and MCP, and all other Pdu enzymes are localized to the MCP.} 
\label{fig:Mode2PduW}
\end{figure}

\begin{figure}[!htp]
\centering
\includegraphics[width=\linewidth]{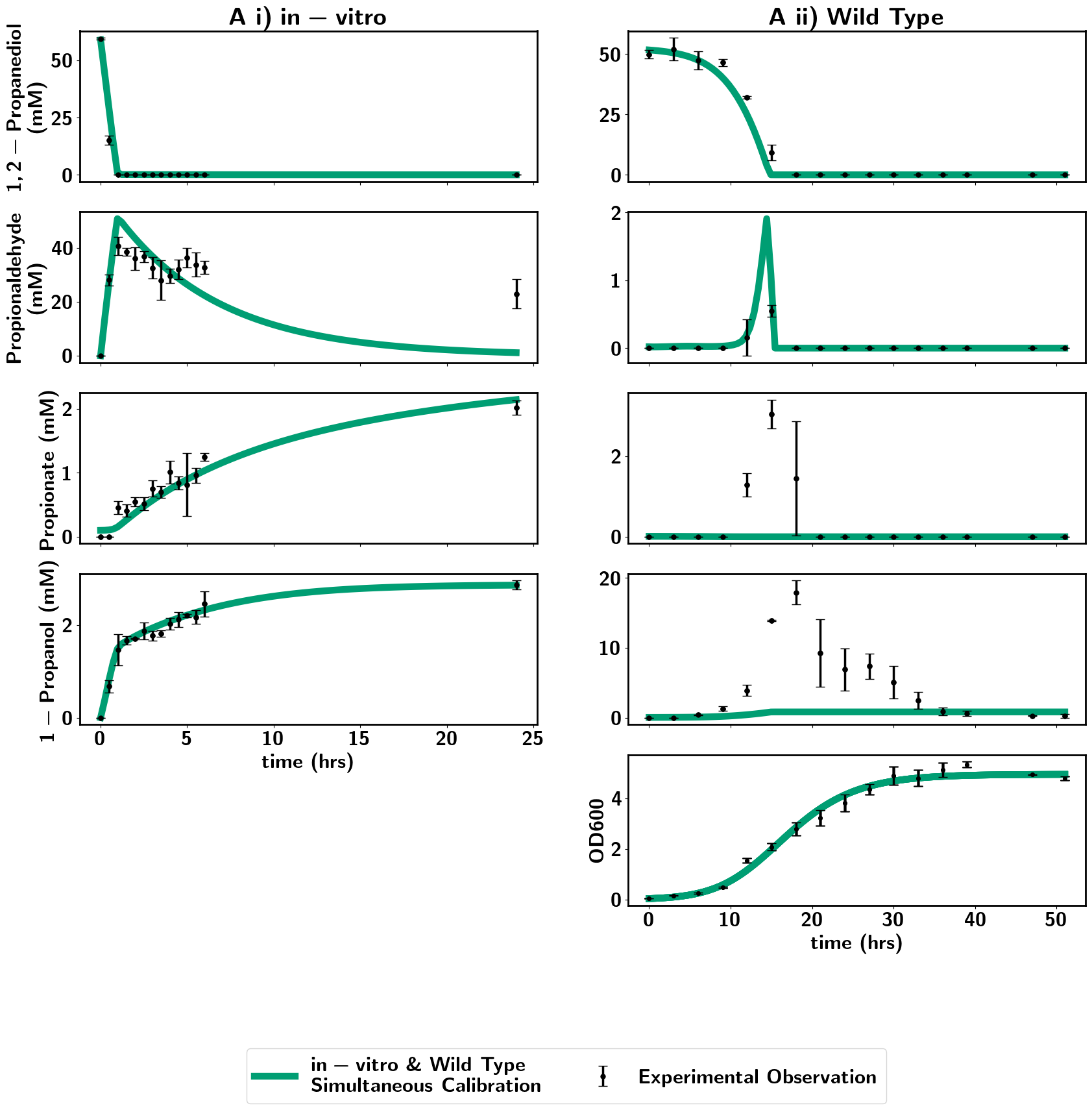}
\caption{Mode 1 fit to in-vitro and WT data. Model assumes PduP and PduW are located in both the cytosol and MCP, and all other Pdu enzymes are localized to the MCP.} 
\label{fig:Mode1PduPPduW}
\end{figure}

\begin{figure}[!htp]
\centering
\includegraphics[width=\linewidth]{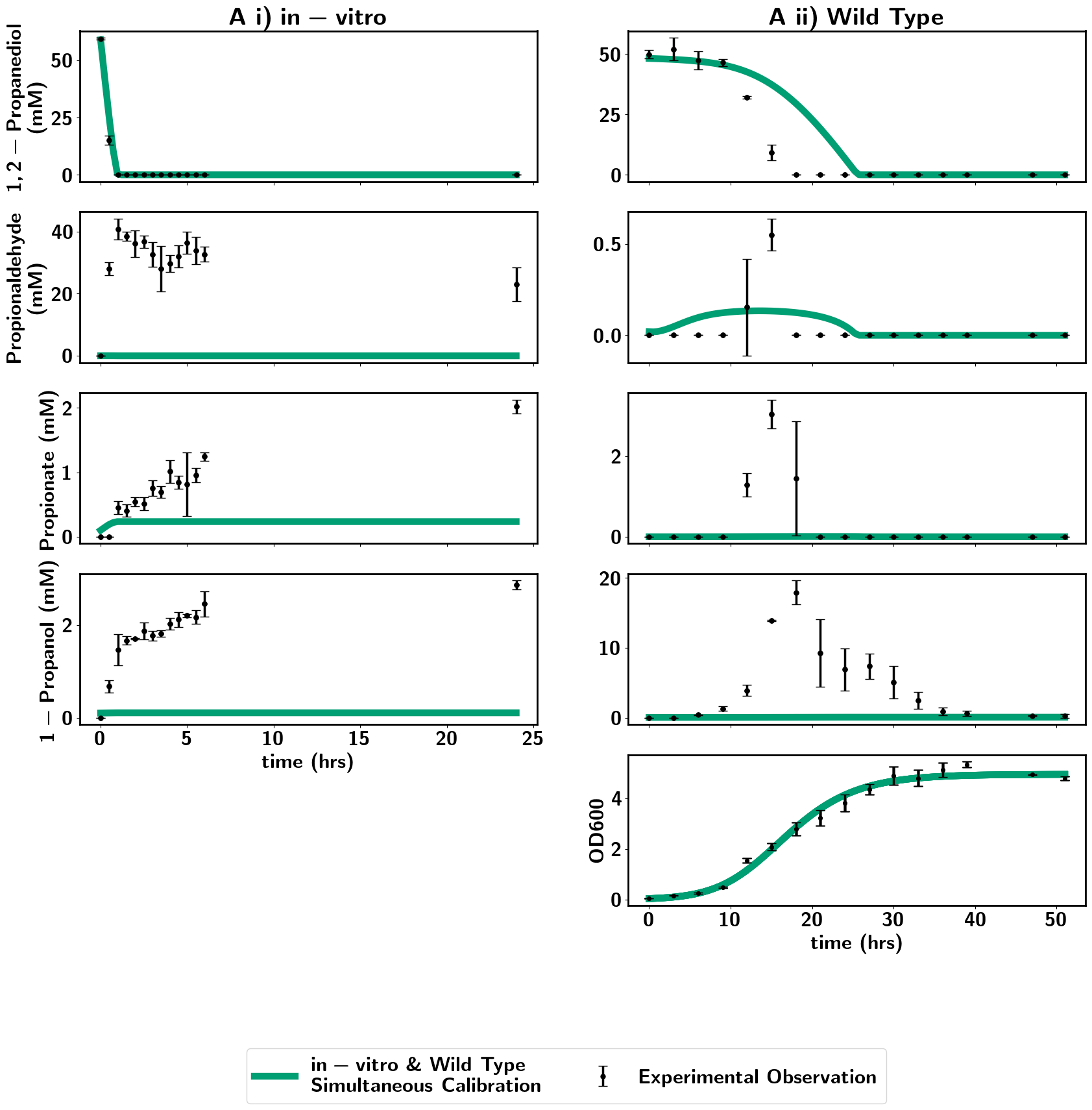}
\caption{Mode 2 fit to in-vitro and WT data. Model assumes PduP and PduW are located in both the cytosol and MCP, and all other Pdu enzymes are localized to the MCP.} 
\label{fig:Mode2PduPPduW}
\end{figure}

\begin{figure}[!htp]
\centering
\includegraphics[width=\linewidth]{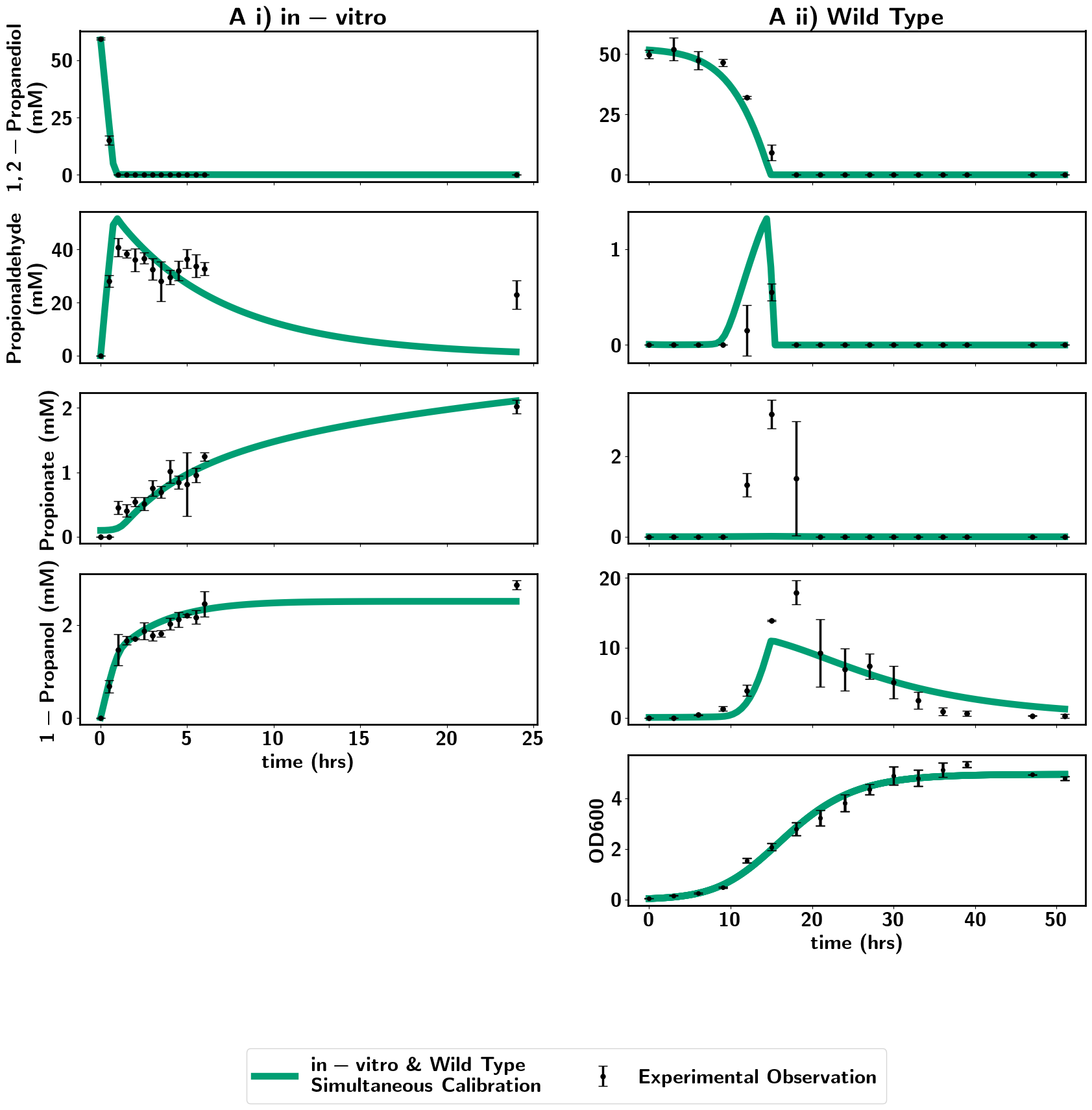}
\caption{Mode 1 fit to in-vitro and WT data. Model assumes PduQ, PduP and PduW are located in both the cytosol and MCP, and all other Pdu enzymes are localized to the MCP.} 
\label{fig:Mode1PduPPduQPduW}
\end{figure}

\begin{figure}[!htp]
\centering
\includegraphics[width=\linewidth]{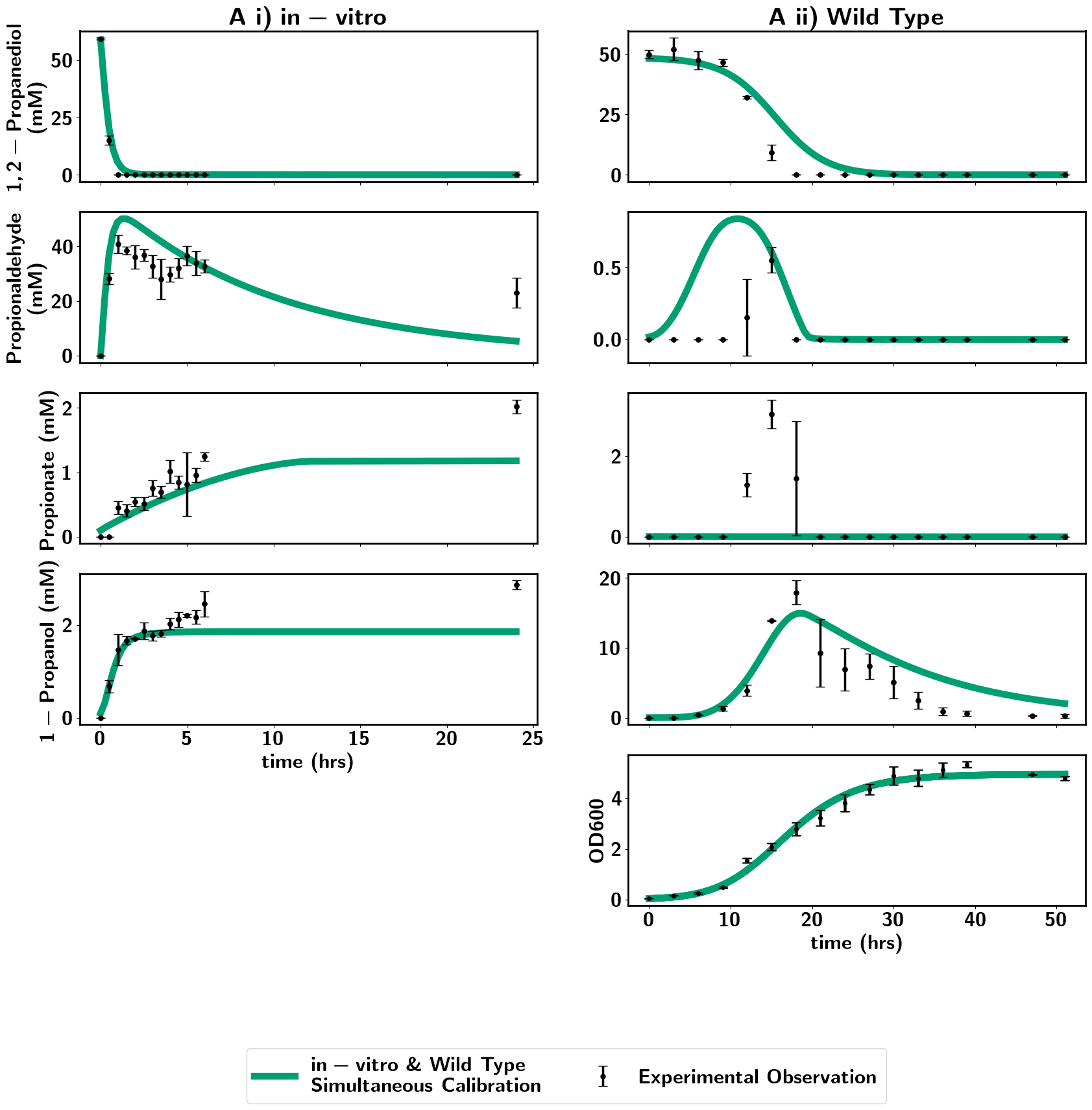}
\caption{Mode 2 fit to in-vitro and WT data. Model assumes PduQ, PduP and PduW are located in both the cytosol and MCP, and all other Pdu enzymes are localized to the MCP.} 
\label{fig:Mode2PduPPduQPduW}
\end{figure}

\begin{figure}[!htp]
\centering
\includegraphics[width=\linewidth]{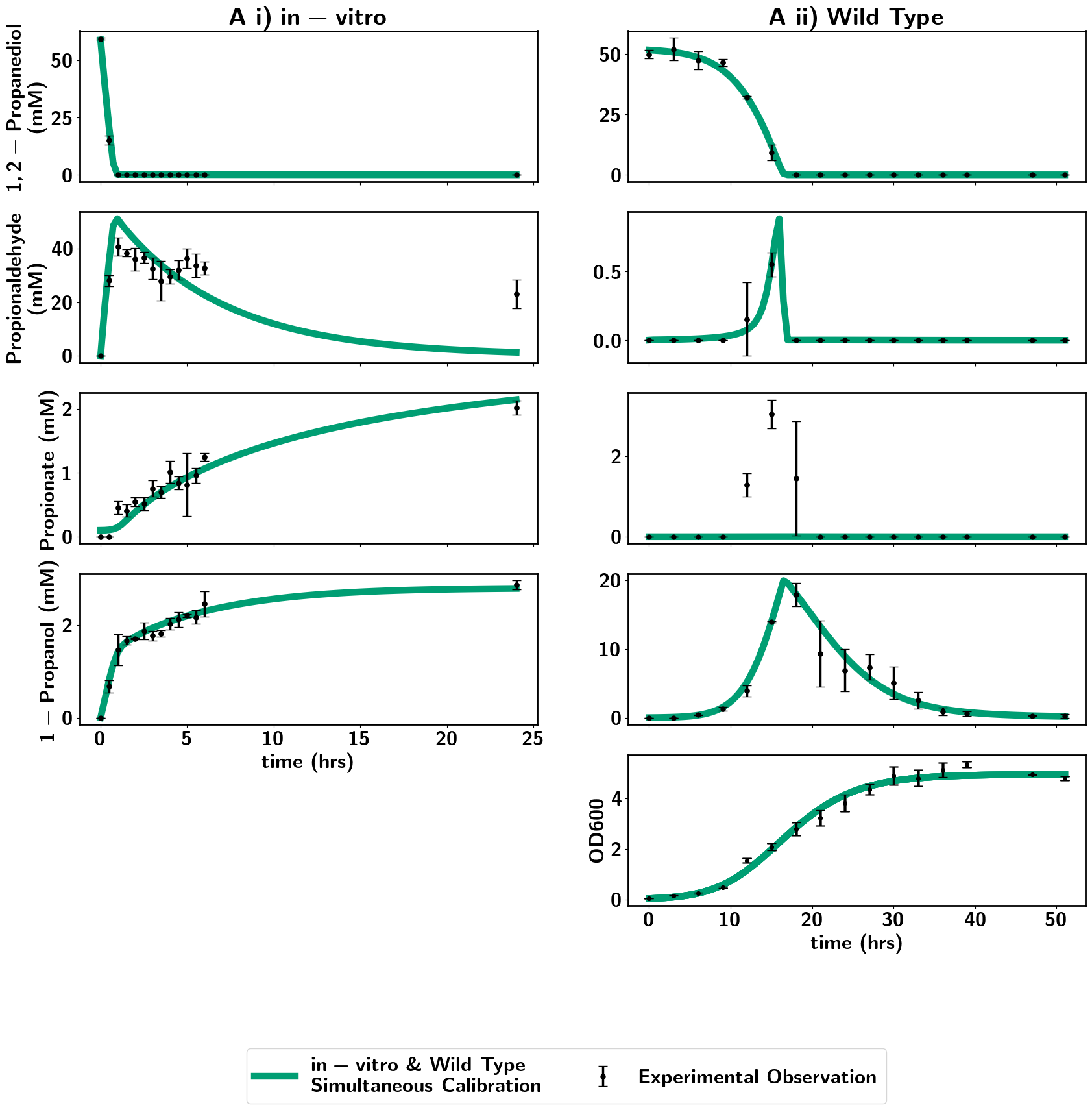}
\caption{Mode 1 fit to in-vitro and WT data. Model assumes PduQ, PduP and PduW are located in both the cytosol and MCP, all other Pdu enzymes are localized to the MCP and the presense of promiscous alcohol dehydrogenase.} 
\label{fig:Mode1PduPPduQAdhPduW}
\end{figure}

\begin{figure}[!htp]
\centering
\includegraphics[width=\linewidth]{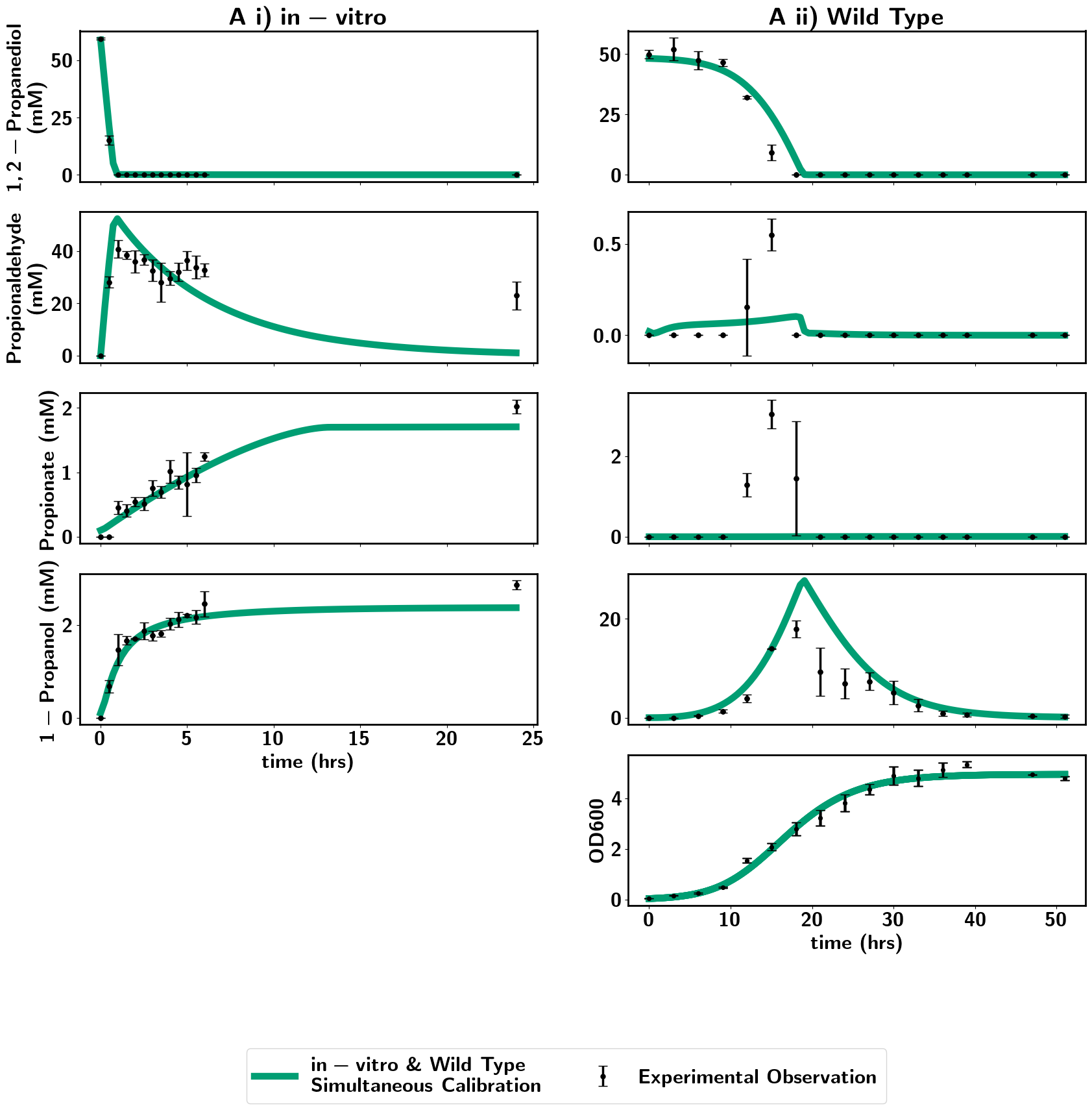}
\caption{Mode 2 fit to in-vitro and WT data. Model assumes PduQ, PduP and PduW are located in both the cytosol and MCP, all other Pdu enzymes are localized to the MCP and the presense of promiscous alcohol dehydrogenase.} 
\label{fig:Mode2PduPPduQAdhPduW}
\end{figure}

\begin{figure}[!htp]
\centering
\includegraphics[width=\linewidth]{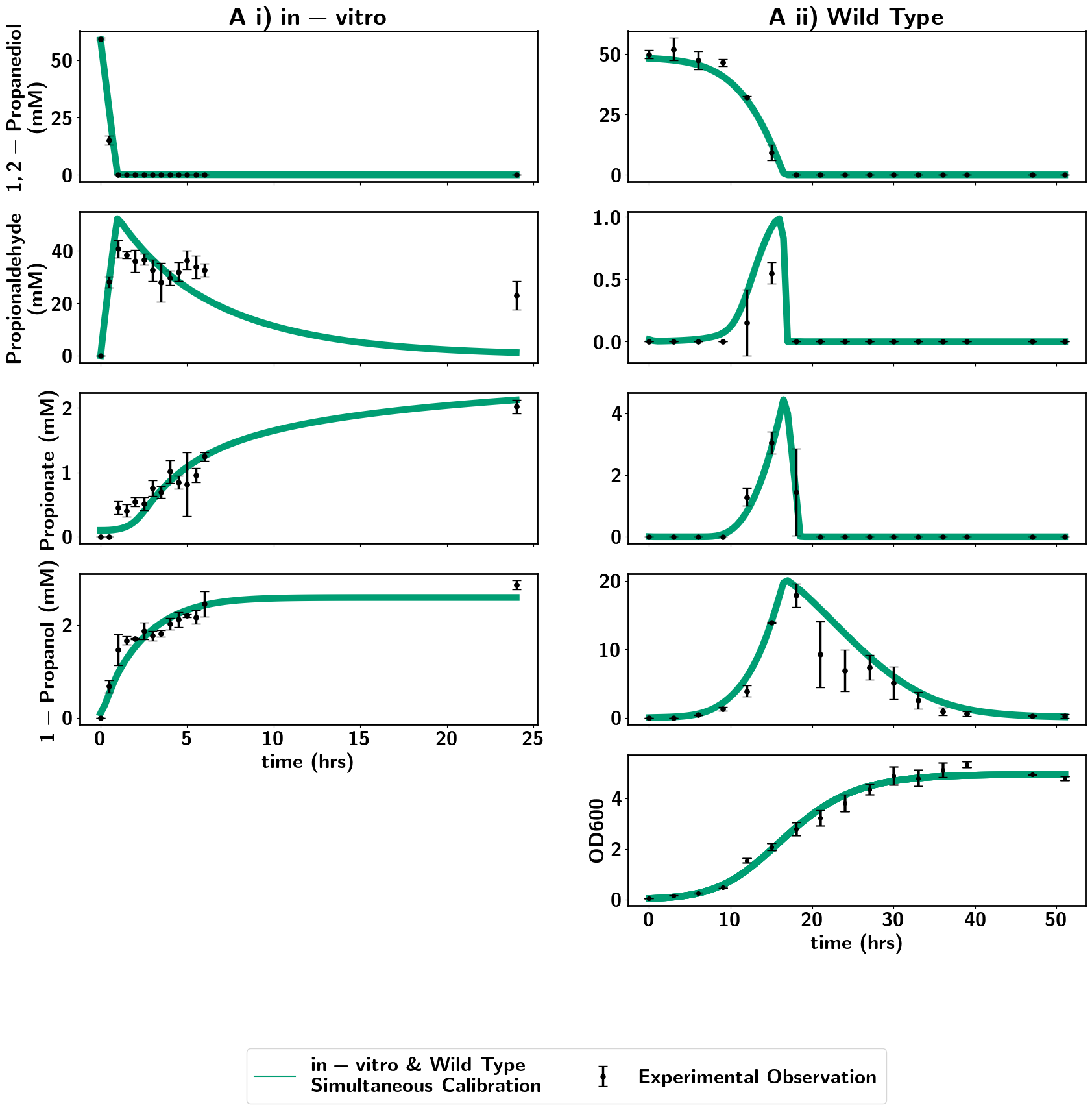}
\caption{Mode 1 fit to in-vitro and WT data. Model assumes PduQ, PduP and PduW are located in both the cytosol and MCP, all other Pdu enzymes are localized to the MCP and the presense of promiscous alcohol dehydrogenase. Propionate with tightened standard deviations of 0.05 for values $>$ 0.01 mM. } 
\label{fig:Mode1PduPPduQAdhPduW_PropionateTightened}
\end{figure}

\begin{figure}[!ht]
\centering
\includegraphics[width=\linewidth]{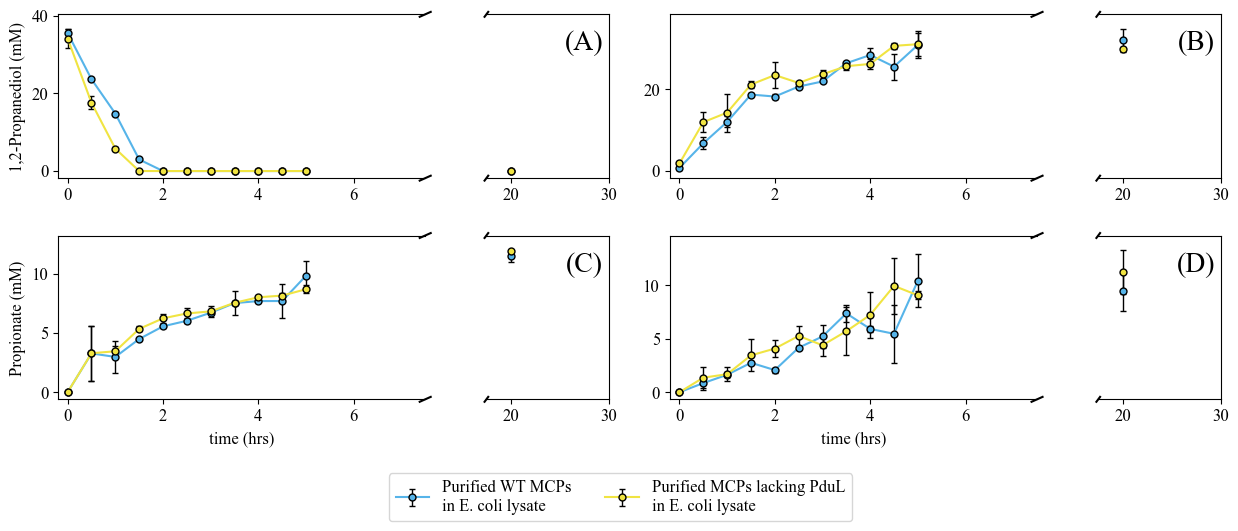}         \caption{Metabolic activity of WT MCPs assayed with 1,2-propanediol in {\it E. coli} lysate, compared to MCPs deficient in the PduL catalytic proteins assayed with 1,2-propanediol in {\it E. coli lysate}. See section \nameref{subsection:MethoddPduLMCPsWTMCPs} for the methodology of the experiment.} 
\label{fig:all_reactant_partition_reactants_in_vitro_WT_dPduL}
\end{figure}

\begin{figure}[!ht]
\centering
\includegraphics[width=0.5\linewidth]{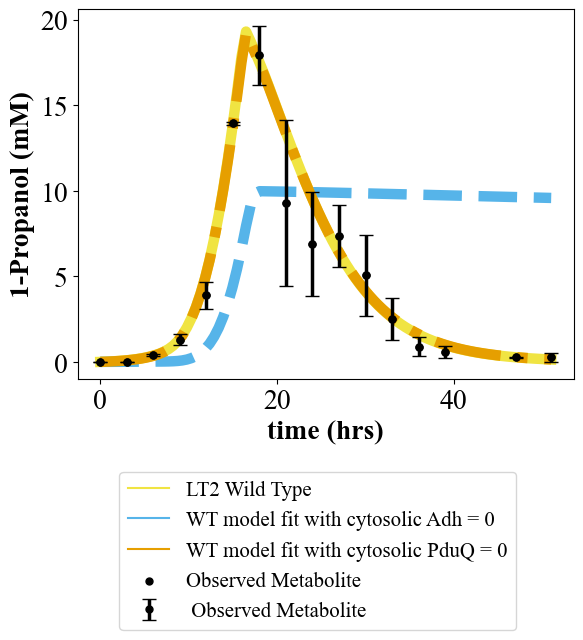}         \caption{WT model fits with cytosolic PduQ = 0 and cytosolic ADH = 0} 
\label{fig:WTdPduQdADH}
\end{figure}

\begin{figure}[!ht]
\centering
\includegraphics[width=\linewidth]{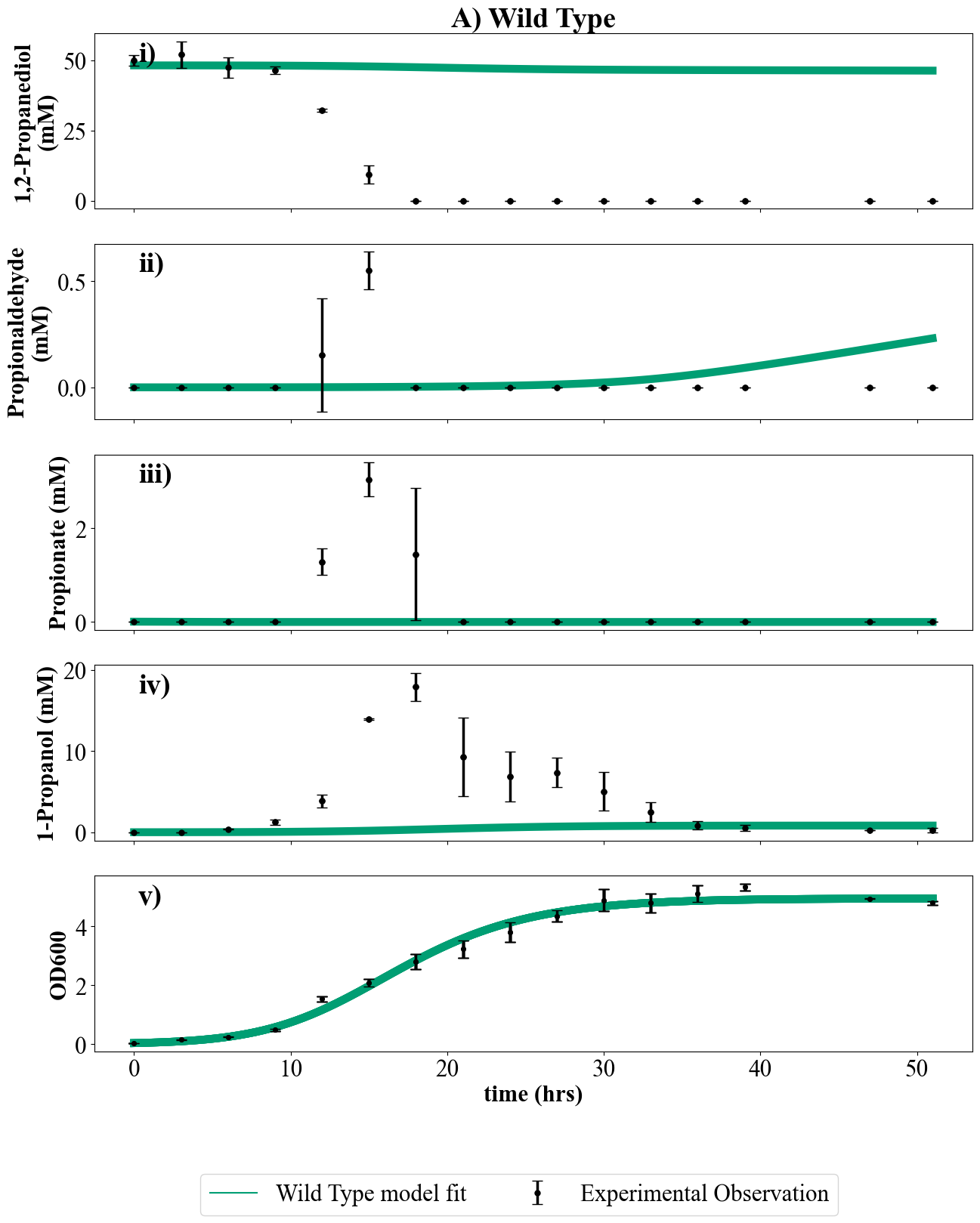}         \caption{Preliminary NUTS WT model optimization assuming encapsulation of all Pdu enzymes except PduW} 
\label{fig:Mode1WTPduW}
\end{figure}

\begin{figure}[!ht]
\centering
\includegraphics[width=\linewidth]{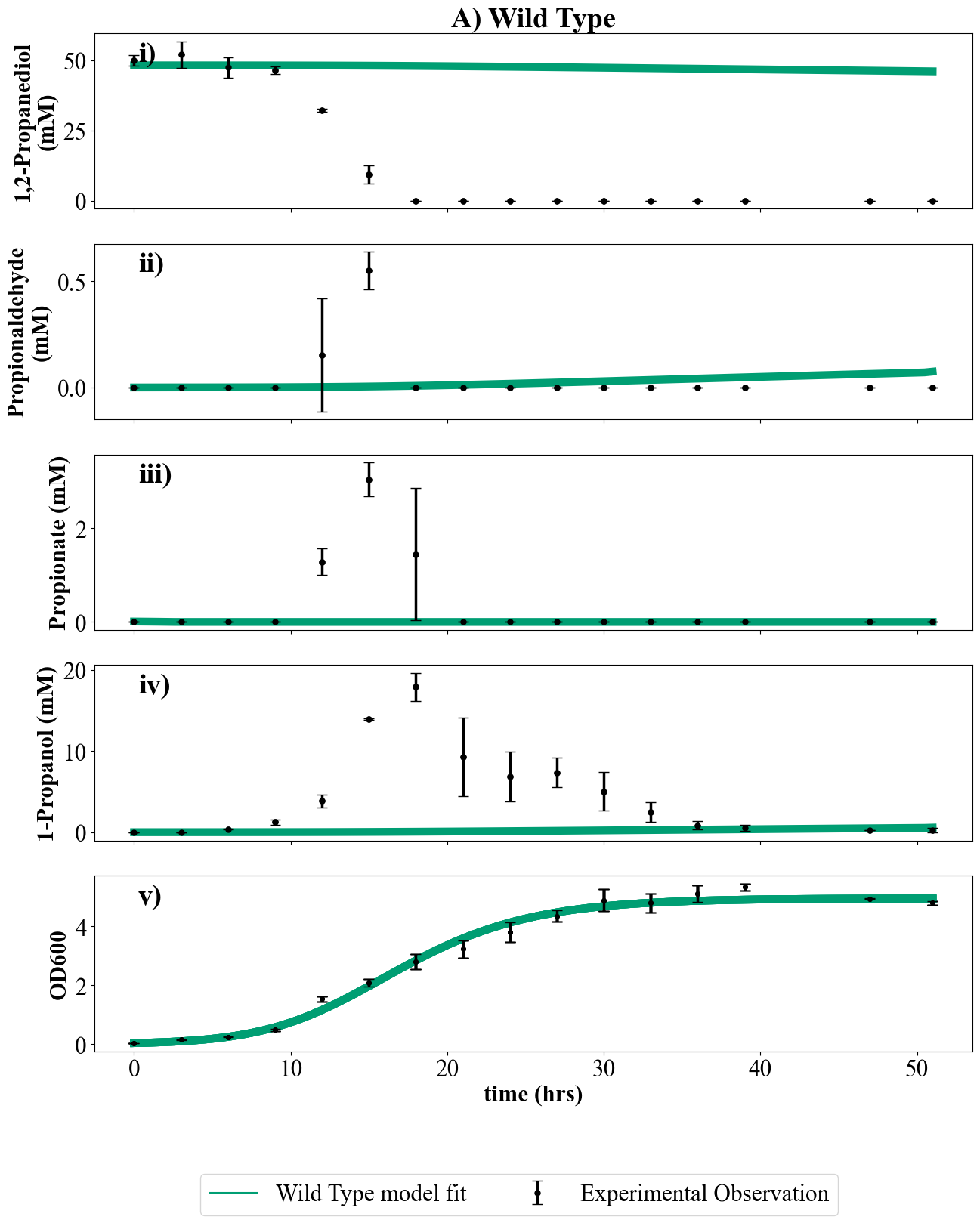}         \caption{Preliminary NUTS WT model optimization assuming encapsulation of all Pdu enzymes  except PduW} 
\label{fig:Mode2WTPduW}
\end{figure}

\begin{figure}[!htp]
\centering
\includegraphics[width=\linewidth]{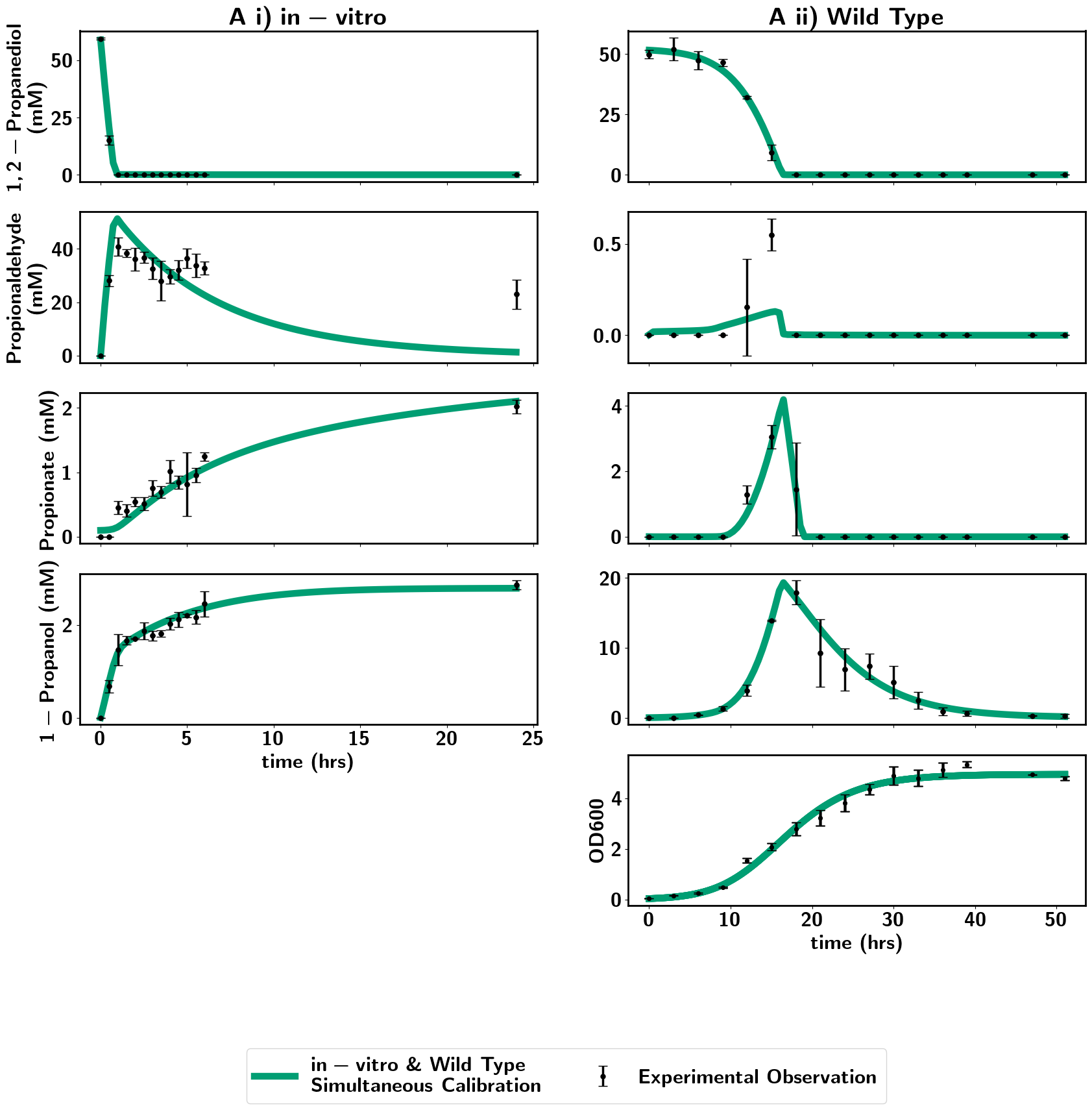}
\caption{Mode 1 fit to in-vitro and WT data. Model assumes PduQ, PduP, PduL and PduW are located in both the cytosol. PduQ cytosolic concentration allowed to beyond 0.5mM.} 
\label{fig:Mode1PduPPduQPduLExpandedPduW}
\end{figure}

\clearpage
\section*{Supplementary Tables}
\renewcommand{\thetable}{S\arabic{table}}
\setcounter{table}{0}
\singlespacing\renewcommand*{\arraystretch}{2}
\begin{center}
 \begin{longtable}{m{8em} >{\raggedright\arraybackslash}m{12.5em} m{7.5em} m{7em}} 
 \toprule
 Parameter& Meaning & Estimated Value & References \\ 
 \midrule
\endfirsthead
\toprule
 Parameter& Meaning & Estimated Value & References \\ 
\midrule
\endhead
\midrule
\multicolumn{4}{r}{Continued on next page} \\
\midrule
\endfoot
\endlastfoot
\bottomrule
 $\ell$ & maximum distance from hemispherical ends of the cells  & $2.47$ $\mu$m & \cite{Smit1975} \\
  $d$ & diameter of the hemispherical ends of the cells & $0.75$ $\mu$m & \cite{Smit1975} \\
    $N_{\text{OD}600}$ & Cell concentration per $\text{OD}_{600}$ & $10^{15}$ cells per m$^3$ & \cite{Volkmer2011} \\

Total NAD pool & Total concentration NAD pool in the cytosol  & $0.5-10$ mM &  \cite{Grose2006,Bennett2009,Park2016}   \\
NADH:NAD+ concentration ratio & Homeostatic NADH:NAD+  ratio &  $[10^{-3},1.3]$  &  \cite{Andersen1977, Leonardo1996, Zhou2011,Wimpenny1972,Liu2019,Bennett2009,Park2016, Grose2006,Husain2008}  \\
Total adenosine pool & Total concentration of adenosine pool in the cytosol &  $1.-12.$ mM  & \cite{Mempin2013,Bennett2009,Park2016}  \\
ATP:ADP concentration ratio & Homeostatic ATP:ADP ratio & $5-120$   &  \cite{Bennett2009,Park2016,Tran1998} \\
AMP concentration   & Homeostatic AMP concentration in the cell &  $0.01-0.4$  mM &  \cite{Bennett2009,Park2016} \\
Total CoA pool & Total concentration of CoA and its deriative products in the cytosol &  $0.1-20$ mM  & \cite{Wolfe2005,CHOHNAN1998,Park2016} \\
Phosphate concentration   & Homeostatic phosphate concentration in the cell &  $1-24$ mM & \cite{Nesmeyanova2000,Park2016,Nausch2004}  \\
Oxaloacetate concentration   & Homeostatic oxaloacetate concentration in the cell &  $10^{-4} - 10^{-3}$ mM & \cite{Park2016} \\
Pyrophosphate concentration   & Homeostatic pyrophosphate concentration in the cytosol &   $\approx 0$ mM &   \cite{Wolfe2005,Baykov2013} \\
Water concentration &  Homeostatic water concentration in the cell & $4 \times 10^4$ mM  &  \\
 $P_{\text{cell, 1,2-PD}}$ & cell membrane permeability of 1,2-propanediol &  $2 \times 10^{-6}$ m/s &  \cite{Orbach1980}\\
 $P_{\text{cell, Propionaldehyde}}$ & cell membrane permeability of propionaldehyde &  &  \\ 
 $P_{\text{cell, 1-Propanol}}$ & cell membrane permeability of 1-propanol & $6.5 \times 10^{-5}$  m/s  & \cite{Brahm1983}\\ 
 $P_{\text{cell, CoA}}$ & cell membrane permeability of CoA and its derivative products & 0 m/s &  \cite{Pietrocola2015, Visser2006}\\
 $P_{\text{cell, }\substack{\text{Propionyl-}\\\text{Phosphate}}}$ & cell membrane permeability of propionyl phosphate &  &  \\
 $P_{\text{cell, Propionate}}$ & cell membrane permeability of propionate &  &  \\ 
$k_{\text{cat, PrpE}}^{\text{propionyl-CoA}}$ & maximum forward rate reaction of the PrpE propionyl-CoA formation reaction &  $33 \pm 2$ s$^{-1}$ & \cite{Horswill2002}  \\
$K^{\substack{\text{propionate}}}_{\text{M, PrpE}}$ & half-max concentration of propionate the PrpE propionyl-CoA formation reaction & $(20 \pm 2) \times 10^{-3}$ mM  & \cite{Horswill2002}  \\
$K^{\substack{\text{CoA}}}_{\text{M, PrpE}}$ & half-max concentration of CoA for the PrpE propionyl-CoA formation reaction & $(215 \pm 31) \times 10^{-3}$ mM  &  \cite{Horswill2002} \\
$K^{\text{ATP}}_{\text{M, PrpE}}$ & half-max concentration of ATP for the PrpE propionyl-CoA formation reaction & $(57 \pm 5) \times 10^{-3}$ mM  & \cite{Horswill2002}  \\
$k_{\text{cat, PrpE}}^{\text{ATP}}$ & maximum rate of the PrpE ATP formation reaction &  $79 \pm 5$ s$^{-1}$ &  \cite{Horswill2002} \\
$K^{\substack{\text{PPi}}}_{\text{M, PrpE}}$ & half-max concentration of PPi for the PrpE ATP formation reaction & $(254 \pm 9)\times 10^{-3}$ mM  &   \cite{Horswill2002}\\
$K_{\text{eq, PrpE}}$ & Dissociation constants for the PrpE reaction & $[1.77, 1.55 \times 10^3] \times 10^{-2}$ & \cite{Beber2021}  \\ 
$k_{\text{cat, PrpC}}^{\text{propionyl-CoA}}$ & maximum reaction rate for the PrpC condensation reaction &  $7.2 \pm 0.2$ s$^{-1}$ &  \cite{Horswill1999a} \\
$K^{\substack{\text{propionyl-CoA}}}_{\text{M, PrpC}}$ & half-max concentration of propionyl-CoA for the PrpC condensation reaction & $(48 \pm 8) \times 10^{-3}$ mM  & \cite{Horswill1999a}  \\
$K^{\substack{\text{OAA}}}_{\text{M, PrpC}}$ & half-max concentration of oxaloacetate for the PrpC condensation reaction & $(12 \pm 2)\times 10^{-3}$ mM  & \cite{Horswill1999a}   \\
$K_{\text{eq, PrpC}}$ & Dissociation constants for the PrpE reaction & $[2.84, 1.71 \times 10^4] \times 10^4$  & \cite{Beber2021}  \\ \hline
\caption{Table of Measured Parameters. See Archer et al. 2024 for MCP, and Pdu and AckA Michaelis-Menten parameter measurements}
 \label{Table:Chapter6AvailableParameters}
\end{longtable}
\end{center}
\doublespacing

\clearpage

{
\singlespacing\renewcommand*{\arraystretch}{3}
\begin{longtable}
{p{.1\textwidth}p{.2\textwidth} p{0.5\textwidth}} \hline

 \toprule
Enzyme & Reaction Description & Michaelis-Menten Parameters  \\ 
 \midrule
\endfirsthead
\toprule
Enzyme & Reaction Description & Michaelis-Menten Parameters  \\ 
\midrule
\endhead
\midrule
\multicolumn{3}{r}{Continued on next page} \\
\midrule
\endfoot
\endlastfoot
\bottomrule
            \multirow{3}{*}{PrpE}  & Propionyl-CoA Formation & 
            $k_{\text{cat, PrpE}}^{\text{propionyl-CoA}} = \frac{k_{5,\text{PrpE}}k_{9,\text{PrpE}}k_{11,\text{PrpE}}}{\splitdfrac{k_{5,\text{PrpE}}k_{11,\text{PrpE}}}{  \splitdfrac{+k_{9,\text{PrpE}}k_{11,\text{PrpE}}}{ + k_{5,\text{PrpE}}k_{9,\text{PrpE}}}}}$
            
            $K^{\text{ATP}}_{\text{M}, \text{PrpE}} =\frac{k_{5,\text{PrpE}}k_{9,\text{PrpE}}k_{11,\text{PrpE}}}{\splitdfrac{k_{1,\text{PrpE}}k_{5,\text{PrpE}}k_{11,\text{PrpE}}}{\splitdfrac{ + k_{1,\text{PrpE}}k_{9}k_{11,\text{PrpE}}}{ + k_{1,\text{PrpE}}k_{5,\text{PrpE}}k_{9,\text{PrpE}}}}}$
            
            $K^{\text{CoA}}_{\text{M}, \text{PrpE}} = \frac{k_{5,\text{PrpE}}(k_{8,\text{PrpE}} + k_{9,\text{PrpE}})k_{11,\text{PrpE}}}{\splitdfrac{k_{7,\text{PrpE}}k_{5,\text{PrpE}}k_{11,\text{PrpE}}}{\splitdfrac{ + k_{7,\text{PrpE}}k_{9,\text{PrpE}}k_{11,\text{PrpE}}}{ + k_{7,\text{PrpE}}k_{5,\text{PrpE}}k_{9,\text{PrpE}}}}}$
            
            $K^{\text{Propionate}}_{\text{M}, \text{PrpE}} = \frac{k_{9,\text{PrpE}}(k_{4,\text{PrpE}} + k_{5,\text{PrpE}})k_{11,\text{PrpE}}}{\splitdfrac{k_{3,\text{PrpE}}k_{5,\text{PrpE}}k_{11,\text{PrpE}}}{ \splitdfrac{+ k_{3,\text{PrpE}}k_{9,\text{PrpE}}k_{11,\text{PrpE}}}{ + k_{3,\text{PrpE}}k_{5,\text{PrpE}}k_{9,\text{PrpE}}}}}$
            \\
             & Propionate Formation & $
            k_{\text{cat, PrpE}}^{\text{propionate}} = \frac{k_{2,\text{PrpE}}k_{4,\text{PrpE}}k_{8,\text{PrpE}}}{k_{2}k_{4} + k_{2,\text{PrpE}}k_{8,\text{PrpE}} + k_{4,\text{PrpE}}k_{8,\text{PrpE}}}$
            
            $K^{\text{PPi}}_{\text{M}, \text{PrpE}} = \frac{k_{2,\text{PrpE}}(k_{4,\text{PrpE}} + k_{5,\text{PrpE}})k_{8,\text{PrpE}}}{\splitdfrac{k_{6,\text{PrpE}}k_{2,\text{PrpE}}k_{4,\text{PrpE}}}{\splitdfrac{ + k_{6,\text{PrpE}}k_{2,\text{PrpE}}k_{8,\text{PrpE}}}{ + k_{6,\text{PrpE}}k_{4,\text{PrpE}}k_{8,\text{PrpE}}}}}$
            
            $K^{ \text{AMP}}_{\text{M},\text{PrpE}} = \frac{k_{2,\text{PrpE}}k_{4,\text{PrpE}}k_{8,\text{PrpE}}}{\splitdfrac{k_{12,\text{PrpE}}k_{2,\text{PrpE}}k_{4,\text{PrpE}}}{\splitdfrac{ + k_{12,\text{PrpE}}k_{2,\text{PrpE}}k_{8}}{ + k_{12,\text{PrpE}}k_{4,\text{PrpE}}k_{8,\text{PrpE}})}}}$
            
            $K^{\text{Propionyl-CoA}}_{\text{M}, \text{PrpE}} = \frac{k_{2,\text{PrpE}}(k_{8,\text{PrpE}} + k_{9,\text{PrpE}})k_{4,\text{PrpE}}}{\splitdfrac{k_{10,\text{PrpE}}k_{2,\text{PrpE}}k_{4,\text{PrpE}}}{\splitdfrac{ + k_{10,\text{PrpE}}k_{2,\text{PrpE}}k_{8,\text{PrpE}}}{ +k_{10,\text{PrpE}}k_{4,\text{PrpE}}k_{8,\text{PrpE}}}}}$\\
            & ATP Formation & $
            k_{\text{cat, PrpE}}^{\text{ATP}} = \frac{k_{2,\text{PrpE}}k_{4,\text{PrpE}}}{k_{2,\text{PrpE}} + k_{4,\text{PrpE}}}$
            
            $K^{\text{PPi}}_{\text{M}, \text{PrpE}} = \frac{k_{2,\text{PrpE}}(k_{4,\text{PrpE}} + k_{5,\text{PrpE}})k_{8,\text{PrpE}}}{\splitdfrac{k_{6,\text{PrpE}}k_{2,\text{PrpE}}k_{4,\text{PrpE}}}{\splitdfrac{+ k_{6,\text{PrpE}}k_{2,\text{PrpE}}k_{8,\text{PrpE}}}{ + k_{6,\text{PrpE}}k_{4,\text{PrpE}}k_{8,\text{PrpE}}}}}$
            
            $K^{\text{Propionyl-AMP}}_{\text{M}, \text{PrpE}} = \frac{k_{2,\text{PrpE}}k_{4,\text{PrpE}}k_{8,\text{PrpE}}}{\splitdfrac{k_{14,\text{PrpE}}k_{2,\text{PrpE}}k_{4,\text{PrpE}}}{\splitdfrac{ + k_{14,\text{PrpE}}k_{2,\text{PrpE}}k_{8,\text{PrpE}}}{ + k_{14,\text{PrpE}}k_{4,\text{PrpE}}k_{8,\text{PrpE}}}}}$ \\\\\hline
            \multirow{2}{*}{PrpC} &  Condensation & $k_{\text{cat, PrpC}}^{\text{condensation}} = \frac{k_{7, \text{PrpC}}k_{9, \text{PrpC}}}{k_{7, \text{PrpC}} + k_{9, \text{PrpC}}}$
            
            $K^{\text{Propionyl-CoA}}_{\text{M}, \text{PrpC}} = \frac{k_{7, \text{PrpC}}k_{9, \text{PrpC}}}{k_{1, \text{PrpC}}(k_{7, \text{PrpC}} + k_{9, \text{PrpC}})}$

            $K^{\text{Oxaloacetate}}_{\text{M}, \text{PrpC}} = \frac{k_{7, \text{PrpC}}k_{9, \text{PrpC}}}{k_{3, \text{PrpC}}(k_{7, \text{PrpC}} + k_{9, \text{PrpC}})}$
            
            $K^{\text{Water}}_{\text{M}, \text{PrpC}} = \frac{(k_{6, \text{PrpC}} + k_{7, \text{PrpC}})k_{9, \text{PrpC}}}{k_{5, \text{PrpC}}(k_{7, \text{PrpC}} + k_{9, \text{PrpC}})}$\\
             & Dehydration &
             $k_{\text{cat, PrpC}}^{\text{hydrolysis}} = \frac{k_{2}k_{4, \text{PrpC}}k_{6, \text{PrpC}}}{\splitdfrac{k_{2, \text{PrpC}}k_{4, \text{PrpC}}}{\splitdfrac{ + k_{2, \text{PrpC}}k_{6, \text{PrpC}}}{ + k_{4, \text{PrpC}}k_{6, \text{PrpC}}}}}$
            
            $K^{\text{CoA}}_{\text{M}, \text{PrpC}} = \frac{k_{2, \text{PrpC}}k_{4, \text{PrpC}}k_{6, \text{PrpC}}}{\splitdfrac{k_{10, \text{PrpC}}k_{2, \text{PrpC}}k_{4, \text{PrpC}}}{\splitdfrac{ + k_{10, \text{PrpC}}k_{4, \text{PrpC}}k_{6, \text{PrpC}}}{ + k_{10, \text{PrpC}}k_{4, \text{PrpC}}k_{6, \text{PrpC}}}}}$
            
            $K^{\text{2-Methylcitrate}}_{\text{M}, \text{PrpC}} = \frac{k_{2, \text{PrpC}}(k_{6, \text{PrpC}} + k_{7, \text{PrpC}})k_{4, \text{PrpC}}}{\splitdfrac{k_{8, \text{PrpC}}k_{2, \text{PrpC}}k_{4, \text{PrpC}}}{\splitdfrac{ + k_{8, \text{PrpC}}k_{2, \text{PrpC}}k_{6, \text{PrpC}}}{ + k_{8, \text{PrpC}}k_{4, \text{PrpC}}k_{6, \text{PrpC}}}}}$
            \\\hline
            
            
            
            
             
            
             \caption{Table of Michaelis-Menten Formulae. See Archer et al. 2024 for Pdu and AckA Michaelis-Menten Formulae}
 \label{Table:Chapter6MMFormulae}
\end{longtable}}

\clearpage

\subsection{Free and Leading Parameters}
%
{ \singlespacing
\begin{longtable}
{p{0.1\textwidth} p{.15\textwidth} p{.15\textwidth} p{.5\textwidth}} 
\toprule
Enzyme & Kinetic Parameters & Free Variables & Leading Variables \\ 
\midrule
\endhead
\midrule
\multicolumn{4}{r}{Continued on next page} \\
\midrule
\endfoot
\endlastfoot
\bottomrule

            PrpE & $k_{1,\text{PrpE}}$, $k_{2,\text{PrpE}}$, $k_{3,\text{PrpE}}$, $k_{4,\text{PrpE}}$, $k_{5,\text{PrpE}}$, $k_{6,\text{PrpE}}$, $k_{7,\text{PrpE}}$, $k_{8,\text{PrpE}}$, $k_{9,\text{PrpE}}$, $k_{10,\text{PrpE}}$, $k_{11,\text{PrpE}}$, $k_{12,\text{PrpE}}$ & $k_{4,\text{PrpE}}$, $k_{5,\text{PrpE}}$, $k_{8,\text{PrpE}}$, $k_{9,\text{PrpE}}$, $k_{10,\text{PrpE}}$, $k_{\text{cat, PrpE}}^{\text{propionyl-CoA}}$, $K^{\text{ATP}}_{\text{M}, \text{PrpE}}$, $K^{\text{CoA}}_{\text{M}, \text{PrpE}}$, $K^{\text{Propionate}}_{\text{M}, \text{PrpE}}$, $k_{\text{cat, PrpE}}^{\text{ATP}}$, $K^{\text{PPi}}_{\text{M}, \text{PrpE}}$, $K_{\text{eq},\text{PrpE}}$  & $\begin{aligned} k_{1,\text{PrpE}} &= \frac{k_{\text{cat, PrpE}}^{\text{propionyl-CoA}}}{K^{\text{ATP}}_{\text{M}, \text{PrpE}}}\\
            k_{2,\text{PrpE}} &= \frac{k_{4,\text{PrpE}} k_{\text{cat, PrpE}}^{\text{ATP}}}{k_{4,\text{PrpE}}-k_{\text{cat, PrpE}}^{\text{ATP}}}\\
            k_{3,\text{PrpE}} &= \frac{k_{\text{cat, PrpE}}^{\text{propionyl-CoA}} (k_{4,\text{PrpE}}+k_{5,\text{PrpE}})}{k_{5,\text{PrpE}} K^{\text{Propionate}}_{\text{M}, \text{PrpE}}}\\
            k_{6,\text{PrpE}} &= \frac{k_{8,\text{PrpE}} k_{\text{cat, PrpE}}^{\text{ATP}} (k_{4,\text{PrpE}}+k_{5,\text{PrpE}})}{k_{4,\text{PrpE}} K^{\text{PPi}}_{\text{M}, \text{PrpE}} (k_{8,\text{PrpE}}+k_{\text{cat, PrpE}}^{\text{ATP}})}\\
            k_{7,\text{PrpE}} &= \frac{k_{\text{cat, PrpE}}^{\text{propionyl-CoA}} (k_{8,\text{PrpE}}+k_{9,\text{PrpE}})}{k_{9,\text{PrpE}} K^{\text{CoA}}_{\text{M}, \text{PrpE}}}\\
            k_{11,\text{PrpE}} &= \frac{k_{5,\text{PrpE}} k_{9,\text{PrpE}} k_{\text{cat, PrpE}}^{\text{propionyl-CoA}}}{\splitdfrac{k_{5,\text{PrpE}} k_{9,\text{PrpE}}}{\splitdfrac{-k_{5,\text{PrpE}} k_{\text{cat, PrpE}}^{\text{propionyl-CoA}}}{-k_{9,\text{PrpE}} k_{\text{cat, PrpE}}^{\text{propionyl-CoA}}}}} \\
            k_{12,\text{PrpE}} &= \frac{\splitdfrac{k_{1,\text{PrpE}}k_{3,\text{PrpE}}k_{5,\text{PrpE}}}{k_{7,\text{PrpE}}k_{9,\text{PrpE}}k_{11,\text{PrpE}}}}{\splitdfrac{k_{2,\text{PrpE}}k_{4,\text{PrpE}}k_{6,\text{PrpE}}}{k_{8,\text{PrpE}}k_{10,\text{PrpE}}K_{\text{eq, PrpE}}}}\end{aligned} $ \\\\
            
            PrpC & $k_{1,\text{PrpC}}$, $k_{2,\text{PrpC}}$, $k_{3,\text{PrpC}}$, $k_{4,\text{PrpC}}$, $k_{5,\text{PrpC}}$, $k_{6,\text{PrpC}}$, $k_{7,\text{PrpC}}$, $k_{8,\text{PrpC}}$, $k_{9,\text{PduP}}$, $k_{10,\text{PrpC}}$   & $k_{2,\text{PrpC}}$, $k_{4,\text{PrpC}}$, $k_{5,\text{PrpC}}$, $k_{6,\text{PrpC}}$, $k_{7,\text{PrpC}}$, $k_{8,\text{PrpC}}$, $K_{\text{eq},\text{PrpC}}$, $k_{\text{cat, PrpC}}^{\text{condensation}}$, $K^{\text{Propionyl-CoA}}_{\text{M}, \text{PrpC}}$, $K^{\text{Oxaloacetate}}_{\text{M}, \text{PrpC}}$ & $k_{1,\text{PrpC}} = \frac{k_{\text{cat, PrpC}}^{\text{condensation}}}{K^{\text{Propionyl-CoA}}_{\text{M}, \text{PrpC}}}$
            
            $k_{3,\text{PrpC}} = \frac{k_{\text{cat, PrpC}}^{\text{condensation}}}{K^{\text{Oxaloacetate}}_{\text{M}, \text{PrpC}}}$
            
            $k_{9,\text{PrpC}} = \frac{k_{7,\text{PrpC}} k_{\text{cat, PrpC}}^{\text{condensation}}}{k_{7,\text{PrpC}}-k_{\text{cat, PrpC}}^{\text{condensation}}}$
            
            $k_{10,\text{PrpC}} = \frac{k_{1,\text{PrpC}}k_{3,\text{PrpC}}k_{5,\text{PrpC}} k_{7,\text{PrpC}}k_{9,\text{PrpC}}}{k_{2,\text{PrpC}}k_{4,\text{PrpC}}k_{6,\text{PrpC}} k_{8,\text{PrpC}}K_{\text{eq, PrpC}}}$ \\\hline
            
            
            
            
            
             
                         \caption{Table of free and leading kinetic parameters. See Archer et al. 2024 for Pdu and AckA leading and free variables.}
 \label{Table:Chapter6ConstrainedKinetics}
\end{longtable}}
\doublespacing
\clearpage

{ \singlespacing
\begin{longtable}
{p{0.1\textwidth} p{.35\textwidth} p{.35\textwidth} }
 \toprule
Enzyme & Inequalities & Free Variables\\ 
 \midrule
\endfirsthead
\toprule
Enzyme & Inequalities & Free Variables \\ 
\midrule
\endhead
\midrule
\multicolumn{2}{r}{Continued on next page} \\
\midrule
\endfoot
\endlastfoot
\bottomrule
 PrpE & $\begin{aligned}k_{5,\text{PrpE}} &> k_{\text{cat,PrpE}}^{\text{Propionyl-CoA}}\\
                          k_{9,\text{PduP}} &> \frac{\text{factor}_{1,\text{PrpE}} }{\text{factor}_{1,\text{PrpE}} -1}\\
                          &\qquad \times k_{\text{cat,PrpE}}^{\text{Propionyl-CoA}}\\
             \text{where } k_{5,\text{PrpE}} &= \text{factor}_{1,\text{PrpE}} \\
             &\times k_{\text{cat,PrpE}}^{\text{Propionyl-CoA}}\\
                          \end{aligned}$  & 
             $\begin{aligned}\text{factor}_{1,\text{PrpE}} &> 1\\
             \text{factor}_{2,\text{PrpE}} &> 1\\
             \text{where \,} k_{9,\text{PduP}} &= \frac{\text{factor}_{1,\text{PrpE}} }{\text{factor}_{1,\text{PrpE}} -1}\\
             &\qquad\times \text{factor}_{2,\text{PrpE}}\\
             &\qquad\times k_{\text{cat,PrpE}}^{\text{Propionyl-CoA}}
             \end{aligned}$   \\\hline
             
             PrpC & $\begin{aligned}k_{7,\text{PrpC}}>k_{\text{cat, PrpC}}^{\text{condensation}}\end{aligned}$  & $\begin{aligned}\text{factor}_{1,\text{PrpC}} &> 1\\
                          \text{where}&\\
             k_{7,\text{PrpC}} &= \text{factor}_{1,\text{PrpC}}\\
             &\qquad \times k_{\text{cat, PrpC}}^{\text{condensation}}\end{aligned}$ \\\hline
                                     \caption{Table of free kinetic parameter constraints. See Table \ref{Table:Chapter6FreeVariableKinetics} for Pdu and AckA leading and free variables.}
 \label{Table:Chapter6FreeVariableKinetics}
\end{longtable}}

\clearpage 
{ \singlespacing
\begin{longtable}
{p{0.1\textwidth} p{.125\textwidth} p{.165\textwidth} p{.5\textwidth}} 
\toprule
Enzyme & Kinetic Parameters & Free Variables & Leading Variables \\ 
\midrule
\endhead
\midrule
\multicolumn{4}{r}{Continued on next page} \\
\midrule
\endfoot
\endlastfoot
\bottomrule

            PrpE & $k_{1,\text{PrpE}}$, $k_{2,\text{PrpE}}$, $k_{3,\text{PrpE}}$, $k_{4,\text{PrpE}}$, $k_{6,\text{PrpE}}$, $k_{7,\text{PrpE}}$, $k_{8,\text{PrpE}}$, $k_{9,\text{PrpE}}$, $k_{10,\text{PrpE}}$, $k_{11,\text{PrpE}}$, $k_{12,\text{PrpE}}$ & $k_{4,\text{PrpE}}$, $k_{5,\text{PrpE}}$, $k_{8,\text{PrpE}}$, $k_{9,\text{PrpE}}$, $k_{10,\text{PrpE}}$, $k_{\text{cat, PrpE}}^{\text{propionyl-CoA}}$, $K^{\text{ATP}}_{\text{M}, \text{PrpE}}$, $K^{\text{CoA}}_{\text{M}, \text{PrpE}}$, $K^{\text{Propionate}}_{\text{M}, \text{PrpE}}$, $k_{\text{cat, PrpE}}^{\text{ATP}}$, $K^{\text{PPi}}_{\text{M}, \text{PrpE}}$, $K_{\text{eq},\text{PrpE}}$, $\text{factor}_{1,\text{PrpE}}$, $\text{factor}_{2,\text{PrpE}}$  & $\begin{aligned} k_{1,\text{PrpE}} &= \frac{k_{\text{cat, PrpE}}^{\text{propionyl-CoA}}}{K^{\text{ATP}}_{\text{M}, \text{PrpE}}}\\
            k_{2,\text{PrpE}} &= \frac{k_{4,\text{PrpE}} k_{\text{cat, PrpE}}^{\text{ATP}}}{k_{4,\text{PrpE}}-k_{\text{cat, PrpE}}^{\text{ATP}}}\\
            k_{3,\text{PrpE}} &= \frac{k_{\text{cat, PrpE}}^{\text{propionyl-CoA}} (k_{4,\text{PrpE}}+k_{5,\text{PrpE}})}{k_{5,\text{PrpE}} K^{\text{Propionate}}_{\text{M}, \text{PrpE}}}\\
            k_{5,\text{PrpE}} &= 10^\text{factor}_{1,\text{PrpE}} k_{\text{cat,PrpE}}^{\text{Propionyl-CoA}}\\ \text{factor}_{1,\text{PrpE}} &> 0\\
            k_{6,\text{PrpE}} &= \frac{k_{8,\text{PrpE}} k_{\text{cat, PrpE}}^{\text{ATP}} (k_{4,\text{PrpE}}+k_{5,\text{PrpE}})}{k_{4,\text{PrpE}} K^{\text{PPi}}_{\text{M}, \text{PrpE}} (k_{8,\text{PrpE}}+k_{\text{cat, PrpE}}^{\text{ATP}})}\\
            k_{7,\text{PrpE}} &= \frac{k_{\text{cat, PrpE}}^{\text{propionyl-CoA}} (k_{8,\text{PrpE}}+k_{9,\text{PrpE}})}{k_{9,\text{PrpE}} K^{\text{CoA}}_{\text{M}, \text{PrpE}}}\\
            k_{9,\text{PduP}} &= \frac{\text{factor}_{2,\text{PrpE}}}{1-10^{-\text{factor}_{1,\text{PrpE}}}}k_{\text{cat,PrpE}}^{\text{Propionyl-CoA}} \\
            \text{factor}_{2,\text{PrpE}} &> 0\\
            k_{11,\text{PrpE}} &= \frac{k_{5,\text{PrpE}} k_{9,\text{PrpE}} k_{\text{cat, PrpE}}^{\text{propionyl-CoA}}}{\splitdfrac{k_{5,\text{PrpE}} k_{9,\text{PrpE}}}{\splitdfrac{-k_{5,\text{PrpE}} k_{\text{cat, PrpE}}^{\text{propionyl-CoA}}}{-k_{9,\text{PrpE}} k_{\text{cat, PrpE}}^{\text{propionyl-CoA}}}}} \\
            k_{12,\text{PrpE}} &= \frac{\splitdfrac{k_{1,\text{PrpE}}k_{3,\text{PrpE}}k_{5,\text{PrpE}}}{k_{7,\text{PrpE}}k_{9,\text{PrpE}}k_{11,\text{PrpE}}}}{\splitdfrac{k_{2,\text{PrpE}}k_{4,\text{PrpE}}k_{6,\text{PrpE}}}{k_{8,\text{PrpE}}k_{10,\text{PrpE}}K_{\text{eq, PrpE}}}}\end{aligned} $ \\\hline
            
            PrpC & $k_{1,\text{PrpC}}$, $k_{2,\text{PrpC}}$, $k_{3,\text{PrpC}}$, $k_{4,\text{PrpC}}$, $k_{5,\text{PrpC}}$, $k_{6,\text{PrpC}}$, $k_{7,\text{PrpC}}$, $k_{8,\text{PrpC}}$, $k_{9,\text{PduP}}$, $k_{10,\text{PrpC}}$   & $k_{2,\text{PrpC}}$, $k_{4,\text{PrpC}}$, $k_{5,\text{PrpC}}$, $k_{6,\text{PrpC}}$, $\text{factor}_{1,\text{PrpC}}$, $k_{8,\text{PrpC}}$, $K_{\text{eq},\text{PrpC}}$, $k_{\text{cat, PrpC}}^{\text{condensation}}$, $K^{\text{Propionyl-CoA}}_{\text{M}, \text{PrpC}}$, $K^{\text{Oxaloacetate}}_{\text{M}, \text{PrpC}}$ & $k_{1,\text{PrpC}} = \frac{k_{\text{cat, PrpC}}^{\text{condensation}}}{K^{\text{Propionyl-CoA}}_{\text{M}, \text{PrpC}}}$
            
            $k_{3,\text{PrpC}} = \frac{k_{\text{cat, PrpC}}^{\text{condensation}}}{K^{\text{Oxaloacetate}}_{\text{M}, \text{PrpC}}}$
            
            $k_{7,\text{PrpC}} = \text{factor}_{1,\text{PrpC}}k_{\text{cat, PrpC}}^{\text{condensation}}$
            
            $\text{factor}_{1,\text{PrpC}} > 1$

            $k_{9,\text{PrpC}} = \frac{k_{7,\text{PrpC}} k_{\text{cat, PrpC}}^{\text{condensation}}}{k_{7,\text{PrpC}}-k_{\text{cat, PrpC}}^{\text{condensation}}}$
            
            $k_{10,\text{PrpC}} = \frac{k_{1,\text{PrpC}}k_{3,\text{PrpC}}k_{5,\text{PrpC}}k_{7,\text{PrpC}}k_{9,\text{PrpC}}}{k_{2,\text{PrpC}}k_{4,\text{PrpC}}k_{6,\text{PrpC}}k_{8,\text{PrpC}}K_{\text{eq, PrpC}}}$ \\\hline
            

            

            
            
             
                         \caption{Table of free and leading kinetic parameters. See Archer et al. 2024 for Pdu and AckA leading and free variables.}
 \label{Table:Chapter6ConstrainedKineticswithFactors}
\end{longtable}}
\subsection*{Text}
\subsubsection*{Methods for assay of WT MCPs and PduL-deficient MCPs in {\it E. coli} lysate and 1,2-propanediol.}
\label{subsection:MethoddPduLMCPsWTMCPs}
Cell-free reactions were performed in 30 $\mu$L in 2 mL Eppendorf tubes and incubated at 30 °C as previously described \cite{Dudley2016}. The standard reaction contained the following components: 200 mM glucose, acetate salts (8mM magnesium acetate, 10mM ammonium acetate, 134 mM potassium acetate), 50 ug/mL kanamycin, 100 mM Bis-Tris, 1,2-propanediol (0.4\%), and 20$\mu$M Ado B12. All reagents and chemicals were purchased from Sigma Aldrich.

Extract concentration for all CFME reactions was 10 mg/mL total protein. The relative levels of each MCP were adjusted to maintain a total MCP concentration of .067 mg/mL. Reactions were quenched by precipitating proteins using 30 $\mu$L of 10\% trichloroacetic acid and centrifuging at 21,000 x g for 10 minutes at 4 °C. The resulting supernatant was then stored at -80 °C until analysis by HPLC.
\end{document}